\documentclass[10pt,graphicx,subfigure,axodraw,cleverref]{article}
\usepackage{listings}
\usepackage{xcolor}
\usepackage{soul}
\usepackage[T1]{fontenc}
\usepackage{bm}
\usepackage{ae,aecompl}
\usepackage{tabularx}
\usepackage{multirow}
\usepackage{cancel}
\usepackage{amsfonts}
\usepackage{amssymb}
\usepackage{epsfig}
\usepackage{float}
\usepackage{amsmath}
\usepackage{amsthm}
\usepackage{slashed}
\usepackage[title]{appendix}
\restylefloat{table}
\usepackage{tabulary}
\usepackage{hyperref}
\usepackage{cleveref}
\usepackage[parfill]{parskip}
\usepackage{hyperref}
\usepackage{titlesec}
\usepackage{enumitem}
\usepackage{tikz}
\usetikzlibrary{shapes.geometric, arrows}
\DeclareMathOperator{\Tr}{Tr}

\titleformat{\paragraph}
{\normalfont\normalsize\bfseries}{\theparagraph}{1em}{}
\titlespacing*{\paragraph}
{0pt}{3.25ex plus 1ex minus .2ex}{1.5ex plus .2ex}

\hypersetup{
	colorlinks=true,
	linkcolor=red,
citecolor=blue,
	filecolor=magenta,
	urlcolor=green,
	pdftitle={Overleaf Example},
}

\newcommand{\beqn}{\begin{eqnarray}}
	\newcommand{\eeqn}{\end{eqnarray}}

\newcommand{\sign}{\text{sgn}}

\newcounter{subeq}
\newcommand{\stags}{
	\addtocounter{equation}{+1}
	\setcounter{subeq}{0}
}
\newcommand{\stag}{%
	\addtocounter{subeq}{1}%
	\theequation\alph{subeq}%
}
\DeclareMathAlphabet{\mathsfit}{T1}{\sfdefault}{\mddefault}{\sldefault}
\SetMathAlphabet{\mathsfit}{bold}{T1}{\sfdefault}{\bfdefault}{\sldefault}
\newcommand{\vect}[1]{\boldsymbol{#1}}
\DeclareMathAlphabet{\mathpzc}{OT1}{pzc}{m}{it}
\DeclareMathAlphabet{\mathbfsf}{\encodingdefault}{\sfdefault}{bx}{sl}

\definecolor{forestgreen(web)}{rgb}{0.13, 0.55, 0.13}
\hypersetup{pdfpagemode=UseNone}
\begin{document}
{\small
\begin{titlepage}

\author{\bf{Azadan Bhagwagar}$^a$\footnote{Email: b00074086@aus.edu} ~and \bf{Raza M. Syed}$^{b,c}$\footnote{
Email: rsyed@aus.edu}\\~\\
$^{a}$\textit{\normalsize Department of Computer Science and Engineering, American University of Sharjah, P.O. Box 26666, Sharjah, UAE} \\
$^{b}$\textit{\normalsize Department of Physics, American University of Sharjah, P.O. Box 26666, Sharjah, UAE\footnote{Permanent address}} \\
$^{c}$\textit{\normalsize Department of Physics, Northeastern University, Boston, MA 02115-5000, USA} }
\title{\bf{Using C++ to Calculate {$\mathsf{SO(10)}$} Tensor Couplings}}

\maketitle

\begin{abstract}
\noindent Model building in $\mathsf{SO(10)}$, which is the leading grand unification framework, often
involves large Higgs representations and their couplings. Explicit calculations of such
couplings is a multi-step process that involves laborious calculations that are time-consuming and error-prone, an issue which only grows as the complexity of the coupling increases. Therefore, there exists an opportunity to leverage the abilities of computer software in order to algorithmically perform these calculations on-demand. This paper outlines the details of such a software, implemented in C++ using in-built libraries. The software is capable of accepting invariant couplings involving an arbitrary number of $\mathsf{SO(10)}$ Higgs tensors, each having up to 5 indices. The output is then produced in \LaTeX, so that it is universally readable and sufficiently expressive. Through the use of this software, $\mathsf{SO(10)}$ coupling analysis can be performed in a way that minimizes calculation time, eliminates errors, and allows for experimentation with couplings that have not been computed before in the literature. Furthermore, this software can be expanded in the future to account for similar Higgs-Spinor coupling analysis, or extended to include further $\mathsf{SO(N)}$ invariant couplings.
\end{abstract}

\end{titlepage}

\section{Introduction}

 The Standard Model of particle physics, including the strong and electroweak interactions comprising on the group $\mathsf{SU(3)_C  \times SU(2)_L \times U(1)_Y}$, is highly successful \cite{Weinberg:1967tq,Glashow:1961tr,Salam:1968rm,Gross:1973id,Politzer:1973fx}. One of the important challenges for particle physicists is to determine the underlying scheme where these forces are all unified. The aim of grand unified theories is three-fold: (1) to accomplish these unifications; (2) to provide an understanding of the three generations of quarks and leptons; (3) to provide an explanation of the hierarchy of their masses and of other properties. A more ambitious goal is to extend these ideas to encompass gravity, which requires the framework of superstring theory.

Grand unified models based on the gauge group $\mathsf{SO(10)}$ \cite{Georgi:1974my,Fritzsch:1974nn} have the most desirable features. They  provide a framework for the unification of electroweak and strong interactions. They also allow for all the quarks and leptons of one generation to reside in a single $\mathsf{16}-$plet irreducible spinor representation. Additionally, this complex representation contains a right-handed singlet state, needed for the generation of neutrino masses via the seesaw mechanism. $\mathsf{SO(10)}$ models also solve - in a relatively natural way - the doublet-triplet splitting  problem without fine tuning. Next, they possess  gauge interactions that conserve parity. $\mathsf{SO(10)}$ are aslo free from all gauge anomalies. Supersymmetric $\mathsf{SO(10)}$  models have the additional feature that they manage to unify the gauge couplings, and solve the hierarchy problem.

The Higgs sector of $\mathsf{SO(10)}$ models is largely unconstrained consisting of numerous possible representations. An avenue to the computation of the $\mathsf{SO(10)}$ couplings is via the decomposition of these in terms of $\mathsf{SU(5)}\times \mathsf{U(1)}$ invariant couplings
and then the further decomposition of them in terms of the $\mathsf{SU(3)_C\times SU(2)_L\times U(1)_Y}$
invariant couplings. The formalism to accomplish this was carried out in \cite{Nath:2001uw,Nath:2001yj,Syed:2004if,Syed:2005gd}.
However, the explicit analyses of the couplings can be laborious and time consuming.
For instance the complexity of computation involving large Higgs representations can
be seen in the works of \cite{Aboubrahim:2020dqw,Aboubrahim:2021phn,Nath:2015kaa}. Thus various models employ elaborate Higgs representations to break the $\mathsf{SO(10)}$ grand unified theory (GUT) symmetry down to the Standard Model (SM) gauge group $\mathsf{SU(3)_C\times SU(2)_L\times U(1)_Y}$. These consist of both small and large Higgs representations of $\mathsf{SO(10)}$ such as $\mathsf{10},~\mathsf{16}+\overline{\mathsf{16}},~\mathsf{45},~\mathsf{54},~\mathsf{120},~\mathsf{126}+\overline{\mathsf{126}},~\mathsf{144}+\overline{\mathsf{144}},
~ \mathsf{210}$ and $\mathsf{560}+\overline{\mathsf{560}}$. This enormous freedom of choosing symmetry breaking patterns allows one to construct $\mathsf{SO(10)}$ models with natural splitting of Higgs doublets and Higgs triplets to accomplish electroweak symmetry breaking
\cite{Aboubrahim:2020dqw,Aboubrahim:2021phn,Nath:2015kaa,Babu:2011tw} and models with a one-scale breaking of $\mathsf{SO(10)}$ GUT symmetry \cite{Babu:2005gx, Babu:2006rp, Nath:2005bx, Nath:2007eg,Nath:2009nf}. However, there are restrictions  on the Higgs content of a GUT model, such as
the strict proton decay limits \cite{Nath:1985ub,Nath:2006ut}. There are two commonly used symmetry breaking paths: one through the $\mathsf{SU(5)}\times \mathsf{U(1)}$ maximal subgroup \cite{Nath:2001uw,Nath:2001yj,Syed:2004if,Syed:2005gd,Mohapatra:1979nn,GrootNibbelink:2000hu} and the other through the $\mathsf{SU(4)\times SU(2)_L\times SU(2)_R}$  maximal subgroup \cite{Aulakh:2002zr,Aulakh:2004hm}.

The Higgs-Higgs interactions, appearing in supersymmetric and non-supersymmetric $\mathsf{SO(10)}$ models, are necessary to break the GUT and electroweak symmetries. Additionally, a thorough study of higher dimensional operators arising from three point, four point and higher Higgs-Higgs Interactions and matter-Higgs interactions is necessary to explore physics beyond the SM. For example, from matter-Higgs interactions, a top-down approach \cite{Nath:2015kaa} has been used to find dimension five, seven and nine $B-L=-2$  operators within the supersymmetric  $\mathsf{SO(10)}$  grand unification framework. This is in stark contrast to the bottom-up approach that exists prominently in the literature. These $B-L$ violating operators are important in the investigation of seesaw neutrino masses, baryogenesis, proton decay and $n-\overline{n}$  oscillations.

In this paper, we use the techniques (see appendix \ref{su(n) reducible tensors}) developed in \cite{Nath:2001uw,Nath:2001yj,Syed:2005gd} for the analysis of such $\mathsf{SO(2N)}$ invariant interactions which allows a full exhibition of the $\mathsf{SU(N)}$ invariant content of tensor representations. In particular, in this paper,  we focus on the analytic determination of $\mathsf{SO(10)}$  tensor interactions in terms of irreducible $\mathsf{SU(5)}$ fields. It would then be very straightforward to expand all the $\mathsf{SU(5)}$ invariants in terms of SM group invariants using the particle assignments. We wish to point out that our approach here is field theoretic rather than group theoretic \cite{Slansky:1981yr,Anderson:1999em} or other technique \cite{He:1990jw,Fukuyama:2004ps}. Our method is specially suited for the computation of $\mathsf{SO(N)}$ tensor couplings.

Manually preforming such tensor calculations can be a time-consuming endeavor that can very easily lead to errors. There are numerous intermediate calculations, each posing the risk of making an error, which may propagate throughout, affecting the final result. Therefore, there exists a compelling argument to use computer software to perform these calculations automatically, as proposed in this paper.

This C++ program allows users to perform Higgs-Higgs $\mathsf{SO(10)}$ coupling calculations rapidly and automatically, reducing the time needed to complete them and eliminating calculation errors.

With this program, the user is able to enter a coupling via a simple text-based interface. Next, the user is provided with the final normalized output, in a \LaTeX format, which can easily be added to a user’s publication. Further, the user is allowed to enter an arbitrary number of tensors in an $\mathsf{SO(10)}$ invariant  and can name their indices and tensors as needed. The maximum number of indices a single tensor can  have is 5.

Furthermore, to ensure the correctness of the algorithm when applied to couplings not previously attempted in the literature, manual hand calculations were evaluated for certain terms and were successfully matched to the program output.

This program can be easily extended to account for Higgs-Spinor couplings \cite{Cardoso:2015gfa}, as well as other $\mathsf{SO(N)}$ couplings.

The code is available to download \href{https://github.com/AHB99/tensor-coupling-program/releases}{here}.

There exists a variety of software packages for Lie algebra related computations used in grand unified models that are based on various classical and exceptional groups.  Here are some examples. LieART 2.0 \cite{ Feger:2019tvk} uses a Mathematica library to perform various  computations such as tensor product decomposition and subalgebra branching of irreducible representations. SARAH 4 \cite{Staub:2013tta} also exploits a Mathematica library that carries out calculations used in the study of supersymmetric and non-supersymmetric grand unified models, particularly, the full two-loop renormalization group equations for a supersymmetric theory. CleGo \cite{Horst:2010qj} package uses OCaml programming language to determine Clebsch–Gordan coefficients of  irreducible tensor product representations of Lie algebras $A-G$. FeynRules 2.0 \cite{Alloul:2013bka} is a Mathematica package that derives Feynman rules from the Lagrangian  of the Standard Model, Minimal Supersymmetric Standard model and their numerous extensions. These software packages rely extensively on standard group theoretic methods and display different degrees of generality.  As mentioned earlier, our C++ code uses the more intuitive field theoretic formalism for the analytic decomposition of the $\mathsf{SO(10)}$ invariants  of arbitrary order in terms of $\mathsf{SU(5)}$ invariants exhibiting their precise tensorial structure and Clebsch–Gordan coefficients. Our computer program is complementary to the existing software packages.

The paper is organized as follows. Section \ref{materials and methods} contains a detailed breakdown of the computer algorithm to provide readers with an overview of how the code is designed.  Section \ref{results} provides sample calculation output and Section \ref{discussion} discusses the significance of our algorithm. To fully appreciate and understand the properties associated with the $\mathsf{SO(N)}$ groups, we provide a thorough presentation of these groups in the appendices. Specifically, we discuss vector and tensor representations of $\mathsf{SO(N)}$ group and aspects of $\mathsf{SO(N)}$ gauge theory in Appendix \ref{so(n) group}. $\mathsf{SO(2N)}$ group algebra in $\mathsf{SU(N)}$ basis, branching rules for $\mathsf{SO(2N)}$ into $\mathsf{SU(N)}\times \mathsf{U(1)}$ irreducible representations and its specialization to $\mathsf{SO(10)}$ case are explained in Appendix \ref{so(2n) group in u(n) basis}. We show explicitly the technique to decompose $\mathsf{SO(2N)}$ tensor invariants in terms of $\mathsf{SU(N)}$ tensor invariants in Appendix \ref{su(n) reducible tensors}. This technique (The Basic Theorem) uses a unique set of reducible $\mathsf{SU(N)}$ tensors in terms of which the $\mathsf{SO(2N)}$ invariants have a straight forward decomposition. The Basic Theorem is specially useful for couplings involving large tensor representations and is central to the computation of any $\mathsf{SO(N)}$ invariant couplings. $\mathsf{SO(10)}$ tensors expressed in terms of $\mathsf{SU(5)}$ irreducible tensors with canonically normalized kinetic energy terms are exhibited in Appendix \ref{so(10) tensors in terms of su(5) tensors}. In Appendix \ref{singlets,doublets and triplets of SM}, we identify $\mathsf{SU(3)_C}\times \mathsf{SU(2)_L}\times U(1)_Y$ singlets, $\mathsf{SU(2)_L}$ doublets and $\mathsf{SU(3)_C}$ triplets in $\mathsf{SU(5)}$ fields.

\section{Materials and Methods}\label{materials and methods}

The program evaluates tensor couplings from input to their final normalized form. Internally, it is executed through various functions of the Product Resolver class, which stores all the intermediate terms during the calculation. It is based on the algorithm developed in Appendix \ref{su(n) reducible tensors}.

In brief, first the user inputs the original tensor coupling. This coupling is expanded to all its unsimplified reducible tensor terms. Then, all the reducible tensor terms are simplified by reordering the indices, accounting for anti-symmetry, and renaming indices if valid. Next, each reducible tensor term is substituted for its corresponding irreducible tensor expression. This expression is then expanded out fully. All Levi-Civita tensors are evaluated, following which the expression is simplified using the Kronecker Deltas and further properties of irreducible tensors. Then, all simplified irreducible tensor terms are further simplified by renaming if valid. Finally, these terms are substituted for their related normalized tensor term expressions, and the final expression is determined, to be output in \LaTeX format. See Figure \ref{fig:M1}.

\tikzstyle{startstop} = [rectangle, draw, fill=red!20,text width=5em, text centered, rounded corners, minimum height=4em]
\tikzstyle{io} = [trapezium, trapezium left angle=70, trapezium right angle=110,  minimum height=0.75cm, draw, fill=blue!20, text width=5em, text badly centered]
\tikzstyle{process} = [rectangle, draw, fill=orange!20,text width=7em, text centered, minimum height=3em]
\tikzstyle{decision} = [diamond, minimum width=3cm, minimum height=1cm, text centered, draw=black, fill=green!30]
\tikzstyle{arrow} = [thick,->,>=stealth]
\begin{figure}[hbt!]
\centering
\begin{tikzpicture}[node distance=2cm]
\node (start) [startstop] {Start};
\node (in1) [io, below of=start] {User input};
\node (pro1) [process,  right of=in1, node distance=3.5cm, fill=orange!50] {Expand to unsimplified reducible form};
\node (pro2) [process, right of=pro1, node distance=3.5cm, fill=orange!50] {Simplify reducible tensors};
\node (pro3) [process, right of=pro2, node distance=3.5cm, fill=orange!50 ] {Convert tensors to irreducible forms};
\node (pro4) [process, below of=pro3] {Expand expression};
\node (pro5) [process, left of=pro4, node distance=3.5cm] {Evaluate Levi-Civita and Kronecker delta symbols};
\node (pro6) [process, left of=pro5, node distance=3.5cm] {Simplify expression};
\node (pro7) [process, below of=pro6, fill=orange!50] {Canonically normalize all irreducible tensors};
\node (out1) [io, left of=pro7, node distance=3.5cm, fill=blue!35] {Output in \LaTeX ~format};
\node (stop) [startstop, below of=out1] {End};
\draw [arrow] (start) -- (in1);
\draw [arrow] (in1) -- (pro1);
\draw [arrow] (pro1) -- (pro2);
\draw [arrow] (pro2) -- (pro3);
\draw [arrow] (pro3) -- (pro4);
\draw [arrow] (pro4) -- (pro5);
\draw [arrow] (pro5) -- (pro6);
\draw [arrow] (pro6) -- (pro7);
\draw [arrow] (pro7) -- (out1);
\draw [arrow] (out1) -- (stop);
\end{tikzpicture}
\caption{Flowchart showing the breakdown of the computer algorithm to compute $\mathsf{SO(10)}$ tensor invariants.} \label{fig:M1}
\end{figure}
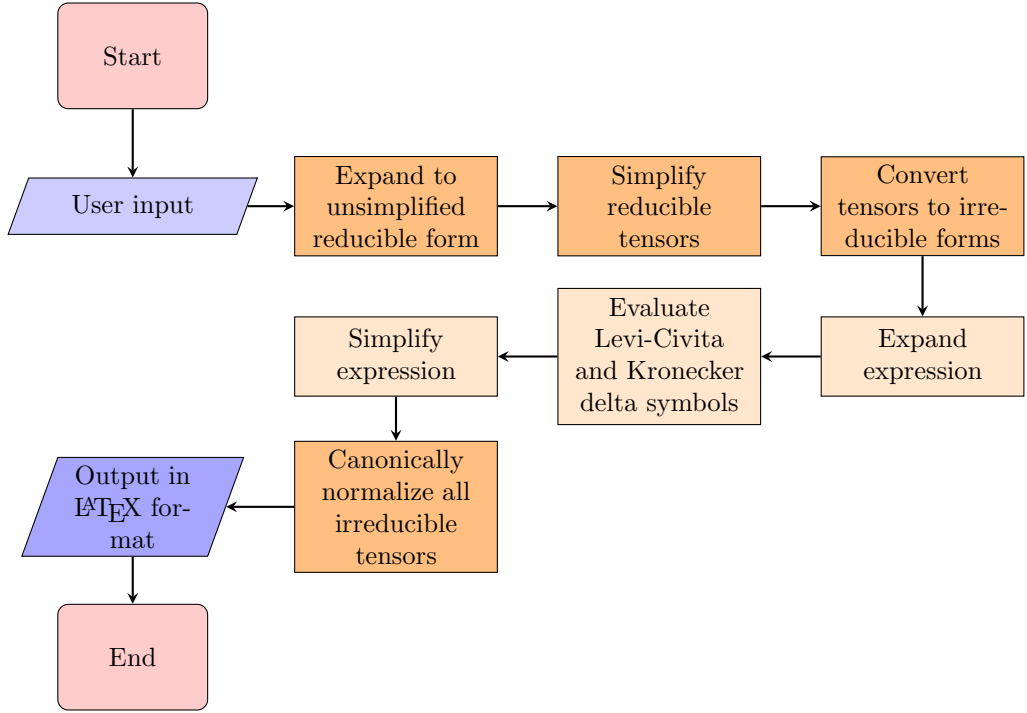

\subsection{\colorbox{blue!20}{Input Phase}}
The user inputs a tensor coupling in the format ``X\_\{/i/j/k\}Y\_\{/i/j/k\}...'' where ``X'' and ``Y'' are labels for tensors and ``/i/j/k'' are labels for index names. The input is parsed using the standard C++ regex library. A regex expression is hard-coded to tokenize the tensors and the indices each. The Product Resolver class uses the following method:

\begin{lstlisting}
	void parseInput(std::string input);
\end{lstlisting}

\subsection{\colorbox{orange!50}{Generation Phase}}

All the initial generated reducible tensor terms can be determined by the following pattern. If one were to take a binary number where each binary digit corresponds to each unique index of the input tensor, and then increment the binary number from 0 to $2^n-1$ (where n is the number of unique indices), each binary number would correspond to a generated term. A term can be developed by the rule where for each binary digit, the corresponding index in that position will have its first occurrence barred if the binary digit is 1, or unbarred if it is 0. The function is as follows:

\begin{lstlisting}
	void generateAllUnsimplifiedReducibleTensorTerms();
\end{lstlisting}

\subsection{\colorbox{orange!50}{Simplification of Reducible Tensor Term Expression Phase}}

Now we have all the raw, generated reducible tensor terms, we may simplify them. This phase is based on the idea of tensor terms having similar “structure”. Two terms have the same “structure” if, assuming they have the same number of tensors, every tensor in every corresponding position has the same label and number of barred and unbarred indices. Only once two tensor terms have the same “structure” can the possibility of a rename operation leading to equality between them be considered. Currently, the generated tensor terms have their indices unordered, with no grouping of barred and unbarred indices. So firstly, each reducible tensor term has its indices reordered, accounting for anti-symmetry, by the below algorithm. The function is as follows:

\begin{lstlisting}
	void simplifyReducibleTensorTerms();
\end{lstlisting}

\subsubsection{\colorbox{orange!20}{Reordering of Reducible Tensor Term Indices}}

The algorithm is a modified implementation of bubble sorting, accounting for grouping of barred/unbarred indices, as well as anti-symmetry, acting on each Tensor. By the end, the tensors will have their barred indices grouped to the left, and unbarred indices grouped to the right, with each group ordered in ascending order of index names.

\subsubsection{\colorbox{orange!20}{Sorting Tensor Terms}}

The tensor terms need to now be sorted in a uniform manner so that it becomes obvious to identify terms with the same “structure”, as defined above. We require sorting to be done first by tensor labels (alphabetically), then by number of indices, and then by number of barred indices

After sorting, the tensor terms are transferred to a separate container of simplified reducible tensor terms, checking to see if an addition or rename operation can be done with any existing simplified term first. The rename operation is detailed below

\subsubsection{\colorbox{orange!20}{Renaming of Reducible Tensor Term Indices (Ignoring Commutativity)} \\\colorbox{orange!20}{Algorithm}}

The algorithm takes two tensor terms, an Attempt Term that we are will be manipulating and renaming, as well as a Source Term, which will not be modified and is what the Attempt Term will be manipulated to match, if possible. It outputs a Renamed Term if the operation was successful, that, once its indices are reordered, will be identical to the Source Term (excluding coefficients, which are to be added). Also, the algorithm makes use of the concept of “zones”, which are either unbarred or barred groups of indices in a tensor. Tensor terms with the same “structure” will also have the same “zones”, and hence when deciding whether a rename operation is possible, we only have to focus on mapping each “zone” of the Attempt Term to the Source Term.

The algorithm assumes the Source Term and Attempt Term have their indices grouped as barred/unbarred and each group is ordered ascendingly, and that their tensors are sorted. The handling of commutativity, for tensor terms that retain identical structure upon permutations, is not considered in this algorithm.

\subsubsection{\colorbox{orange!20}{Renaming of Reducible Tensor Term Indices (Accounting for}\\ \colorbox{orange!20}{Commutativity) Algorithm}}

This algorithm deals with the concept of “ambiguity” of tensor terms. This is a characteristic of a tensor term where it is possible to permute the ordering of the tensors within the term (due to the commutativity of tensors) and still retain identical “structure” (as defined above). This requires special consideration as the above Renaming Algorithm focuses on trying to equate corresponding “zones” in the Source and Attempt Terms based on position, so when such permutations are possible, the algorithm may fail to find a legal rename mapping for one arbitrary permutation while it may have been possible in another.

Hence, we need to exhaust every single possible permutation of “ambiguous” tensor terms. Specifically, we deal with the concept of “ambiguous zones”, which are groups of tensors within a tensor term that cause the “ambiguity” property (there can be multiple “ambiguous zones” within a single tensor term)

\textbf{Ambiguity Detection:}

During the rename operation, if a Source Term (and hence it’s matching structure Attempt Term) is found to be ambiguous, then we must generate all possible permutations of the ambiguous zones of the Attempt Term, attempt to rename them to the Source Term, and if even 1 successful rename exists, that must be chosen and applied. Only failing this do we conclude no possible rename exists. To generate all these permutations of the Attempt Term, we first must locate all such zones.

\textbf{Finding all ambiguous zone locations and sizes:}

Once we have the locations and sizes of all ambiguous zones of a tensor, we can generate all the permutations of the locations of all the ambiguous zones. This will be stored in a vector (the overall container) of vectors (for each zone) of vectors (for each permutation) of integers (the locations). For uniformity, we will consider each zone’s locations start from 0, and offset it with the actual location later.

\textbf{Finding all permutations of the locations for all ambiguous zones:}

Now, given this information, we can generate the tensor terms. However, we must consider the fact that for tensor terms with multiple ambiguous zones, we must generate all combinations of all the possible permutations of the ambiguous zones.
For example, if a tensor term has 2 ambiguous zones, we must not permute the first zone and exhaust all the permutations of the second zone. Once we have exhausted them, only then can we proceed with the next permutation of the first zone, but again pause as we exhaust all the permutations of the second zone. This logic is expanded to the general case in the algorithm below.

\textbf{Finding all permutations of the tensor terms for all ambiguous zones:}

Now we permute the actual tensor terms in the locations to achieve all possible permutations to test out.

\subsubsection{\colorbox{orange!20}{Simplification of Reducible Terms Algorithm}}

Now, we put all the algorithms together to simplify the raw, generated reducible tensor terms. The Product Resolver class contains within itself a vector of the raw, unsimplified terms, as well as the simplified terms.

\begin{enumerate}

	\item Reorder the indices of all terms, alphabetically.
	\item Sort the tensors of all terms, as defined above.
	\item Iterate over all unsimplified terms:
	\begin{enumerate}
		\item Iterate over all simplified terms:
		\begin{enumerate}
				\item If the current unsimplified (Attempt Term) and simplified (Source Term) terms have the same structure, attempt to rename them.
				\item If the rename is successful, reorder the indices of the Re-named Term.
				\item Check if the Renamed Term is identical to the Source Term. (Theoretically, this must always be true, as else the rename operation would not have been successful). If so, add the coefficient of the Renamed Term into the Source Term. Break iteration over the simplified terms.
				
		\end{enumerate}
		\item If all simplified terms were iterated over with no successful matches, add the current unsimplified term to the vector of simplified terms. (It could not be simplified with any existing simplified term).
		
	\end{enumerate}
	\item Erase all simplified tensor terms with a coefficient of 0
\end{enumerate}

By the end of this stage, we have a vector of all simplified, reducible tensor terms.

\subsection{\colorbox{orange!50}{Reduction of Reducible Tensor Terms to Irreducible Tensor Terms Phase}}

In this phase, we will first substitute the simplified, reducible tensors with their corresponding (pre-determined) expressions written in terms of irreducible tensors, Kronecker Deltas, and Levi-Civita tensors. We will then expand out these expressions, expand the Levi-Civita tensors, and then simplify the expression using the properties of Kronecker Deltas, Symmetric-Asymmetric irreducible tensors, and the ability to rename the indices of irreducible tensors and add them.

This phase will make use of the Math Expression class, which resembles a typical algebraic expression, composed of algebraic terms from the Math Expression Term class. The Math Expression Term will serve as a more complex form of the Tensor Term class used in previous phases, as it now will contain various mathematical objects. There are also self-explanatory classes for Deltas, Levi-Civitas, Irreducible Tensors, and Coefficients. The function is as follows:

\begin{lstlisting}
	void fullyReduceTensorTerms();
\end{lstlisting}

\subsubsection{\colorbox{orange!20}{Substitutions of Reducible Tensors for Expressions with Irreducible} \\\colorbox{orange!20}{Tensors Algorithm}}

\begin{enumerate}
	\item Given the Source Reducible Tensor, find the number of indices it contains.
	\item Based on the number of indices, choose the correct substitution sub-category
	\item In the chosen substitution sub-category, given the Source Reducible Tensor, find the number of barred indices it contains
	\item Based on the number of barred indices of this sub-category (of number of indices), return the corresponding substituted Math Expression

\end{enumerate}

The terms are then expanded by multiplication.

\subsubsection{\colorbox{orange!20}{Multiplication of Levi-Civita Tensors Algorithm}}

\begin{enumerate}
\item Iterate for every Math Expression Term in a Math Expression:
	\begin{enumerate}
		\item If there is more than 1 Levi-Civita:
		\begin{enumerate}
			\item If there are 3 or more Levi-Civitas, group the Levis based on number of common indices
			\item Create a temporary Math Expression for the full Delta Expression
			\item Iterate while there exists atleast one pair of Levis:
		\begin{enumerate}
			\item Expand the pair into this temporary.
			\item Delete the pair
		\end{enumerate}
		\item Multiply the original term with the full Delta Expression, and append this to the final result Math Expression.
		\item Set the original term to 0.
		\item Delete all terms with 0 coefficient
		\end{enumerate}
	\end{enumerate}
\end{enumerate}

\subsubsection{\colorbox{orange!20}{Simplification of Expression by Kronecker Deltas Algorithm}}

\begin{enumerate}
	\item Iterate for every term:
	\begin{enumerate}
		\item Sum over the indices of the Deltas.
		\item Check for possibility of the deltas cancelling out the term. If so, move on to the next term.
		\item Rename the indices of the irreducible tensors by using the Deltas.
		\item Solve identical Deltas and modify the coefficient accordingly.
	\end{enumerate}
	\item Erase all terms with coefficient 0.
\end{enumerate}

\subsubsection{\colorbox{orange!20}{Simplification of Expression by Renaming Indices of Irreducible Tensors}\\
 \colorbox{orange!20}{(Accounting for Commutativity)}}

\begin{itemize}
	\item While before the concept of “zones” related to barred versus unbarred indices, now “zones” relates to upper versus lower indices.
	\item While before the concept of “structure” (and hence “ambiguity”) related to whether 2 reducible tensors have the same label, number of indices, and number of barred indices, now “structure” relates to whether 2 irreducible tensors have the same bar state, field, symmetric property, and number of upper and lower indices.
	\item While before the sorting of the reducible tensor terms was by (in ascending order) label, then by number of indices, and then by number of barred indices, now the sorting of the irreducible tensor terms is (in ascending order) by field, then by symmetric property, then by bar state, then by number of upper indices, and then by number of lower indices.
	\item A Single Levi-Civita tensor may survive after simplification and also must be considered in the renaming process. It is considered as its own unique “zone”.
\end{itemize}

\subsubsection{\colorbox{orange!20}{Overall Reduction Phase}}

\begin{enumerate}
	\item 	Iterate for every Math Expression Term:
	\begin{enumerate}
		\item Substitute the reducible tensor terms for the appropriate Math Expressions containing irreducible tensors, Levi-Civitas, and Deltas.
		\item Multiply and expand all the substituted expressions.
		\item Multiply the Levi-Civitas of each term and expand the expression.
		\item Simplify the overall expression by the properties of Kronecker Deltas.
		\item Simplify the overall expression by the property of Symmetric and Asymmetric Irreducible tensors that share 2+ indices.
		\item Simplify the overall expression by the renaming the indices of the expression.
		\item Multiply the overall expression with the original coefficient from the reducible tensor term.
		\item Append this result to the overall Math Expression of the final result.
	\end{enumerate}
	\item Simplify this final result by renaming the indices of the expression
	\begin{enumerate}
		\item (This step is repeated because before it was only considered within each expression that arose from a reducible tensor term, whereas now it is considering all simplified expressions from all the reducible tensor terms.)
	\end{enumerate}
	\item Multiply the overall expression with the coefficient that arose from expanding the initial user input to the reducible tensors (of the form $ \frac{1}{2^(Number Of Indices)} $  )
\end{enumerate}

\subsection{\colorbox{orange!50}{Normalization of Irreducible Tensors to Normalized Tensors Phase}}

In this final phase, we substitute the irreducible tensors with the corresponding normalized tensors, including their coefficients. This is similar to the step in the previous phase where we substituted the reducible tensors for their respective expressions. Each substitution is a unique function due to its coefficients, however the general procedure is outlined below.

\begin{enumerate}
	\item For a given irreducible tensor of a Math Expression Term:
	\item Create a normalized tensor and transfer the indices exactly as they are from the irreducible tensor.
	\item Multiply the whole Math Expression Term by the required coefficient.
	\item Erase the Irreducible Tensor from the Math Expression Term.
\end{enumerate}

This process is repeated for every irreducible tensor in every Math Expression Term in the result expression. No further simplification occurs here because all simplification has been completed before this. Hence, we arrive at our final result. The function is as follows:

\begin{lstlisting}
	void normalizeIrreducibleTensorTerms();
\end{lstlisting}

\subsection{\colorbox{blue!35}{Output Phase}}

Every component in the result, be it a Delta, Normalized tensor, Coefficient, etc., has a representation in \LaTeX designed for it, and hence printing a term simply calls all these functions in turn, and printing an expression calls the printing of each term. The function is as follows:

\begin{lstlisting}
	void printMathExpressionAsLatex();
\end{lstlisting}

\section{\colorbox{blue!20}{Results}}\label{results}

The computer program outputs results in \LaTeX~format that can be easily included in a publication, as it has been done in Sections \ref{quadratic couplings}-\ref{quartic couplings}. The original $\mathsf{SO(10)}$ coupling inputs are mentioned along with their respective processed results in their final format. Some of these couplings have not previously been demonstrated in the literature, proving the generalized capabilities of the software.

The normalizations of $\mathsf{SU(5)}$ fields are displayed in Appendix \ref{so(10) tensors in terms of su(5) tensors}. Here the $H-$fields represent $\mathsf{SU(5)}$ irreducible tensors with canonically normalized kinetic energy terms.

\subsection{Quadratic Couplings}\label{quadratic couplings}

\subsubsection{{\boldmath$\mathsf{126}\times\overline{\mathsf{126}}$}}

$X_{\mu1\mu2\mu3\mu4\mu5}X_{\mu1\mu2\mu3\mu4\mu5}= $
\begin{flushleft}
	\sloppy
	$  +2H^{(\overline{126})}\overline{H}^{(126)} +H^{mp (\overline{126})}_{(S)}H^{(126)}_{mp (S)} +H^{pq (126)}H^{(\overline{126})}_{pq} +\frac{1}{6}H^{ijk (\overline{126})}_{lm}H^{lm (126)}_{ijk} +H^{jk (126)}_{m}H^{m (\overline{126})}_{jk} +2H^{k (\overline{126})}H^{(126)}_{k}  $
\end{flushleft}

\subsection{Cubic Couplings}

\subsubsection{{\boldmath$\mathsf{210} \times \mathsf{210} \times \mathsf{210}$}}

$ X_{\mu1\mu2\mu3\mu4}X_{\mu3\mu4\mu5\mu6}X_{\mu5\mu6\mu1\mu2}= $
\begin{flushleft}
	\sloppy
	$+\frac{1\sqrt{2}}{2}H^{p (210)}H^{(210)}_{n}H^{n (210)}_{p} -\frac{1\sqrt{15}}{5}H^{(210)}H^{p (210)}H^{(210)}_{p} -\frac{1\sqrt{6}}{48}\epsilon^{ijklo}H^{(210)}_{o}H^{m (210)}_{kln}H^{n (210)}_{ijm} +\frac{1\sqrt{2}}{12}\epsilon^{ijklo}H^{(210)}_{o}H^{(210)}_{im}H^{m (210)}_{jkl} +\frac{1\sqrt{6}}{48}\epsilon^{ijklo}H^{(210)}_{o}H^{(210)}_{kl}H^{(210)}_{ij} -\frac{1\sqrt{6}}{48}\epsilon_{ijmno}H^{o (210)}H^{ijk (210)}_{l}H^{lmn (210)}_{k} -\frac{1\sqrt{2}}{12}\epsilon_{ijmno}H^{o (210)}H^{lm (210)}H^{ijn (210)}_{l} +\frac{1\sqrt{6}}{48}\epsilon_{ijmno}H^{o (210)}H^{ij (210)}H^{mn (210)} +\frac{1\sqrt{3}}{6}H^{ijk (210)}_{l}H^{lm (210)}_{kn}H^{n (210)}_{ijm} +\frac{1}{6}H^{ikn (210)}_{l}H^{(210)}_{im}H^{lm (210)}_{kn} -\frac{1\sqrt{2}}{12}H^{m (210)}_{k}H^{ijk (210)}_{n}H^{n (210)}_{ijm} -\frac{1\sqrt{6}}{36}H^{m (210)}_{k}H^{ijk (210)}_{m}H^{(210)}_{ij} +\frac{1\sqrt{2}}{12}H^{l (210)}_{n}H^{ijm (210)}_{l}H^{n (210)}_{ijm} -\frac{1\sqrt{6}}{36}H^{ij (210)}H^{m (210)}_{n}H^{n (210)}_{ijm} +\frac{1\sqrt{2}}{18}H^{ik (210)}H^{m (210)}_{k}H^{(210)}_{im} -\frac{1}{6}H^{ik (210)}H^{jm (210)}_{kn}H^{n (210)}_{ijm} +\frac{5\sqrt{3}}{36}H^{kn (210)}H^{(210)}_{jm}H^{jm (210)}_{kn} +\frac{1\sqrt{3}}{12}H^{ijk (210)}_{l}H^{mn (210)}_{ij}H^{l (210)}_{kmn} -\frac{1\sqrt{15}}{20}H^{(210)}H^{km (210)}H^{(210)}_{km} +\frac{1\sqrt{3}}{36}H^{ij (210)}_{kl}H^{kl (210)}_{mn}H^{mn (210)}_{ij} -\frac{1\sqrt{2}}{6}H^{n (210)}_{i}H^{im (210)}_{kl}H^{kl (210)}_{mn} +\frac{1\sqrt{15}}{60}H^{(210)}H^{mn (210)}_{kl}H^{kl (210)}_{mn} -\frac{1\sqrt{3}}{18}H^{l (210)}_{m}H^{n (210)}_{i}H^{im (210)}_{ln} +\frac{7\sqrt{2}}{54}H^{m (210)}_{n}H^{i (210)}_{m}H^{n (210)}_{i} -\frac{1\sqrt{15}}{60}H^{(210)}H^{m (210)}_{l}H^{l (210)}_{m} -\frac{1\sqrt{15}}{30}H^{(210)}H^{(210)}H^{(210)} +\frac{1\sqrt{3}}{9}H^{ik (210)}_{jl}H^{lm (210)}_{kn}H^{jn (210)}_{im}  $
\end{flushleft}

\subsubsection{{\boldmath$\mathsf{120} \times \mathsf{126} \times \mathsf{210}$}}

$ X_{\mu1\mu2\mu3}Y_{\mu2\mu3\mu4\mu5\mu6}Z_{\mu1\mu4\mu5\mu6}= $
\begin{flushleft}
	\sloppy
	$  +\frac{1\sqrt{15}}{60}H^{(120)}_{mn}H^{q (210)}H^{mn (126)}_{q} +\frac{1\sqrt{5}}{20}H^{(120)}_{nq}H^{q (210)}H^{n (\overline{126})} +\frac{1\sqrt{30}}{480}\epsilon^{ijkop}H^{(120)}_{op}H^{n (210)}_{ilm}H^{lm (126)}_{jkn} +\frac{1\sqrt{10}}{480}\epsilon^{ijkop}H^{(120)}_{op}H^{(210)}_{lm}H^{lm (126)}_{ijk} +\frac{1\sqrt{15}}{120}\epsilon^{ijkop}H^{(120)}_{op}H^{n (210)}_{ijm}H^{m (\overline{126})}_{kn} +\frac{1\sqrt{5}}{240}\epsilon^{ijkop}H^{(120)}_{op}H^{(210)}_{im}H^{m (\overline{126})}_{jk} +\frac{1\sqrt{5}}{240}\epsilon^{ijkop}H^{(120)}_{op}H^{n (210)}_{ijk}H^{(126)}_{n} -\frac{1\sqrt{15}}{240}\epsilon^{ijkop}H^{(120)}_{op}H^{(210)}_{ij}H^{(126)}_{k} -\frac{1\sqrt{5}}{20}H^{(120)}_{nq}H^{n (210)}_{i}H^{iq (\overline{126})}_{(S)} +\frac{1\sqrt{30}}{120}H^{(120)}_{mn}H^{mn (210)}_{ij}H^{ij (126)} +\frac{1\sqrt{5}}{60}H^{(120)}_{nr}H^{n (210)}_{i}H^{ir (126)} +\frac{1\sqrt{6}}{40}H^{(120)}_{qr}H^{(210)}H^{qr (126)} -\frac{1\sqrt{10}}{40}H^{(120)}_{lm}H^{lm (210)}\overline{H}^{(126)} -\frac{1\sqrt{15}}{60}H^{op (120)}_{q}H^{q (210)}H^{(\overline{126})}_{op} +\frac{1\sqrt{15}}{30}H^{j (120)}H^{q (210)}H^{(126)}_{jq (S)} -\frac{1\sqrt{15}}{60}H^{o (120)}H^{q (210)}H^{(\overline{126})}_{oq} -\frac{1\sqrt{30}}{120}H^{ij (120)}_{k}H^{n (210)}_{ilm}H^{klm (\overline{126})}_{jn} +\frac{1\sqrt{10}}{240}H^{jn (120)}_{k}H^{(210)}_{lm}H^{klm (\overline{126})}_{jn} +\frac{1\sqrt{15}}{60}H^{il (120)}_{k}H^{n (210)}_{ilm}H^{km (126)}_{n} +\frac{1\sqrt{5}}{120}H^{il (120)}_{n}H^{n (210)}_{ilm}H^{m (\overline{126})} -\frac{1\sqrt{15}}{72}H^{il (120)}_{m}H^{(210)}_{il}H^{m (\overline{126})} -\frac{1\sqrt{30}}{360}H^{j (120)}H^{n (210)}_{ilm}H^{ilm (\overline{126})}_{jn} +\frac{1\sqrt{5}}{40}H^{n (120)}H^{(210)}_{lm}H^{lm (126)}_{n} +\frac{1\sqrt{15}}{60}H^{ij (120)}_{k}H^{mn (210)}_{il}H^{kl (126)}_{jmn} +\frac{1\sqrt{10}}{120}H^{jm (120)}_{k}H^{n (210)}_{l}H^{kl (126)}_{jmn} +\frac{1\sqrt{5}}{40}H^{jn (120)}_{m}H^{m (210)}_{l}H^{l (\overline{126})}_{jn} -\frac{1\sqrt{6}}{120}H^{jn (120)}_{l}H^{(210)}H^{l (\overline{126})}_{jn} -\frac{1\sqrt{30}}{120}H^{il (120)}_{k}H^{mn (210)}_{il}H^{k (\overline{126})}_{mn} -\frac{1\sqrt{10}}{120}H^{il (120)}_{m}H^{mn (210)}_{il}H^{(126)}_{n} +\frac{1\sqrt{15}}{180}H^{ln (120)}_{m}H^{m (210)}_{l}H^{(126)}_{n} +\frac{1\sqrt{5}}{60}H^{m (120)}H^{n (210)}_{l}H^{l (\overline{126})}_{mn} -\frac{1\sqrt{15}}{40}H^{i (120)}H^{n (210)}_{i}H^{(126)}_{n} -\frac{1\sqrt{2}}{20}H^{n (120)}H^{(210)}H^{(126)}_{n} +\frac{1\sqrt{15}}{360}\epsilon_{jlmno}H^{ij (120)}_{k}H^{lmn (210)}_{i}H^{ko (\overline{126})}_{(S)} -\frac{1\sqrt{5}}{240}\epsilon_{jlmno}H^{jn (120)}_{k}H^{lm (210)}H^{ko (\overline{126})}_{(S)} +\frac{1\sqrt{5}}{80}\epsilon_{jmnop}H^{jn (120)}_{l}H^{lm (210)}H^{op (126)} -\frac{1\sqrt{15}}{360}\epsilon_{jlmno}H^{j (120)}H^{lmn (210)}_{i}H^{io (\overline{126})}_{(S)} -\frac{1\sqrt{5}}{240}\epsilon_{jlmno}H^{j (120)}H^{lm (210)}H^{no (\overline{126})}_{(S)} -\frac{1\sqrt{15}}{360}\epsilon_{lmnop}H^{i (120)}H^{lmn (210)}_{i}H^{op (126)} +\frac{1\sqrt{5}}{240}\epsilon_{lmnop}H^{n (120)}H^{lm (210)}H^{op (126)} -\frac{1\sqrt{15}}{720}\epsilon_{jlmno}H^{i (120)}H^{lmn (210)}_{i}H^{jo (\overline{126})}_{(S)} -\frac{1\sqrt{30}}{30}H^{(120)}_{o}H^{o (210)}H^{(\overline{126})} +\frac{1\sqrt{15}}{240}\epsilon^{jklmo}H^{i (120)}_{jk}H^{n (210)}_{ilm}H^{(126)}_{no (S)} +\frac{1\sqrt{5}}{120}\epsilon^{jklmo}H^{i (120)}_{jk}H^{(210)}_{il}H^{(126)}_{mo (S)} -\frac{1\sqrt{5}}{240}\epsilon^{jklmo}H^{n (120)}_{jk}H^{(210)}_{lm}H^{(126)}_{no (S)} +\frac{1\sqrt{5}}{80}\epsilon^{klmop}H^{i (120)}_{km}H^{(210)}_{il}H^{(\overline{126})}_{op} +\frac{1\sqrt{15}}{360}\epsilon^{jklmo}H^{(120)}_{k}H^{n (210)}_{jlm}H^{(126)}_{no (S)} +\frac{1\sqrt{5}}{240}\epsilon^{jklmo}H^{(120)}_{k}H^{(210)}_{jl}H^{(126)}_{mo (S)} +\frac{1\sqrt{5}}{240}\epsilon^{klmop}H^{(120)}_{k}H^{(210)}_{lm}H^{(\overline{126})}_{op} -\frac{1\sqrt{15}}{360}\epsilon^{jlmop}H^{(120)}_{n}H^{n (210)}_{jlm}H^{(\overline{126})}_{op} +\frac{1\sqrt{15}}{120}H^{i (120)}_{jk}H^{mn (210)}_{il}H^{jkl (\overline{126})}_{mn} +\frac{1\sqrt{10}}{120}H^{m (120)}_{jk}H^{n (210)}_{l}H^{jkl (\overline{126})}_{mn} +\frac{1\sqrt{5}}{40}H^{i (120)}_{kl}H^{n (210)}_{i}H^{kl (126)}_{n} -\frac{1\sqrt{5}}{60}H^{n (120)}_{km}H^{m (210)}_{l}H^{kl (126)}_{n} -\frac{1\sqrt{6}}{120}H^{n (120)}_{kl}H^{(210)}H^{kl (126)}_{n} -\frac{1\sqrt{10}}{120}H^{i (120)}_{mn}H^{mn (210)}_{il}H^{l (\overline{126})} -\frac{1\sqrt{15}}{180}H^{i (120)}_{ln}H^{n (210)}_{i}H^{l (\overline{126})} -\frac{1\sqrt{5}}{60}H^{(120)}_{k}H^{n (210)}_{i}H^{ik (126)}_{n} -\frac{1\sqrt{30}}{120}H^{(120)}_{m}H^{mn (210)}_{il}H^{il (126)}_{n} -\frac{1\sqrt{15}}{40}H^{(120)}_{n}H^{n (210)}_{i}H^{i (\overline{126})} -\frac{1\sqrt{2}}{20}H^{(120)}_{l}H^{(210)}H^{l (\overline{126})} -\frac{1\sqrt{30}}{720}H^{i (120)}_{jk}H^{lmn (210)}_{i}H^{jk (126)}_{lmn} +\frac{1\sqrt{10}}{240}H^{n (120)}_{jk}H^{lm (210)}H^{jk (126)}_{lmn} +\frac{1\sqrt{5}}{60}H^{n (120)}_{kl}H^{lm (210)}H^{k (\overline{126})}_{mn} +\frac{1\sqrt{5}}{120}H^{i (120)}_{lm}H^{lmn (210)}_{i}H^{(126)}_{n} -\frac{1\sqrt{15}}{72}H^{n (120)}_{lm}H^{lm (210)}H^{(126)}_{n} +\frac{1\sqrt{30}}{360}H^{(120)}_{k}H^{lmn (210)}_{i}H^{ik (126)}_{lmn} +\frac{1\sqrt{5}}{40}H^{(120)}_{k}H^{lm (210)}H^{k (\overline{126})}_{lm} +\frac{1\sqrt{15}}{120}H^{(120)}_{m}H^{lm (210)}H^{(126)}_{l} -\frac{1\sqrt{30}}{720}H^{jk (120)}_{i}H^{i (210)}_{lmn}H^{lmn (\overline{126})}_{jk} +\frac{1\sqrt{5}}{60}H^{kl (120)}_{n}H^{(210)}_{lm}H^{mn (126)}_{k} +\frac{1\sqrt{15}}{120}H^{m (120)}H^{(210)}_{lm}H^{l (\overline{126})} +\frac{1\sqrt{15}}{120}H^{jk (120)}_{i}H^{in (210)}_{lm}H^{lm (126)}_{jkn} -\frac{1\sqrt{5}}{60}H^{kl (120)}_{m}H^{n (210)}_{l}H^{m (\overline{126})}_{kn} -\frac{1\sqrt{30}}{120}H^{l (120)}H^{in (210)}_{lm}H^{m (\overline{126})}_{in} +\frac{1\sqrt{15}}{240}\epsilon_{jkmno}H^{jk (120)}_{i}H^{imn (210)}_{l}H^{lo (\overline{126})}_{(S)} +\frac{1\sqrt{5}}{120}\epsilon_{jkmno}H^{jk (120)}_{i}H^{im (210)}H^{no (\overline{126})}_{(S)} -\frac{1\sqrt{30}}{30}H^{o (120)}H^{(210)}_{o}\overline{H}^{(126)} +\frac{1\sqrt{15}}{360}\epsilon^{klmno}H^{j (120)}_{ik}H^{i (210)}_{lmn}H^{(126)}_{jo (S)} -\frac{1\sqrt{15}}{720}\epsilon^{klmno}H^{(120)}_{i}H^{i (210)}_{lmn}H^{(126)}_{ko (S)} +\frac{1\sqrt{15}}{60}H^{j (120)}_{ik}H^{in (210)}_{lm}H^{klm (\overline{126})}_{jn} -\frac{1\sqrt{30}}{120}H^{j (120)}_{in}H^{in (210)}_{lm}H^{lm (126)}_{j} -\frac{1\sqrt{30}}{120}H^{j (120)}_{ik}H^{imn (210)}_{l}H^{kl (126)}_{jmn} +\frac{1\sqrt{15}}{60}H^{j (120)}_{im}H^{imn (210)}_{l}H^{l (\overline{126})}_{jn} -\frac{1\sqrt{15}}{60}H^{q (120)}_{op}H^{(210)}_{q}H^{op (126)} +\frac{1\sqrt{15}}{30}H^{(120)}_{k}H^{(210)}_{q}H^{kq (\overline{126})}_{(S)} -\frac{1\sqrt{15}}{60}H^{(120)}_{o}H^{(210)}_{q}H^{oq (126)} -\frac{1\sqrt{10}}{40}H^{lm (120)}H^{(210)}_{lm}H^{(\overline{126})} -\frac{1\sqrt{5}}{20}H^{lq (120)}H^{n (210)}_{l}H^{(126)}_{nq (S)} +\frac{1\sqrt{30}}{120}H^{lm (120)}H^{ij (210)}_{lm}H^{(\overline{126})}_{ij} +\frac{1\sqrt{5}}{60}H^{lr (120)}H^{j (210)}_{l}H^{(\overline{126})}_{jr} +\frac{1\sqrt{6}}{40}H^{qr (120)}H^{(210)}H^{(\overline{126})}_{qr} +\frac{1\sqrt{30}}{480}\epsilon_{ijkop}H^{op (120)}H^{imn (210)}_{l}H^{jkl (\overline{126})}_{mn} +\frac{1\sqrt{10}}{480}\epsilon_{ijkop}H^{op (120)}H^{mn (210)}H^{ijk (\overline{126})}_{mn} +\frac{1\sqrt{15}}{120}\epsilon_{ijkop}H^{op (120)}H^{ijn (210)}_{l}H^{kl (126)}_{n} +\frac{1\sqrt{5}}{240}\epsilon_{ijkop}H^{op (120)}H^{in (210)}H^{jk (126)}_{n} +\frac{1\sqrt{5}}{240}\epsilon_{ijkop}H^{op (120)}H^{ijk (210)}_{l}H^{l (\overline{126})} -\frac{1\sqrt{15}}{240}\epsilon_{ijkop}H^{op (120)}H^{ij (210)}H^{k (\overline{126})} +\frac{1\sqrt{15}}{60}H^{mn (120)}H^{(210)}_{q}H^{q (\overline{126})}_{mn} +\frac{1\sqrt{5}}{20}H^{nq (120)}H^{(210)}_{q}H^{(126)}_{n}  $
\end{flushleft}

\subsubsection{{\boldmath$\mathsf{210} \times \mathsf{120} \times \mathsf{120}$}}

$ X_{\mu1\mu2\mu3\mu4}X_{\mu1\mu2\mu5}X_{\mu3\mu4\mu5}= $
\begin{flushleft}
	\sloppy
$  +\frac{1\sqrt{6}}{18}H^{no (120)}H^{p (120)}_{no}H^{(210)}_{p} +\frac{1\sqrt{6}}{6}H^{np (120)}H^{(120)}_{n}H^{(210)}_{p} -\frac{1\sqrt{6}}{18}H^{m (120)}_{ij}H^{l (120)}_{km}H^{ijk (210)}_{l} -\frac{1\sqrt{2}}{9}H^{(120)}_{m}H^{m (120)}_{ij}H^{ij (210)} -\frac{1\sqrt{6}}{36}H^{(120)}_{k}H^{l (120)}_{ij}H^{ijk (210)}_{l} +\frac{1\sqrt{2}}{36}H^{(120)}_{j}H^{(120)}_{i}H^{ij (210)} +\frac{1\sqrt{6}}{36}\epsilon_{ijmno}H^{no (120)}H^{lm (120)}_{k}H^{ijk (210)}_{l} -\frac{1\sqrt{2}}{18}\epsilon_{ijmno}H^{no (120)}H^{jm (120)}_{k}H^{ik (210)} -\frac{1\sqrt{6}}{72}\epsilon_{ijmno}H^{l (120)}H^{no (120)}H^{ijm (210)}_{l} +\frac{1\sqrt{3}}{18}H^{kl (120)}_{m}H^{m (120)}_{ij}H^{ij (210)}_{kl} +\frac{1\sqrt{15}}{90}H^{ij (120)}_{m}H^{m (120)}_{ij}H^{(210)} -\frac{1\sqrt{3}}{18}H^{(120)}_{j}H^{kl (120)}_{i}H^{ij (210)}_{kl} +\frac{1\sqrt{2}}{9}H^{(120)}_{j}H^{jl (120)}_{i}H^{i (210)}_{l} -\frac{1\sqrt{3}}{18}H^{l (120)}H^{k (120)}_{ij}H^{ij (210)}_{kl} +\frac{1\sqrt{2}}{9}H^{i (120)}H^{k (120)}_{ij}H^{j (210)}_{k} +\frac{1\sqrt{2}}{6}H^{k (120)}H^{(120)}_{j}H^{j (210)}_{k} -\frac{1\sqrt{15}}{15}H^{j (120)}H^{(120)}_{j}H^{(210)} +\frac{1\sqrt{3}}{18}H^{kl (120)}H^{(120)}_{ij}H^{ij (210)}_{kl} +\frac{2\sqrt{2}}{9}H^{kq (120)}H^{(120)}_{jq}H^{j (210)}_{k} -\frac{1\sqrt{15}}{30}H^{pq (120)}H^{(120)}_{pq}H^{(210)} +\frac{2\sqrt{3}}{9}H^{jm (120)}_{i}H^{l (120)}_{km}H^{ik (210)}_{jl} +\frac{1\sqrt{2}}{9}H^{km (120)}_{i}H^{l (120)}_{km}H^{i (210)}_{l} -\frac{1\sqrt{6}}{18}H^{jm (120)}_{i}H^{kl (120)}_{m}H^{i (210)}_{jkl} -\frac{1\sqrt{6}}{36}H^{l (120)}H^{jk (120)}_{i}H^{i (210)}_{jkl} -\frac{1\sqrt{2}}{9}H^{l (120)}H^{jk (120)}_{l}H^{(210)}_{jk} -\frac{7\sqrt{2}}{36}H^{k (120)}H^{j (120)}H^{(210)}_{jk} +\frac{1\sqrt{6}}{36}\epsilon^{klmno}H^{(120)}_{no}H^{j (120)}_{im}H^{i (210)}_{jkl} +\frac{1\sqrt{2}}{18}\epsilon^{klmno}H^{(120)}_{no}H^{j (120)}_{lm}H^{(210)}_{jk} -\frac{1\sqrt{6}}{72}\epsilon^{klmno}H^{(120)}_{i}H^{(120)}_{no}H^{i (210)}_{klm} +\frac{1\sqrt{6}}{18}H^{kl (120)}_{p}H^{(120)}_{kl}H^{p (210)} +\frac{1\sqrt{6}}{6}H^{l (120)}H^{(120)}_{lp}H^{p (210)}  $
\end{flushleft}

\subsection{Quartic Couplings}\label{quartic couplings}

\subsubsection{{\boldmath$\mathsf{210} \times \mathsf{120} \times \mathsf{120} \times \mathsf{45}$}}

$ X_{\mu1\mu2\mu3\mu4}X_{\mu1\mu2\mu5}X_{\mu3\mu4\mu6}X_{\mu5\mu6}= $
\begin{flushleft}
	\sloppy
	$  +\frac{1\sqrt{3}}{72}\epsilon^{ijklo}H^{(45)}_{mn}H^{m (120)}_{ij}H^{n (120)}_{kl}H^{(210)}_{o} -\frac{1\sqrt{3}}{144}\epsilon^{ijklo}H^{(45)}_{ik}H^{(120)}_{j}H^{(120)}_{l}H^{(210)}_{o} +\frac{1\sqrt{3}}{36}\epsilon_{klnqr}H^{kn (45)}H^{ls (120)}H^{qr (120)}H^{(210)}_{s} -\frac{1\sqrt{3}}{72}\epsilon_{klnqr}H^{ns (45)}H^{kl (120)}H^{qr (120)}H^{(210)}_{s} +\frac{1\sqrt{6}}{36}H^{(45)}_{mn}H^{jm (120)}_{i}H^{ln (120)}_{k}H^{ik (210)}_{jl} +\frac{1\sqrt{30}}{360}H^{(45)}_{mn}H^{km (120)}_{l}H^{ln (120)}_{k}H^{(210)} +\frac{1\sqrt{30}}{60}H^{(45)}_{mn}H^{m (120)}H^{n (120)}H^{(210)} +\frac{1\sqrt{6}}{144}H^{(45)}_{ik}H^{j (120)}H^{l (120)}H^{ik (210)}_{jl} +\frac{1\sqrt{6}}{36}H^{mn (45)}H^{j (120)}_{im}H^{l (120)}_{kn}H^{ik (210)}_{jl} +\frac{1\sqrt{30}}{360}H^{mn (45)}H^{k (120)}_{lm}H^{l (120)}_{kn}H^{(210)} +\frac{1\sqrt{30}}{60}H^{mn (45)}H^{(120)}_{m}H^{(120)}_{n}H^{(210)} +\frac{1\sqrt{6}}{144}H^{jl (45)}H^{(120)}_{i}H^{(120)}_{k}H^{ik (210)}_{jl} +\frac{1\sqrt{3}}{36}\epsilon^{klnqr}H^{(45)}_{kn}H^{(120)}_{ls}H^{(120)}_{qr}H^{s (210)} -\frac{1\sqrt{3}}{72}\epsilon^{klnqr}H^{(45)}_{ns}H^{(120)}_{kl}H^{(120)}_{qr}H^{s (210)} +\frac{1\sqrt{3}}{72}\epsilon_{ijklo}H^{mn (45)}H^{ij (120)}_{m}H^{kl (120)}_{n}H^{o (210)} -\frac{1\sqrt{3}}{144}\epsilon_{ijklo}H^{ik (45)}H^{j (120)}H^{l (120)}H^{o (210)} $
\end{flushleft}

\subsubsection{{\boldmath$\mathsf{210}\times \mathsf{210} \times \mathsf{210} \times \mathsf{210}$}}

$ X_{\mu1\mu2\mu3\mu4}X_{\mu1\mu2\mu3\mu4}X_{\mu5\mu6\mu7\mu8}X_{\mu5\mu6\mu7\mu8}= $
\begin{flushleft}
	\sloppy
	$  +4H^{s (210)}H^{t (210)}H^{(210)}_{s}H^{(210)}_{t} +\frac{4}{3}H^{r (210)}H^{(210)}_{r}H^{mno (210)}_{p}H^{p (210)}_{mno} +4H^{r (210)}H^{mn (210)}H^{(210)}_{r}H^{(210)}_{mn} +2H^{r (210)}H^{(210)}_{r}H^{mn (210)}_{op}H^{op (210)}_{mn} +4H^{r (210)}H^{(210)}_{r}H^{n (210)}_{p}H^{p (210)}_{n} +4H^{(210)}H^{(210)}H^{r (210)}H^{(210)}_{r} +\frac{1}{9}H^{ijk (210)}_{l}H^{mno (210)}_{p}H^{l (210)}_{ijk}H^{p (210)}_{mno} +\frac{2}{3}H^{mn (210)}H^{ijk (210)}_{l}H^{(210)}_{mn}H^{l (210)}_{ijk} +H^{ij (210)}H^{mn (210)}H^{(210)}_{ij}H^{(210)}_{mn} +\frac{1}{3}H^{ijk (210)}_{l}H^{mn (210)}_{op}H^{op (210)}_{mn}H^{l (210)}_{ijk} +\frac{2}{3}H^{n (210)}_{p}H^{p (210)}_{n}H^{ijk (210)}_{l}H^{l (210)}_{ijk} +\frac{2}{3}H^{(210)}H^{(210)}H^{ijk (210)}_{l}H^{l (210)}_{ijk} +H^{ij (210)}H^{(210)}_{ij}H^{mn (210)}_{op}H^{op (210)}_{mn} +2H^{ij (210)}H^{n (210)}_{p}H^{p (210)}_{n}H^{(210)}_{ij} +2H^{(210)}H^{(210)}H^{ij (210)}H^{(210)}_{ij} +\frac{1}{4}H^{ij (210)}_{kl}H^{kl (210)}_{ij}H^{mn (210)}_{op}H^{op (210)}_{mn} +H^{n (210)}_{p}H^{p (210)}_{n}H^{ij (210)}_{kl}H^{kl (210)}_{ij} +H^{(210)}H^{(210)}H^{ij (210)}_{kl}H^{kl (210)}_{ij} +H^{j (210)}_{l}H^{l (210)}_{j}H^{n (210)}_{p}H^{p (210)}_{n} +2H^{(210)}H^{(210)}H^{j (210)}_{l}H^{l (210)}_{j} +H^{(210)}H^{(210)}H^{(210)}H^{(210)}
	  $
\end{flushleft}

\section{Discussion}\label{discussion}

This paper proposes a novel C++ software that can compute $\mathsf{SO(10)}$ Higgs couplings automatically in terms of $\mathsf{SU(5)}$ invariants, based on the algorithm developed in Appendix \ref{su(n) reducible tensors}. These interactions arise in both supersymmetric and non-supersymmetric $\mathsf{SO(10)}$ models. A  top-down approach to calculate two-point, three-point  and higher Higgs-Higgs Interactions is necessary. Such couplings are needed in the breaking of GUT and electroweak symmetries and in the study of higher dimensional operators such as $B-L=-2$ for the exploration of physics beyond the SM.  This is important as these operators  contribute to the understanding of  neutrino masses, baryogenesis, proton decay and $n-\overline{n}$ oscillations.  The stand-alone software exposes a text-based interface to conveniently facilitate the calculation procedure from start to finish. The user enters their required coupling via text, and the program then processes the various intermediate calculations, until it presents the final normalized result in \LaTeX format. The number of indices can go up to 5, but the number of tensors appearing in an $\mathsf{SO(10)}$ invariant is not limited. The reliability of the algorithm has also been confirmed,  as manual calculations were performed and compared with the computer result, and these were successfully matched.

The code has also been made publicly available so that future development can be done to expand the capabilities of the software. Firstly, one such expansion could be the ability of the software to account for any $\mathsf{SO(N)}$ invariant couplings. Currently this is not supported, however the existing algorithms and data structures could be used as a template for such changes. Secondly, it is also possible to expand this software to account for Higgs-Spinor coupling \cite{Cardoso:2015gfa} analysis in the future. This would require the introduction of certain other algorithms and data structures, however the existing codebase will provide a solid foundation for the key algorithms required for the overall process.

We believe that our C++ program will be very useful for particle physicists in general. Researchers will have access to our computer code which will calculate $\mathsf{SO(10)}$ tensor interactions exactly and efficiently within their own models using the top-down approach. The code is available to download \href{https://github.com/AHB99/tensor-coupling-program/releases}{here}.

\vskip 1cm
\noindent{\bf{Funding:}} The research of AB and RMS was supported by the American University of Sharjah’s Faculty Research Grant Award AS1805. The work in this paper was supported, in part, by the Open Access Program of the American University of Sharjah under award OAPCAS-1110-C00005. This paper represents the opinions of the authors and does not mean to represent the position or opinions of the American University of Sharjah.

\noindent {\bf{Acknowledgments:}} It is a pleasure to acknowledge fruitful and illuminating communications with Pran Nath on many aspects of the manuscript.

\vskip 1cm

\begin{appendices}
\section{$\mathsf{SO(N)}$ group \cite{Syed:2005gd}}\label{so(n) group}

\subsection{Vector representation}\label{Vector representation}

Consider a real $\mathsf{N}-$dimensional coordinate space in which a vector $\vect{{\phi}}=\left(\vect{{\phi}}_1,\vect{{\phi}}_1,\dotsc, \vect{{\phi}}_\mathsf{N}\right)$ transforms as
\begin{eqnarray}
	\vect{{\phi}}_{\mu}\longrightarrow\vect{{\phi}}_{\mu}^{\prime}=\bm{\mathcal{O}}_{\mu\nu}\vect{{\phi}}_{\nu}; ~~~~~\mu,~\nu=1,2,\dotsc,\mathsf{N}.
	\label{trans of vec}
\end{eqnarray}
In order for the transformation (\ref{trans of vec}) to leave the length of $\vect{{\phi}}$ invariant, that is, $\vect{{\phi}}^{\prime \top}\vect{{\phi}}^{\prime}=\vect{{\phi}}^\top\vect{{\phi}}$, Eq.(\ref{trans of vec}) gives $\left(\bm{\mathcal{O}}\vect{{\phi}}\right)^\top \left(\bm{\mathcal{O}}\vect{{\phi}}\right)=\vect{{\phi}}^\top\vect{{\phi}}\Rightarrow\vect{{\phi}}^\top\left(\bm{\mathcal{O}}^\top \bm{\mathcal{O}}\right)\vect{{\phi}}=\vect{{\phi}}^\top\vect{{\phi}}$. Therefore, the matrix $\bm{\mathcal{O}}$ must satisfy $\bm{\mathcal{O}}^\top \bm{\mathcal{O}}=\bm{\mathcal{O}} \bm{\mathcal{O}}^\top=\mathbf{1}$.
The set of such length-preserving transformation matrices $\bm{\mathcal{O}}$ represent rotations in $\mathsf{N}-$dimensions and forms a group called orthogonal group $\mathsf{O(N)}$. Taking the determinant of both sides of the last equation
gives $\det\left(\bm{\mathcal{O}}^\top \bm{\mathcal{O}}\right)=\det\left(\mathbf{1}\right)\Rightarrow\left(\det \bm{\mathcal{O}}^\top\right)\left(\det{\bm{\mathcal{O}}}\right)=1\Rightarrow\left(\det \bm{\mathcal{O}}\right)^2=1$. That is $\det \bm{\mathcal{O}}=\pm1$.

The \emph{special orthogonal group} $\mathsf{SO(N)}$ is a group of $\mathsf{N}\times \mathsf{N}$ of real matrices $\bm{\mathcal{O}}$ obeying,
\begin{eqnarray}
	\bm{\mathcal{O}}^\top \bm{\mathcal{O}}=\bm{\mathcal{O}} \bm{\mathcal{O}}^\top=\mathbf{1};\qquad\det \bm{\mathcal{O}} =+1.
	\label{defn of SO(N)}
\end{eqnarray}
Now consider the group element $\bm{\mathcal{O}}(\bm{\mathsf{a}})$ of $\mathsf{SO(N)}$ which differ infinitesimally from the identity:
\begin{eqnarray}
	\bm{\mathcal{O}}(\bm{\mathsf{a}})\approx \mathbf{1}+\bm{\mathsf{a}}\qquad\textnormal{for}~\bm{\mathsf{a}}\ll\mathbf{1}. \label{group element(approx)}
\end{eqnarray}
On using Eq.(\ref{group element(approx)}) in Eq.(\ref{defn of SO(N)}), we get $\displaystyle\mathbf{1}+\bm{\mathsf{a}}^\top+\bm{\mathsf{a}}+ O(\bm{\mathsf{a}}^2)=\mathbf{1}$. Thus $\displaystyle \bm{\mathsf{a}}^\top=-\bm{\mathsf{a}}$, that is
\begin{eqnarray}
	\bm{\mathsf{a}}_{\mu\nu}=-\bm{\mathsf{a}}_{\nu\mu};\qquad (\mu\neq\nu).
	\label{antisym prop of par}
\end{eqnarray}
The real numbers $\bm{\mathsf{a}}_{\mu\nu}$ are the parameters of the group and specify rotation. Since the $\mathsf{N}\times \mathsf{N}$ matrix $\bm{\mathsf{a}}$ is antisymmetric, it has only $\binom{\mathsf{N}}{2} =
\frac{1}{2}\mathsf{N}\left(\mathsf{N}-1\right)$ independent parameters. Making use of Eq.(\ref{antisym prop of par}) in the infinitesimal $\mathsf{SO(N)}$ transformation (\ref{group element(approx)}), we get $ [\bm{\mathcal{O}}(\bm{\mathsf{a}})]_{\mu\nu}\approx \delta_{\mu\nu}+\bm{\mathsf{a}}_{\mu\nu}= \delta_{\mu\nu}+\frac{1}{2}\left(\bm{\mathsf{a}}_{\mu\nu}-\bm{\mathsf{a}}_{\nu\mu}\right)=\delta_{\mu\nu}+\frac{i}{2}\bm{\mathsf{a}}_{\alpha\beta}\left[-i\left(\delta_{\mu\alpha}\delta_{\nu\beta}
-\delta_{\nu\alpha}\delta_{\mu\beta}\right)\right]\equiv \delta_{\mu\nu}+\frac{i}{2}\bm{\mathsf{a}}_{\alpha\beta}\left(\bm{\mathsf{M}}_{\alpha\beta}\right)_{\mu\nu}$. Thus the  $\frac{1}{2}\mathsf{N}\left(\mathsf{N}-1\right)$ generators of the
$\mathsf{SO(N)}$ group in the vector representation are given by  $\mathsf{N}\times \mathsf{N}$ linearly independent matrices $\bm{\mathsf{M}}_{\alpha\beta}$\footnote{\noindent The factor of $\frac{1}{2}$ is chosen for convenience and the reason for inserting $i$ in Eq.(\ref{gen in vec rep}) is because it is more
	convenient in quantum mechanics to use the anti-Hermitian generators ($\bm{\mathsf{M}}_{\alpha\beta}^{\dagger}=-\bm{\mathsf{M}}_{\alpha\beta}$) rather than antisymmetric ($\bm{\mathsf{M}}_{\alpha\beta}^\top=-\bm{\mathsf{M}}_{\alpha\beta}$). Then group $\bm{\mathcal{O}}(\bm{\mathsf{a}})$ in Eq.(\ref{group element(exact)}) is unitary ($\bm{\mathcal{O}}^{\dagger}=\bm{\mathcal{O}}^{-1}$). Of course this does not change the fact that $\mathsf{SO(N)}$ is a real Lie algebra.}:
\begin{eqnarray}
	\Big(\bm{\mathsf{M}}_{\alpha\beta}\Big)_{\mu\nu}=-i\left(\delta_{\mu\alpha}\delta_{\nu\beta}
	-\delta_{\nu\alpha}\delta_{\mu\beta}\right);\qquad 1\leq \mu < \nu \leq \mathsf{N}
	\label{gen in vec rep}
\end{eqnarray}
Note that the matrices $\bm{\mathsf{M}}_{\alpha\beta}$ are antisymmetric:  $\left(\bm{\mathsf{M}}_{\alpha\beta}\right)_{\mu\nu}=-\left(\bm{\mathsf{M}}_{\alpha\beta}\right)_{\nu\mu}\Rightarrow \bm{\mathsf{M}}_{\alpha\beta}^\top=-\bm{\mathsf{M}}_{\beta\alpha}$. Hence, necessarily traceless: $\left(\bm{\mathsf{M}}_{\alpha\beta}\right)_{\mu\mu}=0\Rightarrow \Tr(\bm{\mathsf{M}}_{\alpha\beta})=0$. Eq.(\ref{gen in vec rep}) also shows that the only non-vanishing elements of the  matrix $\bm{\mathsf{M}}_{\alpha\beta}$ are $-i$ and $+i$ at the intersection of the $\alpha^{\textnormal{th}}$ row, $\beta^{\textnormal{th}}$ column ($\alpha\neq\beta$) and $\beta^{\textnormal{th}}$ row, $\alpha^{\textnormal{th}}$ column, respectively. The commutation relation satisfied by the generators $\bm{\mathsf{M}}_{\alpha\beta}$ can be easily calculated using Eq.(\ref{gen in vec rep}):
\begin{eqnarray}
	\Big[\bm{\mathsf{M}}_{\alpha\beta}, \bm{\mathsf{M}}_{\gamma\rho} \Big]=-i\Big(\delta_{\beta\gamma}\bm{\mathsf{M}}_{\alpha\rho}+ \delta_{\alpha\rho}\bm{\mathsf{M}}_{\beta\gamma}
	-\delta_{\alpha\gamma}\bm{\mathsf{M}}_{\beta\rho}-\delta_{\beta\rho}\bm{\mathsf{M}}_{\alpha\gamma}\Big).
	\label{commut rel}
\end{eqnarray}
If we rewrite Eq.(\ref{commut rel}) as
\begin{eqnarray}
	\Big[\bm{\mathsf{M}}_{\alpha\beta}, \bm{\mathsf{M}}_{\gamma\rho} \Big]=if_{\alpha\beta,\gamma\rho}^{\sigma\lambda}~\bm{\mathsf{M}}_{\sigma\lambda},
	\label{Lie algebra}
\end{eqnarray}
then we see that the structure constants $f_{\alpha\beta,\gamma\rho}^{\sigma\lambda}$ are given by
\begin{eqnarray}
	f_{\alpha\beta,\gamma\rho}^{\sigma\lambda}=\delta_{\sigma\beta}\Big(\delta_{\alpha\gamma}\delta_{\rho\lambda}-\delta_{\alpha\rho}\delta_{\gamma\lambda}\Big)
	+\delta_{\sigma\alpha}\Big(\delta_{\beta\rho}\delta_{\gamma\lambda}-\delta_{\beta\gamma}\delta_{\rho\lambda}\Big).
	\label{struc const}
\end{eqnarray}
We now obtain finite transformation for the group element from the infinitesimal transformation: $\bm{\mathcal{O}}(\bm{\mathsf{a}})\approx \mathbf{1}+\frac{i}{2}\bm{\mathsf{a}}_{\alpha\beta}\bm{\mathsf{M}}_{\alpha\beta}\equiv r(\bm{\mathsf{a}})$ using $ \bm{\mathcal{O}}(\bm{\mathsf{a}})= \lim_{n\longrightarrow\infty}\left[r(\frac{\bm{\mathsf{a}}}{n})\right]^n=
\lim_{n\longrightarrow\infty}\left[\mathbf{1}+\frac{i}{2}\left(\frac{\bm{\mathsf{a}}_{\alpha\beta}}{n}\right)\bm{\mathsf{M}}_{\alpha\beta}\right]^n$. Therefore,
\begin{eqnarray}
	\displaystyle \bm{\mathcal{O}}(\bm{\mathsf{a}})= \displaystyle \exp\left\{{\frac{i}{2}\sum_{1\leq\alpha<\beta}^N\bm{\mathsf{a}}_{\alpha\beta}\bm{\mathsf{M}}_{\alpha\beta}}\right\}.
	\label{group element(exact)}
\end{eqnarray}
Note that traceless and antisymmetry conditions satisfied by the generators $\bm{\mathsf{M}}_{\alpha\beta}$ follow immediately from Eqs.(\ref{defn of SO(N)}) and (\ref{group element(exact)}) irrespective of its representation as given by Eq.(\ref{gen in vec rep}): $1=\det \bm{\mathcal{O}}=e^{\Tr(\ln \bm{\mathcal{O}})}=
e^{\Tr(\ln [{e}^{\frac{i}{2}\bm{\mathsf{a}}_{\alpha\beta}\bm{\mathsf{M}}_{\alpha\beta}}])} ={e}^{\frac{i}{2}\bm{\mathsf{a}}_{\alpha\beta}\Tr (\bm{\mathsf{M}}_{\alpha\beta})}\Rightarrow \Tr(\bm{\mathsf{M}}_{\alpha\beta})=0$ and $\mathbf{1}= \bm{\mathcal{O}}^\top \bm{\mathcal{O}}=[{e}^{\frac{i}{2}\bm{\mathsf{a}}_{\alpha\beta}\bm{\mathsf{M}}_{\alpha\beta}}]^\top {e}^{\frac{i}{2}\bm{\mathsf{a}}_{\alpha\beta}\bm{\mathsf{M}}_{\alpha\beta}}=
{e}^{\frac{i}{2}\bm{\mathsf{a}}_{\alpha\beta}\bm{\mathsf{M}}_{\alpha\beta}^\top} {e}^{\frac{i}{2}\bm{\mathsf{a}}_{\alpha\beta}\bm{\mathsf{M}}_{\alpha\beta}}=
{e}^{\frac{i}{2}\bm{\mathsf{a}}_{\alpha\beta}(\bm{\mathsf{M}}_{\alpha\beta}^\top+\bm{\mathsf{M}}_{\alpha\beta})}\Rightarrow \bm{\mathsf{M}}_{\alpha\beta}^\top+\bm{\mathsf{M}}_{\alpha\beta}=\mathbf{0}$.

\subsection{Tensor representation}
In general, we define an $\mathsf{SO(N)}$ rank$-p$  tensor $\vect{\mathit{T}}_{\mu_1\mu_2\dotso\mu_p}$,
having $\mathsf{N}^p$ components, to transform as a product of $p$ ordinary vectors,
$\vect{{\phi}}_{\mu_i}$:
\begin{eqnarray}
	\vect{\mathit{T}}_{\mu_1\mu_2\dotso\mu_p}&\equiv& \vect{{\phi}}_{\mu_1}\otimes\vect{{\phi}}_{\mu_2}\otimes\dotsm\otimes\vect{{\phi}}_{\mu_p},\\
	\vect{\mathit{T}}_{\mu_1\mu_2\dotso\mu_p}\longrightarrow T_{\mu_1\mu_2\dotso\mu_p}^{\prime}&=&\bm{\mathcal{O}}_{\mu_1\nu_1}\bm{\mathcal{O}}_{\mu_2\nu_2}\dotsm \bm{\mathcal{O}}_{\mu_p\nu_p}\vect{\mathit{T}}_{\nu_1\nu_2\dotso\nu_p}
\end{eqnarray}

\subsubsection{Isotropic tensors}
\begin{itemize}
	\item $2^{\textnormal{nd}}-$\textsc{rank identity tensor (kronecker symbol)}: $\delta_{\mu\nu}=\delta^{\mu\nu}=\delta^{\mu}_{\nu}$~\footnote{Recall form Appendix {\ref{Vector representation}} that $\mathsf{SO(N)}$ transformations preserve the scalar product: $\vect{{\phi}}^\top \vect{\psi}$. This invariant can be
		written using upper (or equally with lower) indices as $\vect{{\phi}}^{\mu}\vect{\psi}^{\mu}=\vect{{\phi}}^{\mu}\delta_{\mu\nu}\vect{\psi}^{\nu}\equiv \vect{{\phi}}^{\mu}g_{\mu\nu}\vect{\psi}^{\nu}$, where $g_{\mu\nu}$ is a metric tensor and is defined through $g_{\mu\nu}=e_{\mu}e_{\nu}$ for a set of basis vectors $e_{\mu}$ in an $\mathsf{N}-$ dimensional space. This implies that the metric tensor $g_{\mu\nu}=\delta_{\mu\nu}=e_{\mu}e_{\nu}$ corresponds to the orthogonal group. Further, the metric tensor
		can be used to raise or lower indices of vectors/tensors: $\vect{{\phi}}_{\mu}=g_{\mu\nu}\vect{{\phi}}^{\nu}$, $\vect{{\phi}}^{\mu}=g^{\mu\nu}\vect{{\phi}}_{\nu}$. Therefore, $\vect{{\phi}}_{\mu}=\vect{{\phi}}^{\mu}$. In other words the covariant and contravariant vectors/tensors coincide for orthonormal basis. Hence, we do not distinguish between superscripts and subscripts.}\\
	
	\noindent The Kronecker symbol is defined through
	\begin{displaymath}
		\delta_{\mu\nu}= \delta_{\nu\mu} = \left\{
		\begin{array}{ll}
			1 &\textnormal{\textsf{if}}~\mu=\nu,\\
			0 &\textnormal{\textsf{if}}~\mu\neq\nu.
		\end{array}
		\right.
	\end{displaymath}
	It is invariant under $\mathsf{SO(N)}$ transformation: $\delta_{\mu\nu}\longrightarrow\delta_{\mu\nu}'=\bm{\mathcal{O}}_{\mu\alpha}\bm{\mathcal{O}}_{\nu\beta}\delta_{\alpha\beta}=
	\bm{\mathcal{O}}_{\mu\alpha}\bm{\mathcal{O}}_{\nu\alpha}
	=(\bm{\mathcal{O}}\bm{\mathcal{O}}^\top)_{\mu\nu}=\delta_{\mu\nu}$.\\
	
	\item $\mathsf{N}^{\textnormal{th}}-$\textsc{rank alternating tensor (levi-civita symbol)}: $\epsilon_{\mu_1\mu_2\dotso\mu_{\mathsf{N}}}=\epsilon^{\mu_1\mu_2\dotso\mu_{\mathsf{N}}}$\\
	
	\noindent The completely antisymmetric Levi-Civita symbol is defined through
	\begin{displaymath}
		\epsilon_{\mu_1\mu_2...\mu_{\mathsf{N}}} = \left\{
		\begin{array}{ll}
			+1&{\textnormal{for even permutation of indices},}\\
			-1&{\textnormal{for odd permutation of indices},}\\
			0&{\textnormal{if any two indices equal}.}
		\end{array}
		\right.
	\end{displaymath}
	As a consequence of antisymmetry of the Levi-Civita symbol we have
	\begin{subequations}
		\begin{align}
			\epsilon_{\mu_1\mu_2\dotso\mu_{\mathsf{N}}}~\epsilon^{\nu_1\nu_2\dotso\nu_{\mathsf{N}}}&=
			\delta^{\nu_1}_{[\mu_1}\dotsm\delta^{\nu_{\mathsf{N}}}_{\mu_{\mathsf{N}}]},   \label{levi-civita1}\\
			\epsilon_{\alpha_1\dotso\alpha_m\mu_1\mu_2\dotso\mu_{{\mathsf{N}}-m}}~\epsilon^{\alpha_1\dotso\alpha_m\nu_1\nu_2\dotso\nu_{{\mathsf{N}}-m}}
			&=m!~\delta^{\nu_1}_{[\mu_1}\dotsm\delta^{\nu_{{\mathsf{N}}-m}}_{\mu_{{\mathsf{N}}-m}]}. \label{levi-civita2}
		\end{align}
	\end{subequations}
	Using Eq.(\ref{levi-civita2}), we define the determinant of a $\mathsf{N}\times \mathsf{N}$  matrix $\bm{\mathsf{A}}$ as
	\begin{align*}
		\det \bm{\mathsf{A}}=\frac{1}{{\mathsf{N}}!}\epsilon_{\mu_1\mu_2\dotso\mu_{\mathsf{N}}}
		\epsilon^{\nu_1\nu_2\dotso\nu_{\mathsf{N}}}\bm{\mathsf{A}}^{\mu_1}_{\nu_1}\dotsm \bm{\mathsf{A}}^{\mu_{\mathsf{N}}}_{\nu_{\mathsf{N}}}.
	\end{align*}
	Finally, multiplying both sides of this last equation by
	$\epsilon^{\mu_1\mu_2\dotso\mu_\mathsf{N}}$ and using Eq.(\ref{levi-civita2}), we get
	\begin{eqnarray}
		\epsilon^{\nu_1\nu_2\dotso\nu_\mathsf{N}}\bm{\mathsf{A}}^{\mu_1}_{\nu_1}\dotsm \bm{\mathsf{A}}^{\mu_\mathsf{N}}_{\nu_\mathsf{N}}=\epsilon^{\mu_1\mu_2\dotso\mu_\mathsf{N}}\det \bm{\mathsf{A}}.
		\label{det in terms of epsilon}
	\end{eqnarray}
	Using Eq.(\ref{det in terms of epsilon}), the alternating symbol can be also shown to be  an invariant of the $\mathsf{SO(N)}$ group: $\epsilon_{\mu_1\mu_2\dotso\mu_\mathsf{N}}\longrightarrow\epsilon_{\mu_1\mu_2\dotso\mu_\mathsf{N}}'
	=\bm{\mathcal{O}}_{\mu_1\nu_1}\bm{\mathcal{O}}_{\mu_2\nu_2}
	\dotsm \bm{\mathcal{O}}_{\mu_\mathsf{N}\nu_\mathsf{N}}\epsilon_{\nu_1\nu_2\dotso\nu_{\mathsf{N}}}\\
	=\det \bm{\mathcal{O}}~\epsilon_{\mu_1\mu_2\dotso\mu_\mathsf{N}}=\epsilon_{\mu_1\mu_2\dotso\mu_\mathsf{N}}$.\\
	
	\item\textsc{Other $\mathsf{SO(N)}-$invariants}\\
	
	\noindent One may now construct \cite{Appleby:1987} various $r^{\textnormal{th}}-$rank invariant tensors ${\vect{\mathit{I}}}_{\mu_1\mu_2\dotso\mu_{r}}$ from the linear combination of  the sum of the products of Kronecker symbols and the alternating tensor. For example, the most general invariant tensors for the case when  $r$ is even, take the form
	\begin{align*}
		{\vect{\mathit{I}}}^{(1)}_{\mu_1\mu_2\dotso\mu_{r}}&=\sum_{\sigma\in S_r}A_{\sigma}~\delta_{\sigma(\mu_1)\sigma(\mu_2)}
		\delta_{\sigma(\mu_3)\sigma(\mu_4)}\dotsm \delta_{\sigma(\mu_{r-1})\sigma(\mu_r)}; \quad\textnormal{for}~\substack{r= \textnormal{even}\\ \mathsf{N}=\textnormal{odd}},\\
		{\vect{\mathit{I}}}^{(2)}_{\mu_1\mu_2\dotso\mu_{\mathsf{N}}}&=\displaystyle\sum_{\sigma\in S_{\mathsf{N}}}A_{\sigma}~\delta_{\sigma(\mu_1)\sigma(\mu_2)}
		\delta_{\sigma(\mu_3)\sigma(\mu_4)}\dotsm \delta_{\sigma(\mu_{\mathsf{N}-1})\sigma(\mu_{\mathsf{N}})}+B\epsilon_{\mu_1\mu_2\dotso\mu_\mathsf{N}};\quad\textnormal{for}~\substack{r= \textnormal{even}\\ \mathsf{N}=\textnormal{even}},
\end{align*}
	where the summation is over the set $S_r$ of all $r!$ permutations $\sigma$ of $r$ indices and $\sigma(\mu_1)\sigma(\mu_2)\dotso  \sigma(\mu_r)$ represent a permutation of $\mu_1, \mu_2,\dots,\mu_r$. For the special case $r=4$, we get 4! permutations of $\mu_1, \mu_2,\mu_3,\mu_4$ out of which only three unique quadratic product of Kronecker deltas can be formed. Hence, ${\vect{\mathit{I}}}^{(1)}_{\alpha\beta\gamma\rho}=a_1\delta_{\alpha\beta}\delta_{\gamma\rho}
	+a_2 \delta_{\alpha\gamma}\delta_{\beta\rho} +a_3\delta_{\alpha\rho}\delta_{\beta\gamma}$ and  ${\vect{\mathit{I}}}^{(2)}_{\alpha\beta\gamma\rho}=a_1 \delta_{\alpha\beta}\delta_{\gamma\rho}+a_2\delta_{\alpha\gamma}\delta_{\beta\rho} +a_3~ \delta_{\alpha\rho}\delta_{\beta\gamma}+B\epsilon_{\alpha\beta\gamma\rho}$. Here $a_i$'s are  linear combinations of $4!$ $A_{\sigma}$'s. In a similar fashion, one can form general invariant tensors when $r$ is odd.
\end{itemize}

\subsubsection{Irreducibility}
Contraction of a tensor with Kronecker symbol (trace operation) and Levi-Civita symbol play an important role in constructing irreducible tensors. A tensor is reducible, if through a contraction operation, a new non-vanishing tensor (generally of smaller rank) can be formed.

If a tensor $\vect{\mathit{T}}^{(1)}_{\mu_1\mu_2\mu_3\dotso\mu_r}$ is reducible because it has nonzero trace, say over indices $\mu_1$ and $\mu_2$, then we may contract it with a Kronecker symbol over those two indices,
\begin{align*}
	\delta_{\mu_1\mu_2}\vect{\mathit{T}}^{(1)}_{\mu_1\mu_2\mu_3\dotso\mu_r}=\vect{\mathit{T}}_{\mu_1\mu_1\mu_3\dotso\mu_r}\equiv  \vect{\mathit{\hat{T}}}^{(1)}_{\mu_3\mu_4\dotso\mu_r},
\end{align*}
leading to a tensor $\vect{\mathit{\hat{T}}}^{(1)}_{\mu_3\mu_4\dotso\mu_r}$ of rank $r-2$. Here the first two indices have been contracted and summed over but the trace operation can be applied to any pair. A tensor is traceless if the contraction with a Kronecker symbol of any pair of indices vanishes. Also, a tensor with all $\delta-$contracted indices ($ \vect{\mathit{{T}}}_{\mu_1\mu_1\mu_2\mu_2\dotso\mu_r\mu_r}$) is an $\mathsf{SO(N)}-$invariant scalar (singlet).

On the other hand, if a tensor $\vect{\mathit{T}}^{(2a)}_{\mu_1\mu_2\mu_3\nu_4\nu_5\dotso\nu_r}$ is reducible because it is not symmetric with respect to some
of its indices, say $\mu_1$,~$\mu_2$ and $\mu_3$ or a tensor $\vect{\mathit{T}}^{(2b)}_{\mu_1\mu_2\dotso\mu_r}$ is reducible because it is not symmetric with respect to any of its indices, then we may contract these tensors with a Levi-Civita symbol over those indices,
\begin{align*}
	\epsilon_{\mu_1\mu_2\mu_3\mu_4\dotso\mu_\mathsf{N}}\vect{\mathit{T}}^{(2a)}_{\mu_1\mu_2\mu_3\nu_4\nu_5\dotso\nu_r}\equiv \vect{\mathit{\hat {T}}}^{(2a)}_{\mu_4\mu_5\dotso\mu_{\mathsf{N}}\nu_4\nu_5\dotso\nu_r},~~~~~~~~\epsilon_{\mu_1\mu_2\dotso\mu_\mathsf{N}}\vect{\mathit{T}}^{(2b)}_{\mu_1\mu_2\dotso\mu_r}
	\equiv\vect{\mathit{\hat {T}}}^{(2b)}_{\mu_{r+1}\mu_{r+2}\dotso\mu_\mathsf{N}},
\end{align*}
leading to a tensor $\vect{\mathit{\hat {T}}}^{(2a)}_{\mu_4\mu_5\dotso\mu_{\mathsf{N}}\nu_4\nu_5\dotso\nu_r}$ of rank $\mathsf{N}+r-6$ with mixed symmetry (antisymmetric in $\mu_4,\mu_5,\dotsc\mu_{\mathsf{N}}$ and symmetric in $\nu_4,\nu_5,\dotsc\nu_r$) and  a completely antisymmetric tensor $\vect{\mathit{\hat {T}}}^{(2b)}_{\mu_{r+1}\mu_{r+2}\dotso\mu_\mathsf{N}}$ of rank $\mathsf{N}-r$, respectively. Additionally, a tensor with all $\epsilon-$contracted indices ($\vect{\mathit{\hat{T}}}\equiv \epsilon_{\mu_1\mu_2\mu_3\mu_4\dotso\mu_\mathsf{N}}\vect{\mathit{T}}_{\mu_1\mu_2\dotso\mu_\mathsf{N}}$) is an $\mathsf{SO(N)}-$invariant scalar.

Therefore,
\begin{itemize}
	\item an irreducible tensor $\vect{\mathit{{T}}}_{\mu_1\mu_2\mu_3\dotso\mu_r}$ is completely traceless, that is
	\begin{eqnarray}
		\underbrace{\delta_{\mu_1\mu_2}{\vect{\mathit{T}}}_{\mu_1\mu_2\mu_3\dotso\mu_r}=0,~~\delta_{\mu_2\mu_3}{\vect{\mathit{T}}}_{\mu_1\mu_2\mu_3\dotso\mu_r}=0,~~
		\delta_{\mu_{1}\mu_3}{\vect{\mathit{T}}}_{\mu_1\mu_2\mu_3\dotso\mu_r}=0,~~\dotso}_{\frac{1}{2}r(r-1)~\textnormal{trace conditions}},
		\label{traceless cond}
	\end{eqnarray}

	and
\\
	\item in view of contraction with Levi-Civita tensor, the tensors $\vect{\mathit{T}}^{(2a)}_{\mu_1\mu_2\mu_3\nu_4\nu_5\dotso\nu_r}$ and $\vect{\mathit{T}}^{(2b)}_{\mu_1\mu_2\dotso\mu_r}$ are irreducible, if they are symmetric with respect to the indices on which the sum has been performed,  so that
	\begin{eqnarray}
		\epsilon_{\mu_1\mu_2\mu_3\mu_4\dotso\mu_\mathsf{N}}\vect{\mathit{T}}^{(2a)}_{\mu_1\mu_2\mu_3\nu_4\nu_5\dotso\nu_r}=0, ~~\epsilon_{\mu_1\mu_2\dotso\mu_\mathsf{N}}\vect{\mathit{T}}^{(2b)}_{\mu_1\mu_2\dotso\mu_r}=0.
	\end{eqnarray}
\end{itemize}
Since completely antisymmetric tensors automatically satisfy Eq.(\ref{traceless cond}), they are the first class of irreducible tensors. The second being completely symmetric and traceless tensors and finally traceless tensors with mixed symmetry.

\subsubsection{Completely antisymmetric tensors}
A $2^{\textnormal{nd}}-$rank antisymmetric tensor
$\boldsymbol{\varPhi}_{\mu\nu}^{({A})}\left(=-\boldsymbol{\varPhi}_{\nu\mu}^{({A})}\right)$ is defined through
\begin{eqnarray}
	\boldsymbol{\varPhi}_{\mu\nu}^{({A})}=\Big(\vect{{\phi}}_{\mu}\otimes
	\vect{{\phi}}_{\nu}\Big)\Big|_{{\textnormal{antisymmetric}}}
	=\frac{1}{2!}\Big(\vect{{\phi}}_{\mu}\otimes\vect{{\phi}}_{\nu}-\vect{{\phi}}_{\nu}\otimes\vect{{\phi}}_{\mu}\Big).
	\label{2nd rank antisymmetric tensor}
\end{eqnarray}
Using Eq.(\ref{trans of vec}) in Eq.(\ref{2nd rank antisymmetric tensor}) gives the transformation law for the second rank antisymmetric tensor:
\begin{subequations}
	\begin{align}
		\boldsymbol{\varPhi}_{\mu\nu}^{({A})}\longrightarrow \boldsymbol{\varPhi}_{\mu\nu}^{({A})'}&=\frac{1}{2}\Big(\bm{\mathcal{O}}_{\mu\lambda}\bm{\mathcal{O}}_{\nu\rho}-\bm{\mathcal{O}}_{\mu\rho}\bm{\mathcal{O}}_{\nu\lambda}
		\Big)\boldsymbol{\varPhi}_{\lambda\rho}^{({A})}, \label{trans of 2nd rank antisymmetric tensor(exact)}\\
		&=\bm{\mathcal{O}}_{\mu\rho}\boldsymbol{\varPhi}_{\rho\lambda}^{({A})}\bm{\mathcal{O}}_{\lambda\nu}^\top, \label{trans of 2nd rank antisymmetric tensor(condensed exact)}\\
		&\approx\boldsymbol{\varPhi}_{\mu\nu}^{({A})}+\bm{\mathsf{a}}_{\mu\rho}\boldsymbol{\varPhi}_{\rho\nu}^{({A})}
		+\bm{\mathsf{a}}_{\nu\lambda}\boldsymbol{\varPhi}_{\mu\lambda}^{({A})},\label{trans of 2nd rank antisymmetric tensor(approx)}
	\end{align}
\end{subequations}
where we have made usage of Eq.(\ref{group element(approx)}) in obtaining Eq.(\ref{trans of 2nd rank antisymmetric tensor(approx)}).  Note that this is also the adjoint representation of the group. This is because number of group generators matches the dimensionality
$\binom{\mathsf{N}}{2}=\frac{1}{2}\mathsf{N}(\mathsf{N}-1)$ of the second rank antisymmetric tensor representation.

To find the the generators in the adjoint representation, we use Eq.(\ref{group element(approx)}) in Eq.(\ref{trans of 2nd rank antisymmetric tensor(exact)}) to obtain $\boldsymbol{\varPhi}_{\mu_1\mu_2}^{({A})'}\longrightarrow\boldsymbol{\varPhi}_{\mu_1\mu_2}^{({A})}=1/2[(\delta_{\mu_1\nu_1} \bm{\mathsf{a}}_{\mu_2\nu_2}
+\delta_{\mu_2\nu_2}\bm{\mathsf{a}}_{\mu_1\nu_1})-(\nu_1\leftrightarrow \nu_2)]\boldsymbol{\varPhi}_{\nu_1\nu_2}^{({A})}\equiv \bm{\mathsf{a}}_{\alpha\beta}/2(\bm{\mathsf{M}}^{({A})}_{\alpha\beta})_{\mu_1\mu_2,\nu_1\nu_2}\boldsymbol{\varPhi}_{\nu_1\nu_2}^{({A})}$, where,
\begin{eqnarray}
	\Big(\bm{\mathsf{M}}^{({A})}_{\alpha\beta}\Big)_{\mu_1\mu_2,\nu_1\nu_2}&=&\frac{1}{2}\Bigg\{\bigg[\delta_{\mu_1\nu_1}\Big(\delta_{\alpha\mu_2}\delta_{\beta\nu_2}
	-\delta_{\alpha\nu_2}\delta_{\beta\mu_2}\Big)+\delta_{\mu_2\nu_2}\Big(\delta_{\alpha\mu_1}\delta_{\beta\nu_1}
	-\delta_{\alpha\nu_1}\delta_{\beta\mu_1}\Big)\bigg]\nonumber\\
&&\qquad-\bigg[\nu_1\leftrightarrow \nu_2\bigg]\Bigg\},
	\label{gen in 2nd rank tensor rep}
\end{eqnarray}
are the generators  in the adjoint representation.

In a similar fashion one can define completely antisymmetric tensors of higher rank. In general, an $r^{\textnormal{th}}-$rank antisymmetric tensor of dimensionality
$\binom{\mathsf{N}}{r}$  can be formed from the antisymmetric product of $\vect{{\phi}}$'s as
\begin{align*}
	\boldsymbol{\varPhi}^{({A})}_{\mu_1\mu_2\dotso\mu_r}&=\frac{1}{r!}\sum_{\sigma\in S_r}\sign(\sigma)~ \vect{{\phi}}_{\sigma(\mu_1)}\otimes\vect{{\phi}}_{\sigma(\mu_2)}\otimes\dotsm \vect{{\phi}}_{\sigma(\mu_r)}
\end{align*}
and with the transformation law in various useful forms given by
\begin{subequations}
	\begin{align}
		\boldsymbol{\varPhi}^{{({A})}}_{\mu_1\mu_2\dotso\mu_r}\longrightarrow \boldsymbol{\varPhi}^{({A})'}_{\mu_1\mu_2\dotso\mu_r}
		&=\frac{1}{r!}\left[\sum_{\sigma\in S_r}\sign(\sigma)~\bm{\mathcal{O}}_{{\mu_1}{\sigma(\nu_1)}}\bm{\mathcal{O}}_{{\mu_2}{\sigma(\nu_2)}}\dotsm \bm{\mathcal{O}}_{{\mu_r}{\sigma(\nu_r)}}\right]\boldsymbol{\varPhi}^{({
				A})}_{\nu_1\nu_2\dotso\nu_r},\\
		&=\bm{\mathcal{O}}_{\mu_1\nu_1}\bm{\mathcal{O}}_{\mu_2\nu_2}\dotsm  \bm{\mathcal{O}}_{\mu_r\nu_r}\boldsymbol{\varPhi}^{({
				A})}_{\nu_1\nu_2\dotso\nu_r},\label{trans of r^th rank antisymmetric tensor(condensed exact)}\\
		&\approx \boldsymbol{\varPhi}^{({A})}_{\mu_1\mu_2\dotso\mu_r}+\sum_{i=1}^r
		\bm{\mathsf{a}}_{\mu_i\nu_i}\boldsymbol{\varPhi}^{({A})}_{\mu_1\mu_2\dotso\nu_i\dotso\mu_{r-1}\mu_r}.
	\end{align}
\end{subequations}
Finally, the generators in the $r^{\textnormal{th}}-$rank
antisymmetric representation is given by
\begin{align*}
	\left(\bm{\mathsf{M}}^{(A)}_{\alpha\beta}\right)_{\mu_1\mu_2\dotso\mu_r,\nu_1\nu_2\dotso\nu_r}=\frac{1}{r!}\left\{\left[\sum_{i=1}^r
	~\prod_{\stackrel{j=1}{i\neq j}}^r
	\delta_{\mu_j\nu_j}\left(\delta_{\alpha\mu_i}\delta_{\beta\nu_i}-\delta_{\alpha\nu_i}\delta_{\beta\mu_i}\right)\right]
	-\left[\nu_i \leftrightarrow \nu_j \right]\right\}.
\end{align*}
For $\mathsf{N}$ even ($\mathsf{N}=2m$), a tensor of rank $m$, $\boldsymbol{\varPhi}^{({A})}_{\nu_1\nu_2\dotso\nu_m}$ can be expressed in terms of another tensor of rank $m$, ${}^{*}\!{\boldsymbol{\varPhi}}^{({A})}_{\mu_1\mu_2\dotso\mu_m}$, called the dual, through ${}^{*}\!{\boldsymbol{\varPhi}}^{({A})}_{\mu_1\mu_2\dotso\mu_m}\sim
\epsilon_{\mu_1\mu_2\dotso\mu_m,\nu_1\nu_2\dotso\nu_m}\boldsymbol{\varPhi}^{({A})}_{\nu_1\nu_2\dotso\nu_m}$. Both ${\boldsymbol{\varPhi}}^{({A})}_{\mu_1\mu_2\dotso\mu_m}$
and ${}^{*}\!{\boldsymbol{\varPhi}}^{({A})}_{\mu_1\mu_2\dotso\mu_m}$ are not irreducible tensors under $\mathsf {SO}(2m)$. Now,  if we define $\boldsymbol{\varOmega}^{(\pm)}_{\mu_1\mu_2\dotso\mu_m}\equiv\frac{1}{2}\left[{\boldsymbol{\varPhi}}^{({A})}_{\mu_1\mu_2\dotso\mu_m}\pm {}^{*}\!{\boldsymbol{\varPhi}}^{({A})}_{\mu_1\mu_2\dotso\mu_m}\right]$, then $\boldsymbol{\varOmega}^{(\pm)}_{\mu_1\mu_2\dotso\mu_m}$ are irreducible tensors. Complete formulation of this subtlety is as follows: The  tensor $\boldsymbol{\varPhi}^{({
		A})}_{\mu_1\mu_2...\mu_{m}}$ of dimension $\binom{{2m}}{m}$ splits
into two irreducible tensors
$\boldsymbol{\varOmega}^{({+})}_{\mu_1\mu_2\dotso\mu_{m}}$ and
$\boldsymbol{\varOmega}^{({-})}_{\mu_1\mu_2\dotso\mu_{m}}$ each of dimension
$\frac{1}{2}\binom{{2m}}{m}$ according to the $\mathsf {SO}(2m)$ invariant
decomposition of a tensor of rank $m$,
\begin{subequations}
	\begin{align}
		\boldsymbol{\varPhi}^{({
				A})}_{\nu_1\nu_2...\nu_{m}}&=\boldsymbol{\varOmega}^{({+})}_{\nu_1\nu_2\dotso\nu_{m}}+\boldsymbol{\varOmega}^{({-})}_{\nu_1\nu_2\dotso\nu_{m}},\\ \label{decompositions}
		\textnormal{where,}~~~~~~~~\boldsymbol{\varOmega}^{({\pm})}_{\mu_1\mu_2\dotso\mu_{m}}&=\frac{1}{2}\left(\delta_{\mu_1\nu_1}
		\delta_{\mu_2\nu_2}\dotso\delta_{\mu_{m}\nu_{m}}\pm
		\frac{i^{m^2}}{m!}\epsilon_{\mu_1\mu_2\dotso\mu_m\nu_1\nu_2\dotso\nu_m}\right)\boldsymbol{\varPhi}^{({
				A})}_{\nu_1\nu_2\dotso\nu_{m}},\\
		\textnormal{satisfying,}~~~~~~~~\boldsymbol{\varOmega}^{({\pm})}_{\mu_1\mu_2\dotso\mu_{m}}&=\pm
		\frac{i^{m^2}}{m!}\epsilon_{\mu_1\mu_2\dotso\mu_m\nu_1\nu_2\dotso\nu_m}\boldsymbol{\varOmega}^{({\pm})}_{\nu_1\nu_2\dotso\nu_{m}}.\label{duality conditions}
	\end{align}
\end{subequations}
\sloppy From Eq.(\ref{duality conditions}), we see that for $m=\textnormal{even}$, the tensors $\boldsymbol{\varOmega}^{({+})}_{\mu_1\mu_2\dotso\mu_{m}}$ and  $\boldsymbol{\varOmega}^{({-})}_{\mu_1\mu_2\dotso\mu_{m}}$  are real and satisfy self and anti-self duality conditions, respectively. While for $m=\textnormal{odd}$, the tensors are complex conjugates of each other. To show the validity of the  results (\ref{decompositions})-(\ref{duality conditions}), one can start with ${}^{*}\!{\boldsymbol{\varPhi}}^{({A})}_{\mu_1\mu_2\dotso\mu_m}\equiv
\alpha~\epsilon_{\mu_1\mu_2\dotso\mu_m,\nu_1\nu_2\dotso\nu_m}\boldsymbol{\varPhi}^{({A})}_{\nu_1\nu_2\dotso\nu_m}$, where $\alpha$ needs to be determined. Next compute the dual of the dual tensor: ${}^{**}\!{\boldsymbol{\varPhi}}^{({A})}_{\mu_1\mu_2\dotso\mu_m} = \alpha^2 \epsilon_{\mu_1\mu_2\dotso\mu_m\nu_1\nu_2\dotso\nu_m}\epsilon_{\nu_1\nu_2\dotso\nu_m\lambda_1\lambda_2\dotso\lambda_m}\boldsymbol{\varPhi}^{({A})}_{\lambda_1\lambda_2\dotso\lambda_m}= \alpha^2 (-1)^{m^2}(m!)^2\boldsymbol{\varPhi}^{({A})}_{\mu_1\mu_2\dotso\mu_m}$, where we have made use of Eqs.(\ref{levi-civita1}) and (\ref{levi-civita2}). Now write, $\boldsymbol{\varPhi}^{({A})}_{\mu_1\mu_2\dotso\mu_m}= \frac{1}{2}({\boldsymbol{\varPhi}}^{({A})}_{\mu_1\mu_2\dotso\mu_m}+{}^{*}\!{\boldsymbol{\varPhi}}^{({A})}_{\mu_1\mu_2\dotso\mu_m})+
\frac{1}{2}({\boldsymbol{\varPhi}}^{({A})}_{\mu_1\mu_2\dotso\mu_m}-{}^{*}\!{\boldsymbol{\varPhi}}^{({A})}_{\mu_1\mu_2\dotso\mu_m})\equiv \boldsymbol{\varOmega}^{({+})}_{\mu_1\mu_2\dotso\mu_{m}} +
\boldsymbol{\varOmega}^{({-})}_{\mu_1\mu_2\dotso\mu_{m}}$, then ${}^{*}\!\boldsymbol{\varOmega}^{(\pm)}_{\mu_1\mu_2\dotso\mu_m}=\alpha~\epsilon_{\mu_1\mu_2\dotso\mu_m,\nu_1\nu_2\dotso\nu_m}\boldsymbol{\varOmega}^{(\pm)}_{\nu_1\nu_2\dotso\nu_m}=
\pm\frac{1}{2}[\alpha^2 (-1)^{m^2}(m!)^2\boldsymbol{\varPhi}^{({A})}_{\mu_1\mu_2\dotso\mu_m}\pm {}^{*}\!\boldsymbol{\varPhi}^{({A})}_{\mu_1\mu_2\dotso\mu_m}]$. Finally, requiring
$\alpha^2 (-1)^{m^2}(m!)^2=1$ gives $\alpha=\frac{i^{m^2}}{m!}$.

\subsubsection{Completely symmetric and traceless tensors}
We begin by illustrating second rank symmetric and traceless tensor, $\boldsymbol{\varPhi}_{\mu\nu}^{({S})}\left(\equiv\boldsymbol{\varPhi}_{\nu\mu}^{({S})}\right)$
\begin{equation}
	\boldsymbol{\varPhi}_{\mu\nu}^{({S})}=\Big(\vect{{\phi}}_{\mu}\otimes
	\vect{{\phi}}_{\nu}\Big)\Big|_{{{\textnormal{symmetric}}}\choose
		{\textnormal{traceless}}}=\frac{1}{2!}\Big(\vect{{\phi}}_{\mu}\otimes\vect{{\phi}}_{\nu}+\vect{{\phi}}_{\nu}\otimes\vect{{\phi}}_{\mu}\Big)
	-\frac{\delta_{\mu\nu}}{\mathsf{N}}\vect{{\phi}}_{\lambda}\otimes\vect{{\phi}}_{\lambda}.  \label{2nd rank symmetric tensor}
\end{equation}
Here $\vect{{\phi}}_{\lambda}\otimes\vect{{\phi}}_{\lambda}(\equiv \boldsymbol{\varPhi}_{\lambda\lambda}^{({\cal
		S})})$ is the singlet of $\mathsf{SO(N)}$ group and the dimensionality of
$\boldsymbol{\varPhi}_{\mu\nu}^{({S})}$ is $\binom{\mathsf{N}+1}{2}-1=\frac{1}{2}(\mathsf{N}-1)(\mathsf{N}+2)$.

As before one can easily write down the transformation law for the second rank symmetric and traceless tensor:
\begin{subequations}
	\begin{align}
		\boldsymbol{\varPhi}_{\mu_1\mu_2}^{({S})}\longrightarrow \boldsymbol{\varPhi}_{\mu\nu}^{({\cal S})'}&=\frac{1}{2}\Big(\bm{\mathcal{O}}_{\mu\lambda}\bm{\mathcal{O}}_{\nu\rho}+\bm{\mathcal{O}}_{\mu\rho}\bm{\mathcal{O}}_{\nu\lambda}
		\Big)\boldsymbol{\varPhi}_{\lambda\rho}^{({S})} \label{trans of 2nd rank symmetric tensor(exact)},\\
		&=\bm{\mathcal{O}}_{\mu\rho}\boldsymbol{\varPhi}_{\rho\lambda}^{({S})}\bm{\mathcal{O}}_{\lambda\nu}^\top, \label{trans of 2nd rank symmetric tensor(condensed exact)}\\
		&\approx\boldsymbol{\varPhi}_{\mu\nu}^{({S})}+\bm{\mathsf{a}}_{\mu\rho}\boldsymbol{\varPhi}_{\rho\nu}^{({S})}
		+\bm{\mathsf{a}}_{\nu\lambda}\boldsymbol{\varPhi}_{\mu\lambda}^{({S})}.\label{trans of 2nd rank symmetric tensor(approx)}
	\end{align}
\end{subequations}
Using Eq.(\ref{trans of 2nd rank symmetric tensor(exact)}), one can find the generators in the
$2^{\textnormal{nd}}-$rank symmetric representation as
\begin{eqnarray}
	\displaystyle\Big(\bm{\mathsf{M}}^{({S})}_{\alpha\beta}\Big)_{\mu_1\mu_2,\nu_1\nu_2}&=&\displaystyle\frac{1}{2}\Bigg\{\bigg[\delta_{\mu_1\nu_1}
	\Big(\delta_{\alpha\mu_2}\delta_{\beta\nu_2}
	-\delta_{\alpha\nu_2}\delta_{\beta\mu_2}\Big)+\delta_{\mu_2\nu_2}
	\Big(\delta_{\alpha\mu_1}\delta_{\beta\nu_1}-\delta_{\alpha\nu_1}\delta_{\beta\mu_1}\Big)\bigg]\nonumber\\
	&&\qquad+\bigg[\nu_1\leftrightarrow \nu_2\bigg]\Bigg\}.
\end{eqnarray}

In general, an $r^{\textnormal{th}}-$rank symmetric and traceless tensor of dimensionality $\binom{\mathsf{N}+r-1}{r}-\binom{\mathsf{N}+r-3}{r-2}$ can be formed from the
symmetric product of $\vect{{\phi}}$'s as follows
\begin{align}
	\boldsymbol{\varPhi}^{({S})}_{\mu_1\mu_2\dotso\mu_r}&=\frac{1}{r!}\sum_{\sigma\in S_r} \vect{{\phi}}_{\sigma(\mu_1)}\vect{{\phi}}_{\sigma(\mu_2)}\dotsm \vect{{\phi}}_{\sigma(\mu_r)}\nonumber\\
	&+\kappa_1\bigg[\delta_{\mu_1\mu_2}~{\vect{{\phi}}}_{\mu}{\vect{{\phi}}}_{\mu}{\vect{{\phi}}}_{\mu_3}{\vect{{\phi}}}_{\mu_4}\dotsm{\vect{{\phi}}}_{\mu_r}
		+\delta_{\mu_1\mu_3}~{\vect{{\phi}}}_{\mu}{\vect{{\phi}}}_{\mu_2}{\vect{{\phi}}}_{\mu}{\vect{{\phi}}}_{\mu_4}\dotsm{\vect{{\phi}}}_{\mu_r}\nonumber\\
&\quad\qquad+\dotsb+\delta_{\mu_{r-1}\mu_r}~{\vect{{\phi}}}_{\mu_1}{\vect{{\phi}}}_{\mu_2}\dotsm{\vect{{\phi}}}_{\mu_{r-2}}{\vect{{\phi}}}_{\mu}{\vect{{\phi}}}_{\mu}\bigg]
~\longleftarrow{\scriptstyle\binom{r}{2}~\textnormal{single-trace terms}}\nonumber\\
	&+\kappa_2\bigg[\delta_{\mu_1\mu_2}\delta_{\mu_3\mu_4}~{\vect{{\phi}}}_{\mu}{\vect{{\phi}}}_{\mu}{\vect{{\phi}}}_{\nu}{\vect{{\phi}}}_{\nu}{\vect{{\phi}}}_{\mu_5}
		\dotsm{\vect{{\phi}}}_{\mu_r}+\dotsb\nonumber\\
		&\quad\qquad+
		\delta_{\mu_{r-3}\mu_{r-2}}\delta_{\mu_{r-1}\mu_r}~{\vect{{\phi}}}_{\mu_1}{\vect{{\phi}}}_{\mu_2}\dotsm{\vect{{\phi}}}_{\mu_{r-4}}
		{\vect{{\phi}}}_{\mu}{\vect{{\phi}}}_{\mu}{\vect{{\phi}}}_{\nu}{\vect{{\phi}}}_{\nu}\bigg]
	~\longleftarrow{\scriptstyle\frac{1}{2}\binom{r}{2}\times \binom{r-2}{2}~ \textnormal{double-trace terms}}\nonumber\\
	&~~~~~~~~~~~~~~~~~\vdots~~~~~~~~~~~~~~~~~~~~~~~~~~~~~~~~~~~~~~~~~~~~~~~~~~~~~\vdots\label{rth rank symmetric traceless tensor}
\end{align}
where, for conciseness, we have completely dropped the outer product symbol  and the numerical coefficients $\kappa_i$'s ensure that the tensor $\boldsymbol{\varPhi}^{({S})}_{\mu_1\mu_2\dotso\mu_r}$ is completely traceless. We now illustrate Eq.(\ref{rth rank symmetric traceless tensor}) by means of $3^{\textnormal{rd}}$ and $4^{\textnormal{th}}$
rank symmetric and traceless tensors of $\mathsf{SO(N)}$ with dimensionality $\frac{1}{6}\mathsf{N}(\mathsf{N}-1)(\mathsf{N}+4)$ and $\frac{1}{24}\mathsf{N}(\mathsf{N}-1)(\mathsf{N}+1)(\mathsf{N}+6)$, respectively. Explicit expressions are given by
\begin{subequations}
	\begin{align}
		\boldsymbol{\varPhi}^{({S})}_{\mu_1\mu_2\mu_3}&=\frac{1}{3!}\Big(\vect{{\phi}}_{\mu_1}
		\vect{{\phi}}_{\mu_2}\vect{{\phi}}_{\mu_3}\Big)\Big|_{\textnormal{symmetric}}-\frac{1}{\mathsf{N}+2}\bigg[
		\delta_{\mu_1\mu_2}\vect{{\phi}}_{\nu}
		\vect{{\phi}}_{\nu}\vect{{\phi}}_{\mu_3} +\delta_{\mu_1\mu_3}\vect{{\phi}}_{\nu}
		\vect{{\phi}}_{\mu_2}\vect{{\phi}}_{\nu}\nonumber\\
&~~~~~+\delta_{\mu_2\mu_3}\vect{{\phi}}_{\mu_1}
		\vect{{\phi}}_{\nu}\vect{{\phi}}_{\nu}\bigg],\\
		\boldsymbol{\varPhi}^{({
				S})}_{\mu_1\mu_2\mu_3\mu_4}&=\frac{1}{4!}\Big(\vect{{\phi}}_{\mu_1}
		\vect{{\phi}}_{\mu_2}\vect{{\phi}}_{\mu_3}\vect{{\phi}}_{\mu_4}\Big)\Big|_{\textnormal{symmetric}}
		-\frac{1}{\mathsf{N}+4}\bigg[\delta_{\mu_1\mu_2}\vect{{\phi}}_{\nu}\vect{{\phi}}_{\nu}
		\vect{{\phi}}_{\mu_3}\vect{{\phi}}_{\mu_4} +\delta_{\mu_1\mu_3}\vect{{\phi}}_{\nu}\vect{{\phi}}_{\mu_2}\vect{{\phi}}_{\nu}\vect{{\phi}}_{\mu_4}\nonumber\\
&~~~~~+\delta_{\mu_1\mu_4}\vect{{\phi}}_{\nu}\vect{{\phi}}_{\mu_2}
		\vect{{\phi}}_{\mu_3}\vect{{\phi}}_{\nu}
		+\delta_{\mu_2\mu_3}\vect{{\phi}}_{\mu_1}\vect{{\phi}}_{\nu}
		\vect{{\phi}}_{\nu}\vect{{\phi}}_{\mu_4}+\delta_{\mu_2\mu_4}\vect{{\phi}}_{\mu_1}\vect{{\phi}}_{\nu}
		\vect{{\phi}}_{\mu_3}\vect{{\phi}}_{\nu}\nonumber\\
&~~~~~+\delta_{\mu_3\mu_4}\vect{{\phi}}_{\mu_1}\vect{{\phi}}_{\mu_2}
		\vect{{\phi}}_{\nu}\vect{{\phi}}_{\nu}\bigg]+\frac{1}{(\mathsf{N}+2)(\mathsf{N}+4)}\bigg[\delta_{\mu_1\mu_2}\delta_{\mu_3\mu_4}\vect{{\phi}}_{\mu}\vect{{\phi}}_{\mu}
		\vect{{\phi}}_{\nu}\vect{{\phi}}_{\nu}\nonumber\\
		&~~~~~+\delta_{\mu_1\mu_3}\delta_{\mu_2\mu_4}\vect{{\phi}}_{\mu}\vect{{\phi}}_{\nu}
		\vect{{\phi}}_{\mu}\vect{{\phi}}_{\nu}+\delta_{\mu_1\mu_4}\delta_{\mu_2\mu_3}\vect{{\phi}}_{\mu}\vect{{\phi}}_{\nu}
		\vect{{\phi}}_{\nu}\vect{{\phi}}_{\mu}\bigg].
	\end{align}
\end{subequations}

The transformation law for an $r^{\textnormal{th}}-$rank symmetric tensor in various useful forms take the form
\begin{subequations}
	\begin{align}
		\boldsymbol{\varPhi}^{({ S})}_{\mu_1\mu_2\dotso\mu_r}\longrightarrow \boldsymbol{\varPhi}^{({ S})'}_{\mu_1\mu_2\dotso\mu_r}
		&=\frac{1}{r!}\left[\sum_{\sigma\in S_r}\bm{\mathcal{O}}_{{\mu_1}{\sigma(\nu_1)}}\bm{\mathcal{O}}_{{\mu_2}{\sigma(\nu_2)}}\dotsm \bm{\mathcal{O}}_{{\mu_r}{\sigma(\nu_r)}}\right]\boldsymbol{\varPhi}^{({
				S})}_{\nu_1\nu_2\dotso\nu_r},\\
		&=\bm{\mathcal{O}}_{\mu_1\nu_1}\bm{\mathcal{O}}_{\mu_2\nu_2}\dotsm  \bm{\mathcal{O}}_{\mu_r\nu_r}\boldsymbol{\varPhi}^{({
				S})}_{\nu_1\nu_2\dotso\nu_r},\label{trans of r^th rank symmetric tensor(condensed exact)}\\
		&\approx \boldsymbol{\varPhi}^{({S})}_{\mu_1\mu_2\dotso\mu_r}+\sum_{i=1}^r
		\bm{\mathsf{a}}_{\mu_i\nu_i}\boldsymbol{\varPhi}^{({S})}_{\mu_1\mu_2\dotso\nu_i\dotso\mu_{r-1}\mu_r},
	\end{align}
\end{subequations}
while the generators in this representation are given by
\begin{align*}
	\left(\bm{\mathsf{M}}^{{(S)}}_{\alpha\beta}\right)_{\mu_1\mu_2\dotso\mu_r,\nu_1\nu_2\dotso\nu_r}=\frac{1}{r!}\left\{\left[\sum_{i=1}^r
	~\prod_{\stackrel{j=1}{i\neq j}}^r
	\delta_{\mu_j\nu_j}\left(\delta_{\alpha\mu_i}\delta_{\beta\nu_i}-\delta_{\alpha\nu_i}\delta_{\beta\mu_i}\right)\right]
	+\left[\nu_i \leftrightarrow \nu_j \right]\right\}.
\end{align*}

\subsection{Elements of $\mathsf{SO(N)}$ gauge theory}

\subsubsection{Global symmetries}
The real group parameters $\bm{\mathsf{a}}_{\alpha\beta}$ are independent of space-time coordinate, $x^{{\mathpzc{A}}}$.\\
\begin{itemize}
	\item\textsc{Scalar boson in the vector representation}\\
	
	Introduce a set of $\mathsf{N}$ scalar bosonic fields by means of an $\mathsf{N}$
	dimensional column vector, $\vect{{\phi}}(x)$:\\
	\begin{eqnarray}\label{so(n) bosonic Column Vector}
		\vect{{\phi}}(x)=
		\begin{pmatrix}
			\vect{{\phi}}_1(x)\\\
			\vect{{\phi}}_2(x)\\
			\vdots\\
			\vect{{\phi}}_{\mathsf{N}}(x)
		\end{pmatrix}
	\end{eqnarray}
	Of course, the transformation law is the same as before (see Eqs.(\ref{trans of vec})and
	(\ref{group element(exact)})):
	\begin{equation}
		\vect{{\phi}}(x)\longrightarrow \vect{{\phi}}'(x)=\bm{\mathcal{O}}\vect{{\phi}}(x)\label{trans of global bosonic vector rep}
	\end{equation}
	and in terms of its components, Eq.(\ref{group element(approx)}) in Eq.(\ref{trans of global bosonic vector rep}), gives
	\begin{equation}
		\vect{{\phi}}_{\mu}(x)\longrightarrow \vect{{\phi}}_{\mu}'(x)\approx\vect{{\phi}}_{\mu}(x)+\bm{\mathsf{a}}_{\mu\nu}\vect{{\phi}}_{\nu}(x)
	\end{equation}
	
	Kinetic energy term for the  $\mathsf{N}$ real scalar bosonic fields appearing in the Lagrangian is given by
	\begin{eqnarray}\label{KE for the global bosonic vector rep}
		\begin{split}
			{\mathcal L}_{\textnormal{KE}}^{\vect{{\phi}}} &=\frac{1}{2}\sum_{\mu=1}^{\mathsf{N}}
			\partial_{\mathpzc{A}}\vect{{\phi}}_{\mu}(x)\partial^{\mathpzc{A}}\vect{{\vect{{\phi}}}}_{\mu}(x)\\
			&=\frac{1}{2}\partial_{\mathpzc{A}}\vect{{\phi}}(x)^\top\partial^{\mathpzc{A}}\vect{{\phi}}(x)
		\end{split}
	\end{eqnarray}
	It is invariant under {global rotations}, since
	${\mathcal L}_{\textnormal{KE}}^{\vect{{\phi}}\prime}=\frac{1}{2}\partial_{\mathpzc{A}}\vect{{\phi}}'(x)^\top\partial^{\mathpzc{A}}\vect{{\phi}}'(x)
	=\frac{1}{2}\partial_{\mathpzc{A}}[\vect{{\phi}}(x)^\top\bm{\mathcal{O}}^\top]\partial^{\mathpzc{A}}[\bm{\mathcal{O}}\vect{{\phi}}(x)]
	= \frac{1}{2}\partial_{\mathpzc{A}}\vect{{\phi}}(x)^\top\partial^{\mathpzc{A}}\vect{{\phi}}(x)={\mathcal L}_{\textnormal{KE}}^{\vect{{\phi}}}$. Here ${\mathpzc{A}}$ is the Dirac index (${\mathpzc{A}}=0-3$) and we are using the metric $\eta=diag(1,-1,-1,-1)$.
	
	One can also add to the Lagrangian the self-interaction terms. The most general fourth-order invariant couplings take the form
	\begin{eqnarray}\label{self int for the global bosonic vector rep}
		\begin{split}
			{\mathcal L}_{\textnormal{self-int}}^{\vect{{\phi}}} &=\sum_{\mu=1}^{\mathsf{N}}\left[\lambda_1 \vect{{\phi}}_{\mu}\vect{{\phi}}_{\mu}+\lambda_2
			\left(\vect{{\phi}}_{\mu}\vect{{\phi}}_{\mu}\right)^2\right]\\
			&= \lambda_1\vect{{\phi}}^\top\vect{{\phi}}+\lambda_2\left(\vect{{\phi}}^\top\vect{{\phi}}\right)^2
		\end{split}
	\end{eqnarray}
	\\
	
	\item\textsc{Scalar boson in the $2^{\textnormal{nd}}-$rank (adjoint) antisymmetric tensor representation}\\
	
	Recall that is also the \emph{adjoint} representation of the group. Thus, it implies that we have $\frac{1}{2}{\mathsf{N}}({\mathsf{N}}-1)$  vector gauge bosons denoted by $\vect{\mathcal{G}}^{\mathpzc{A}}_{\mu\nu}$ having the global transformation law (\ref{trans of 2nd rank antisymmetric tensor(approx)}):
	\begin{equation}
		\vect{\mathcal{G}}_{\mu\nu}^{\mathpzc{A}}\longrightarrow
		\vect{\mathcal{G}}_{\mu\nu}^{{\mathpzc{A}}'}=\vect{\mathcal{G}}_{\mu\nu}^{\mathpzc{A}}+\bm{\mathsf{a}}_{\mu\rho}
		\vect{\mathcal{G}}_{\rho\nu}^{\mathpzc{A}}+\bm{\mathsf{a}}_{\nu\lambda}\vect{\mathcal{G}}_{\mu\lambda}^{\mathpzc{A}}
	\end{equation}
	
	In analogy with Eq.(\ref{KE for the global bosonic vector rep}), we define the Lagrangian for ${\mathsf{N}}$ scalar bosonic fields
	\begin{eqnarray}\label{KE for the global bosonic 2nd rank antisymmetric tensor rep}
		\begin{split}
			{\mathcal L}_{\textnormal{KE}}^{\boldsymbol{\varPhi}^{(A)}}&=\frac{1}{4}\sum_{\mu\neq\nu=1}^{\mathsf{N}}\partial_{\mathpzc{A}}\boldsymbol{\varPhi}_{\mu\nu}^{(A)}\partial^{\mathpzc{A}}
			\boldsymbol{\varPhi}_{\mu\nu}^{(A)}\\
			&=\frac{1}{4}\Tr\left(\partial_{\mathpzc{A}}\boldsymbol{\varPhi}^{(A)\top}\partial^{\mathpzc{A}}\boldsymbol{\varPhi}^{(A)}\right)
		\end{split}
	\end{eqnarray}
	and it is easily seen to be invariant under the global transformation (\ref{trans of 2nd rank antisymmetric tensor(condensed exact)}).
	
	The most general invariant quartic self-couplings in the Lagrangian take the form
	\begin{eqnarray}\label{self int for the global bosonic 2nd rank antisymmetric tensor rep}
		\begin{split}
			{\mathcal L}_{\textnormal{self-int}}^{\boldsymbol{\varPhi}^{(A)}} &=\lambda_1 \boldsymbol{\varPhi}_{\mu\nu}^{(A)}\boldsymbol{\varPhi}_{\mu\nu}^{(A)}+\lambda_2
			\left(\boldsymbol{\varPhi}_{\mu\nu}^{(A)}\boldsymbol{\varPhi}_{\mu\nu}^{(A)}\right)^2
			+\lambda_2 \boldsymbol{\varPhi}_{\mu\nu}^{(A)}\boldsymbol{\varPhi}_{\nu\rho}^{(A)}
			\boldsymbol{\varPhi}_{\rho\sigma}^{(A)}\boldsymbol{\varPhi}_{\sigma\mu}^{(A)}\\
			&=\lambda_1\Tr\left(\boldsymbol{\varPhi}^{(A)^2}\right)+\lambda_2\left[\Tr\left(\boldsymbol{\varPhi}^{(A)^2}\right)\right]^2
			+\lambda_3\Tr\left(\boldsymbol{\varPhi}^{(A)^4}\right)
		\end{split}
	\end{eqnarray}
	
	Self-interaction terms for the vector gauge bosons  $\vect{\mathcal{G}}^{\mathpzc{A}}$ and its interactions with scalar bosons will be dealt in the next subsection.
	\\
	
	\item\textsc{Scalar boson in the $2^{\textnormal{nd}}-$rank symmetric and traceless tensor representation}\\
	
	In this case, the globally invariant kinetic energy and self-interaction terms in the Lagrangian are given by given by
	\begin{equation}\label{KE for the global bosonic 2nd rank symmetric tensor rep}
		{\mathcal L}_{\textnormal{KE}}^{\boldsymbol{\varPhi}^{(S)}}=\frac{1}{4}\Tr\left(\partial_{\mathpzc{A}}\boldsymbol{\varPhi}^{({S})\top}\partial^{\mathpzc{A}}\boldsymbol{\varPhi}^{({S})}\right)
	\end{equation}
	\begin{equation}\label{self int for the global bosonic 2nd rank symmetric tensor rep}
		{\mathcal L}_{\textnormal{self-int}}^{\boldsymbol{\varPhi}^{(S)}} =\lambda_1 \boldsymbol{\varPhi}_{\mu\nu}^{(S)}\boldsymbol{\varPhi}_{\mu\nu}^{(S)}+\lambda_2
		\left(\boldsymbol{\varPhi}_{\mu\nu}^{(S)}\boldsymbol{\varPhi}_{\mu\nu}^{(S)}\right)^2
		+\lambda_2 \boldsymbol{\varPhi}_{\mu\nu}^{(S)}\boldsymbol{\varPhi}_{\nu\rho}^{(S)}
		\boldsymbol{\varPhi}_{\rho\sigma}^{(S)}\boldsymbol{\varPhi}_{\sigma\mu}^{(S)}
	\end{equation}
	\\
	
	\item\textsc{Scalar boson in the general $r^{\textnormal{th}}-$rank tensor representation}\\
	
	The invariant Lagrangian under global transformations is given by
	\begin{equation}
		{\mathcal L}_{\textnormal{KE}}^{\boldsymbol{\varPhi}^{(A,S)}}=\frac{1}{2r!}
		\partial_{\mathpzc{A}}\boldsymbol{\varPhi}^{(A,S)}_{\mu_1\mu_2...\mu_r}\partial^{\mathpzc{A}}\boldsymbol{\varPhi}^{(A,S)}_{\mu_1\mu_2...\mu_r}
	\end{equation}
	
\end{itemize}

\subsubsection{Local symmetries}
The $\mathsf{SO(N)}$ group parameters $\bm{\mathsf{a}}_{\alpha\beta}$ are functions of space-time coordinate, $x^{{\mathpzc{A}}}$.\\
\begin{itemize}
	\item \textsc{Scalar boson in the vector representation}\\
	
	This time the scalar fields introduced through Eq.(\ref{so(n) bosonic Column Vector}) must
	transform as
	\begin{subequations}
		\begin{align}
			\vect{{\phi}}_{\mu}(x)\longrightarrow   \vect{{\phi}}_{\mu}'(x)&=\bm{\mathcal{O}}_{\mu\nu}(x)\vect{{\phi}}_{\nu}(x)\label{trans of local bosonic vector rep}\\
			\bm{\mathcal{O}}_{\mu\nu}(x)&=\left(e^{\frac{i}{2}\bm{\mathsf{a}}_{\alpha\beta}(x)\bm{\mathsf{M}}_{\alpha\beta}}\right)_{\mu\nu}.
		\end{align}
	\end{subequations}
	The kinetic energy for $\vect{\phi}$'s given by Eq.(\ref{KE for the global bosonic vector rep}), ${\mathcal L}_{\textnormal{KE}}^{\vect{{\phi}}}=\frac{1}{2}\sum_{\mu=1}^{\mathsf{N}}\partial_{\mathpzc{A}}\vect{{\phi}}_{\mu}(x)
	\partial^{\mathpzc{A}}\vect{{\phi}}_{\mu}(x)$ is no longer invariant under the \emph{local rotations }(\ref{trans of local bosonic vector rep}),  because
	$\partial_{\mathpzc{A}} \vect{{\phi}}'(x)=\bm{\mathcal{O}}(x)\partial_{\mathpzc{A}} \vect{{\phi}}(x)+[\partial_{\mathpzc{A}}
	\bm{\mathcal{O}}(x)]\vect{{\phi}}(x)\neq \bm{\mathcal{O}}(x)\partial_{\mathpzc{A}} \vect{{\phi}}(x)$. As in QED, we modify the Lagrangian by replacing the differential operator $\partial^{\mathpzc{A}}$ by the gauge covariant derivative $\mathbb{D}_{\mathpzc{A}}$, where
	\[\partial^{\mathpzc{A}}\longrightarrow\mathbb{D}^{\mathpzc{A}}_{\alpha\beta} \equiv \delta_{\alpha\beta} \partial^{\mathpzc{A}}-\frac{ig}{2}\vect{\mathcal{G}}_{\mu\nu}^{\mathpzc{A}}(x)\left(\bm{\mathsf{M}}_{\mu\nu}\right)_{\alpha\beta},\]
	and require $\mathbb{D}_{\mathpzc{A}}\vect{{\phi}}(x)$ to transform like $\vect{{\phi}}(x)$:
	\begin{equation}\label{trans of local covariantderivatve and bosonic vector}
		\mathbb{D}_{\mathpzc{A}}\vect{{\phi}}(x)\longrightarrow \Big[\mathbb{D}_{\mathpzc{A}}\vect{{\phi}}(x)\Big]'=\mathbb{D}_{\mathpzc{A}}'\vect{{\phi}}'(x)=\bm{\mathcal{O}}(x)\Big[\mathbb{D}_{\mathpzc{A}}\vect{{\phi}}(x)\Big].
	\end{equation}
	Note that $\mathbb{D}_{\mathpzc{A}}$ is to be understood as a ${\mathsf{N}} \times {\mathsf{N}}$ matrix carrying
	a Dirac index, ${\mathpzc{A}}$, and operating on the ${\mathsf{N}}$ component scalar
	bosonic field, $\vect{{\phi}}(x)$. Here $g$ is a coupling constant between scalar bosons and vector gauge bosons. Further, using the local gauge transformation (\ref{trans of local bosonic vector rep}) in Eq.(\ref{trans of local covariantderivatve and bosonic vector}) gives
	\begin{eqnarray}\label{trans of local covariantderivatve}
		\mathbb{D}_{\mathpzc{A}}\longrightarrow \mathbb{D}_{\mathpzc{A}}'=\bm{\mathcal{O}}(x)\mathbb{D}_{\mathpzc{A}}\bm{\mathcal{O}}^\top(x).
	\end{eqnarray}
	
	There are $\frac{1}{2}{\mathsf{N}}({\mathsf{N}}-1)$ group generators and we introduce one vector gauge boson, $\vect{\mathcal{G}}_{\mu\nu}^{\mathpzc{A}}(x)$ for each and define
	\begin{subequations}\label{covariantderivatve and bosonic vector in terms of Lie valued gauge}
		\begin{align}
			\mathbb{D}^{\mathpzc{A}}\vect{{\phi}}(x)&=\left[\mathbf{1}\partial^{\mathpzc{A}}-\frac{ig}{2}{\widehat{\vect{\mathcal{G}}}}^{\mathpzc{A}}(x)\right]\vect{{\phi}}(x),
			\label{covariantderivatve and bosonic vector in terms of Lie valued gauge part a}\\
			{\textnormal{where}}\quad{\widehat{\vect{\mathcal{G}}}}^{\mathpzc{A}}(x)&\equiv \bm{\mathsf{M}}_{\mu\nu}\vect{\mathcal{G}}_{\mu\nu}^{\mathpzc{A}}(x),
		\end{align}
	\end{subequations}
	in terms of Lie-algebra valued gauge field, $\widehat{\vect{\mathcal{G}}}^{\mathpzc{A}}(x)$. In Eq.(\ref{covariantderivatve and bosonic vector in terms of Lie valued gauge part a}), $\mathbf{1}$ represents $\mathsf{N}\times \mathsf{N}$ identity matrix. Next, we determine the transformation law for the vector gauge boson. Substituting Eqs.(\ref{trans of local bosonic vector rep}) and (\ref{covariantderivatve and bosonic vector in terms of Lie valued gauge part a}) in Eq.(\ref{trans of local covariantderivatve and bosonic vector}) and using the fact $\bm{\mathcal{O}}(x)\bm{\mathcal{O}}(x)^\top=\mathbf{1}$ which implies $[\partial_{\mathpzc{A}} \bm{\mathcal{O}}(x)]\bm{\mathcal{O}}(x)^\top=-\bm{\mathcal{O}}(x)\partial_{\mathpzc{A}} \bm{\mathcal{O}}(x)^\top$, we get
	\begin{equation}\label{trans of local Lie valued gauge field}
		{\widehat{\vect{\mathcal{G}}}}_{\mathpzc{A}}(x)\longrightarrow{\widehat{\vect{\mathcal{G}}}}'_{\mathpzc{A}}(x)=\bm{\mathcal{O}}(x)\left[
		{\widehat{\vect{\mathcal{G}}}}_{\mathpzc{A}}(x)+\mathbf{1}\frac{4i}{g}\partial_{\mathpzc{A}}\right]\bm{\mathcal{O}}(x)^\top.
	\end{equation}
	One could equivalently start from Eq.(\ref{trans of local covariantderivatve}) and derive Eq.(\ref{trans of local Lie valued gauge field}). We now work out the infinitesimal form of Eq.(\ref{trans of local Lie valued gauge field}).  Using Eqs.(\ref{group element(approx)}) and (\ref{gen in vec rep}) in Eq.(\ref{trans of local Lie valued gauge field}), we obtain after
	some algebra, the local transformation law for the vector bosons
	\begin{subequations}
		\begin{align}
			\vect{\mathcal{G}}_{\mu\nu}^{\mathpzc{A}}(x)\longrightarrow
			\vect{\mathcal{G}}_{\mu\nu}^{'{\mathpzc{A}}}(x)&=\vect{\mathcal{G}}_{\mu\nu}^{\mathpzc{A}}(x)+\bm{\mathsf{a}}_{\mu\rho}(x)\vect{\mathcal{G}}_{\rho\nu}^{\mathpzc{A}}(x)
			+\bm{\mathsf{a}}_{\nu\lambda}(x)\vect{\mathcal{G}}_{\mu\lambda}^{\mathpzc{A}}(x)\nonumber\\
			&\quad+\frac{2}{g}\partial^{\mathpzc{A}}\bm{\mathsf{a}}_{\mu\nu}(x),\label{infinitesimal trans of gauge field}\\
			{\textnormal{with}}\quad\vect{\mathcal{G}}_{\mu\nu}^{\mathpzc{A}}(x)&=-\vect{\mathcal{G}}_{\nu\mu}^{\mathpzc{A}}(x).
		\end{align}
	\end{subequations}
	In analogy with QED, we define the {field strength tensor},
	$\vect{\mathcal{F}}_{\mu\nu}^{{\mathpzc{AB}}}(x)$ as
	\begin{subequations}\label{definition of field strength tensor}
		\begin{align}
			\Big[\mathbb{D}_{\mathpzc{A}},\mathbb{D}_{\mathpzc{B}}\Big]&=-\frac{ig}{2}{\widehat{\vect{\mathcal{F}}}}_{{\mathpzc{AB}}}(x),
			\label{definition of field strength tensor part a}\\
			{\textnormal{where}}\quad{\widehat{\vect{\mathcal{F}}}}^{{\mathpzc{AB}}}(x)&=\bm{\mathsf{M}}_{\mu\nu}\vect{\mathcal{F}}_{\mu\nu}^{{\mathpzc{AB}}}(x).
		\end{align}
	\end{subequations}
	Substituting Eqs.(\ref{covariantderivatve and bosonic vector in terms of Lie valued gauge}) into Eqs.(\ref{definition of field strength tensor}) and together with Eq.(\ref{gen in vec rep}), we get
	\begin{eqnarray}\label{field strength tensor in terms of gauge field}
		\vect{\mathcal{F}}_{\mu\nu}^{{\mathpzc{AB}}}(x)=\partial^{\mathpzc{A}}\vect{\mathcal{G}}_{\mu\nu}^{\mathpzc{B}}(x)-\partial^{\mathpzc{B}}\vect{\mathcal{G}}_{\mu\nu}^{\mathpzc{A}}(x)
		-g\Big[\vect{\mathcal{G}}_{\mu\sigma}^{\mathpzc{A}}(x)\vect{\mathcal{G}}_{\sigma\nu}^{\mathpzc{B}}(x)
		-\vect{\mathcal{G}}_{\mu\sigma}^{\mathpzc{B}}(x)\vect{\mathcal{G}}_{\sigma\nu}^{\mathpzc{A}}(x)\Big].
	\end{eqnarray}
	To find the transformation law for $\vect{\mathcal{F}}^{{\mathpzc{AB}}}$ we left and right
	multiply Eq.(\ref{definition of field strength tensor part a}) by $\bm{\mathcal{O}}(x)$ and $\bm{\mathcal{O}}^\top(x)$, respectively, to obtain
	\begin{equation}\label{trans of  Lie valued field strength tensor}
		{\widehat{\vect{\mathcal{F}}}}'_{{\mathpzc{AB}}}(x)=\bm{\mathcal{O}}(x){\widehat{\vect{\mathcal{F}}}}_{{\mathpzc{AB}}}(x)\bm{\mathcal{O}}^\top{x},
	\end{equation}
	where we have use of Eq.(\ref{trans of local covariantderivatve}). Note that Eq.(\ref{trans of  Lie valued field strength tensor})
	is in the form of Eq.(\ref{trans of local Lie valued gauge field}), hence the corresponding infinitesimal
	result (\ref{infinitesimal trans of gauge field}) applies without the derivative term:
	\begin{equation}\label{infinitesimal trans of field strength tensor}
		\vect{\mathcal{F}}_{\mu\nu}^{{\mathpzc{AB}}}(x)\longrightarrow
		\vect{\mathcal{F}}_{\mu\nu}^{'{\mathpzc{AB}}}(x)=\vect{\mathcal{F}}_{\mu\nu}^{{\mathpzc{AB}}}(x)
		+\bm{\mathsf{a}}_{\mu\rho}(x)\vect{\mathcal{F}}_{\rho\nu}^{{\mathpzc{AB}}}(x)+\bm{\mathsf{a}}_{\nu\lambda}(x)\vect{\mathcal{F}}_{\mu\lambda}^{{\mathpzc{AB}}}(x).
	\end{equation}
	
	After the introduction of local transformation the Lagrangian (\ref{KE for the global bosonic vector rep}) must be replaced by
	\begin{eqnarray}
		\begin{split}
			{\mathcal L}&= \frac{1}{2}\mathbb{D}_{\mathpzc{A}}\vect{{\phi}}^\top(x)\mathbb{D}^{\mathpzc{A}}\vect{{\phi}} (x)\\
			&=\frac{1}{2}\partial_{\mathpzc{A}}\vect{{\phi}}_{\mu}\partial^{\mathpzc{A}}\vect{{\phi}}_{\mu}-
			g\left(\partial_{\mathpzc{A}}\vect{{\phi}}_{\mu}\right)\vect{\mathcal{G}}_{\mu\nu}^{{\mathpzc{A}}}\vect{{\phi}}_{\nu}
			-\frac{g^2}{2}\vect{{\phi}}_{\mu}\vect{\mathcal{G}}_{{\mathpzc{A}}\mu\nu}\vect{\mathcal{G}}_{\nu\sigma}^{{\mathpzc{A}}}\vect{{\phi}}_{\sigma}.
		\end{split}
	\end{eqnarray}
	
	In order to define the system including the new gauge field, $\vect{\mathcal{G}}_{\mu\nu}^{\mathpzc{A}}(x)$, it is necessary to include a kinetic energy
	term for $\vect{\mathcal{G}}_{\mu\nu}^{\mathpzc{A}}(x)$:
	\begin{equation}
		{\mathcal L}=-\frac{1}{4}\vect{\mathcal{F}}_{{\mathpzc{AB}}\mu\nu}(x)\vect{\mathcal{F}}_{\mu\nu}^{{\mathpzc{AB}}}(x)
		+\frac{1}{2}\mathbb{D}_{\mathpzc{A}}\vect{{\phi}}^\top(x)\mathbb{D}^{\mathpzc{A}}\vect{{\phi}}
		(x).
	\end{equation}
	\\
	\item \textsc{Scalar boson in the $2^{\textnormal{nd}}-$rank antisymmetric tensor representation}\\
	
	Recall from Eq.(\ref{trans of 2nd rank antisymmetric tensor(condensed exact)}) that the second rank antisymmetric tensor,
	$\boldsymbol{\varPhi}^{({A})}_{\mu\nu}$ transforms as $\boldsymbol{\varPhi}'^{({
			A})}=\bm{\mathcal{O}}\boldsymbol{\varPhi}^{({A})} \bm{\mathcal{O}}^\top$. Then just as in the case of the
	vector representation  we want the $\mathbb{D}_{\mathpzc{A}}\boldsymbol{\varPhi}^{({A})}$ to
	transform like $\boldsymbol{\varPhi}^{({A})}$:
	\begin{equation}\label{covariantderivatve and antisymmetric tensor in terms of Lie valued gauge}
		\mathbb{D}_{\mathpzc{A}}\boldsymbol{\varPhi}^{({A})}\longrightarrow
		\Big[\mathbb{D}_{\mathpzc{A}}\boldsymbol{\varPhi}^{({A})}\Big]'=\bm{\mathcal{O}}(x)\Big[\mathbb{D}_{\mathpzc{A}}\boldsymbol{\varPhi}^{({
				A})}\Big]\bm{\mathcal{O}}^\top(x)
	\end{equation}
	Then the covariant derivative, $\mathbb{D}_{\mathpzc{A}}\boldsymbol{\varPhi}^{({A})}$ in terms of
	{Lie-valued} gauge fields, which has the transformation
	property (\ref{covariantderivatve and antisymmetric tensor in terms of Lie valued gauge}), is given by
	\begin{equation}
		\mathbb{D}_{\mathpzc{A}}\boldsymbol{\varPhi}^{({A})}=\partial_{\mathpzc{A}}\boldsymbol{\varPhi}^{({
				A})}-\frac{ig}{2}\Big[{\widehat{\vect{\mathcal{G}}}}_{\mathpzc{A}}\boldsymbol{\varPhi}^{({
				A})}+{\widehat{\vect{\mathcal{G}}}}_{\mathpzc{A}}^\top\boldsymbol{\varPhi}^{({A})}\Big]
	\end{equation}
	Inserting the generators, we find the expression for the covariant derivative to be
	\begin{equation}
		\left(\mathbb{D}^{\mathpzc{A}}\boldsymbol{\varPhi}^{({A})}\right)_{\mu\nu}=\partial^{\mathpzc{A}}\boldsymbol{\varPhi}^{({
				A})}_{\mu\nu}-g\Big[{{\vect{\mathcal{G}}}}^{\mathpzc{A}}_{\mu\sigma}\boldsymbol{\varPhi}^{({
				A})}_{\sigma\nu}-{{\vect{\mathcal{G}}}}^{\mathpzc{A}}_{\nu\sigma}\boldsymbol{\varPhi}^{({
				A})}_{\sigma\mu}\Big]
	\end{equation}
	
	The total Lagrangian is
	\begin{equation}
		{\mathcal
			L}=-\frac{1}{4}\vect{\mathcal{F}}_{{\mathpzc{AB}}\mu\nu}\vect{\mathcal{F}}_{\mu\nu}^{{\mathpzc{AB}}}+\frac{1}{4}\Tr\left(\mathbb{D}_{\mathpzc{A}}\boldsymbol{\varPhi}^{({
				A})\top}\mathbb{D}^{\mathpzc{A}}\boldsymbol{\varPhi}^{({A})}\right)
	\end{equation}
	\\
	
	\item \textsc{Scalar boson in the $2^{\textnormal{nd}}-$rank symmetric tensor representation}\\
	
	Recall from Eq.(\ref{trans of 2nd rank symmetric tensor(condensed exact)}) that the $2^{\textnormal{nd}}-$rank symmetric tensor,
	$\boldsymbol{\varPhi}^{({S})}$, transforms as $\boldsymbol{\varPhi}'^{({
			S})}=\bm{\mathcal{O}}\boldsymbol{\varPhi}^{({S})} \bm{\mathcal{O}}^\top$. Hence, the results for this case
	will be identical to that for the $2^{\textnormal{nd}}-$rank antisymmetric
	tensor case. Therefore
	\begin{equation}
		\left(\mathbb{D}^{\mathpzc{A}}\boldsymbol{\varPhi}^{({S})}\right)_{\mu\nu}=\partial^{\mathpzc{A}}\boldsymbol{\varPhi}^{({
				S})}_{\mu\nu}-g\left({{\vect{\mathcal{G}}}}^{\mathpzc{A}}_{\mu\sigma}\boldsymbol{\varPhi}^{({
				S})}_{\sigma\nu}-{{\vect{\mathcal{G}}}}^{\mathpzc{A}}_{\nu\sigma}\boldsymbol{\varPhi}^{({
				S})}_{\sigma\mu}\right)
	\end{equation}
	\begin{equation}
		{\mathcal L}=
-\frac{1}{4}\vect{\mathcal{F}}_{{\mathpzc{AB}}\mu\nu}\vect{\mathcal{F}}_{\mu\nu}^{{\mathpzc{AB}}}+\frac{1}{4}\Tr\left(\mathbb{D}_{\mathpzc{A}}\boldsymbol{\varPhi}^{({
				S})\top}\mathbb{D}^{\mathpzc{A}}\boldsymbol{\varPhi}^{({S})}\right)
	\end{equation}
	and of course
	\begin{equation}
		\mathbb{D}_{\mathpzc{A}}\boldsymbol{\varPhi}^{({S})}\longrightarrow
		\Big[\mathbb{D}_{\mathpzc{A}}\boldsymbol{\varPhi}^{({S})}\Big]'=\bm{\mathcal{O}}(x)\Big[\mathbb{D}_{\mathpzc{A}}\boldsymbol{\varPhi}^{({
				S})}\Big]\bm{\mathcal{O}}(x)^\top
	\end{equation}
	\\
	\item \textsc{Scalar boson in the general $r^{\textnormal{th}}-$rank tensor representation}\\
	
	Recall from Eqs.(\ref{trans of r^th rank antisymmetric tensor(condensed exact)}) and (\ref{trans of r^th rank symmetric tensor(condensed exact)}) the transformation law for an arbitrary
	antisymmetric and symmetric tensor of rank $r$: $ \boldsymbol{\varPhi}^{({
			A},{
			S})'}_{\mu_1\mu_2...\mu_r}=\bm{\mathcal{O}}_{\mu_1\nu_1}\bm{\mathcal{O}}_{\mu_2\nu_2}...\bm{\mathcal{O}}_{\mu_r\nu_r}\boldsymbol{\varPhi}^{({
			A,{S}})}_{\nu_1\nu_2...\nu_r}$. Hence, we require that the
	corresponding covariant derivative, $\mathbb{D}_{\mathpzc{A}}\boldsymbol{\varPhi}^{({A},{\cal
			S})'}_{\mu_1\mu_2...\mu_r}$, transforms as
	\begin{equation}
		\left(\mathbb{D}_{\mathpzc{A}}\boldsymbol{\varPhi}^{({A},{
				S})'}\right)_{\mu_1\mu_2...\mu_r}=\bm{\mathcal{O}}_{\mu_1\nu_1}\bm{\mathcal{O}}_{\mu_2\nu_2}...\bm{\mathcal{O}}_{\mu_r\nu_r}
		\left(\mathbb{D}_{\mathpzc{A}}\boldsymbol{\varPhi}^{({
				A},{\cal S})'}\right)_{\nu_1\nu_2...\nu_r}
	\end{equation}
	The expression for the covariant derivative is then given by
	\begin{equation}
		\left(\mathbb{D}^{\mathpzc{A}}\boldsymbol{\varPhi}^{({A},{
				S})}\right)_{\mu_1\mu_2...\mu_r}=\partial^{\mathpzc{A}}\boldsymbol{\varPhi}^{({A},{
				S})'}_{\mu_1\mu_2...\mu_r} -g\sum_P(-1)^{\delta_P}
		\vect{\mathcal{G}}^{\mathpzc{A}}_{{\mu_1}_{P(1)}\nu} \boldsymbol{\varPhi}^{({A},{
				S})'}_{\nu{\mu_2}_{P(2)}{\mu_3}_{P(3)}...{\mu_r}_{P(r)}}
	\end{equation}
	For example, in the case of $3^{\textnormal{rd}}-$rank tensor, the above result
	takes the form
	\begin{equation}
		\left(\mathbb{D}^{\mathpzc{A}}\boldsymbol{\varPhi}^{({A},{
				S})}\right)_{\mu_1\mu_2\mu_3}=\partial^{\mathpzc{A}}\boldsymbol{\varPhi}^{({A},{
				S})'}_{\mu_1\mu_2\mu_r} -g\left( \vect{\mathcal{G}}^{\mathpzc{A}}_{{\mu_1}\nu} \boldsymbol{\varPhi}^{({
				A},{\cal S})'}_{\nu{\mu_2}{\mu_3}}-\vect{\mathcal{G}}^{\mathpzc{A}}_{{\mu_2}\nu} \boldsymbol{\varPhi}^{({
				A},{\cal S})'}_{\nu{\mu_1}{\mu_3}}+\vect{\mathcal{G}}^{\mathpzc{A}}_{{\mu_3}\nu} \boldsymbol{\varPhi}^{({
				A},{\cal S})'}_{\nu{\mu_1}{\mu_2}}\right)
	\end{equation}
\end{itemize}
\section{$\mathsf{SO(2N)}$ group in a $\mathsf{U(N)}$ basis}\label{so(2n) group in u(n) basis}

\subsection{Complete embedding of $\mathsf{U(N)}$ into $\mathsf{SO(2N)}$ \cite{Syed:2005gd}}\label{embed u(n) into so(2n)}
\noindent Let $\vect{\sigma}$ and $\vect{\lambda}$ given by
\begin{equation}
	\vect{\sigma}=\vect{a}+i\vect{b} \quad\textnormal{and}\quad \vect{\lambda}=\vect{c}+i\vect{d},
\end{equation}
be $\mathsf{N}-$dimensional complex column vectors of the $\mathsf{U(N)}$ group
where $\vect{a}$, $\vect{b}$, $\vect{c}$ and $\vect{d}$ are real vectors. Then the $\mathsf{U(N)}$
group transformations,
\begin{equation}
	\begin{split}
		\vect{{\sigma}}'=\bm{\mathcal{U}}\vect{\sigma},&\quad \vect{\lambda}'=\bm{\mathcal{U}}\vect{\lambda},\\
		\bm{\mathcal{U}}=e^{i\bm{\mathsf{b}}\cdot \bm{\mathsf{M}}_{\mathsf{U(N)}}};\quad \bm{\mathcal{U}}^{\dagger}\bm{\mathcal{U}}=\bm{\mathcal{U}}&\bm{\mathcal{U}}^{\dagger}={\bf 1};\quad
		\bm{\mathsf{M}}_{\mathsf{U(N)}}=\bm{\mathsf{M}}_{\mathsf{U(N)}}^{\dagger},
	\end{split}
\end{equation}
leaves the
following scalar products invariant:
\begin{subequations}
	\begin{align}
		\vect{\sigma}^{\dagger}\vect{\sigma}&=\vect{a}^\top \vect{a}+\vect{b}^\top \vect{b}, \label{u(n) invariant 1}\\
		\vect{\lambda}^{\dagger}\vect{\lambda}&=\vect{c}^\top \vect{c}+\vect{d}^\top \vect{d},\\
		\vect{\lambda}^{\dagger}\vect{\sigma}&=\vect{c}^\top \vect{a}+\vect{d}^\top \vect{b}+i\left(\vect{c}^\top \vect{b}-\vect{d}^\top \vect{a}\right).\label{u(n) invariant 3}
	\end{align}
\end{subequations}
Now, define two $\mathsf{2N}-$dimensional real column vectors as follows:
\begin{equation}\label{so(2n) basis}
	\vect{\Sigma}=
	\begin{pmatrix}
		\vect{a}\\ \vect{b}
	\end{pmatrix}
	,
	\quad
	\vect{\Lambda}=
	\begin{pmatrix}
		\vect{c}\\\vect{d}
	\end{pmatrix}
	.
\end{equation}
The $\mathsf{U(N)}$ invariants Eqs.(\ref{u(n) invariant 1})-(\ref{u(n) invariant 3}) can now be expressed in terms of
$\vect{\Sigma}$ and $\vect{\Lambda}$:
\begin{subequations}
	\begin{align}
		\vect{\Sigma}^\top\vect{\Sigma}&=\vect{a}^\top \vect{a}+\vect{b}^\top \vect{b}, \label{u(n)invariant 1 (real)}\\
		\vect{\Lambda}^\top\vect{\Lambda}&=\vect{c}^\top \vect{c}+\vect{d}^\top \vect{d},\\
		\vect{\Lambda}^\top\vect{\Sigma}&=\vect{c}^\top \vect{a}+\vect{d}^\top \vect{b}, \label{u(n)invariant 3 (real)}\\
		\vect{\Lambda}^\top \bm{\mathsf{J}}\vect{\Sigma} &=\vect{c}^\top \vect{b}-\vect{d}^\top \vect{a}, \label{u(n)invariant 4 (real)}
		\quad
		\bm{\mathsf{J}}=
		\begin{pmatrix}
			{\bf 0}&{\bf 1}\\
			-{\bf 1}&{\bf 0}
		\end{pmatrix}
	\end{align}
\end{subequations}
Next, consider the $\mathsf{SO(2N)}$ group acting on the real $\mathsf{2N}$
dimensional vectors $\vect{\Sigma}$ and $\vect{\Lambda}$. Then, the $\mathsf{SO(2N)}$ group
transformations
\begin{equation}
	\begin{split}
		\vect{\Sigma}'=\bm{\mathcal{O}}\vect{\Sigma},&\quad \vect{\Lambda}'=\bm{\mathcal{O}}\vect{\Lambda},\\
		\bm{\mathcal{O}}=e^{i \bm{\mathsf{a}}\cdot \bm{\mathsf{M}}_{\mathsf{SO(2N)}}};\quad \bm{\mathcal{O}}^\top \bm{\mathcal{O}}=\bm{\mathcal{O}}&\bm{\mathcal{O}}^\top={\bf 1};\quad
		\bm{\mathsf{M}}_{\mathsf{SO(2N)}}=-\bm{\mathsf{M}}_{\mathsf{SO(2N)}}^\top,
	\end{split}
\end{equation}
leaves the following scalar products invariant:
\begin{equation}
	\vect{\Sigma}^\top\vect{\Sigma};\quad \vect{\Lambda}^\top\vect{\Lambda};\quad \vect{\Lambda}^\top\vect{\Sigma} \label{so(2n) invariants}
\end{equation}
Since the $\mathsf{SO(2N)}$ invariants in (\ref{so(2n) invariants})
are also $\mathsf{U(N)}$ invariants (see Eqs.(\ref{u(n)invariant 1 (real)})-(\ref{u(n)invariant 3 (real)})), $\mathsf{U(N)}$ is a
``natural'' subgroup of $\mathsf{SO(2N)}$.

Note that since $\bm{\mathcal{O}}\in \mathsf{SO(2N)}$, the antisymmetric generators
$\bm{\mathsf{M}}_{\mathsf{SO(2N)}}$ in the basis of Eq.(\ref{so(2n) basis}), can be written as
\begin{eqnarray}\label{so(2n) matrix in terms of u(n)(initial)}
	\bm{\mathsf{M}}_{\mathsf{SO (2N)}}&=&i
	\begin{pmatrix}
		\bm{\mathsf{A}} &\bm{\mathsf{B}}\\
		-\bm{\mathsf{B}}^\top&\bm{\mathsf{C}}
	\end{pmatrix}
\end{eqnarray}
where $\bm{\mathsf{A}}$ and $\bm{\mathsf{C}}$ are real antisymmetric
($\bm{\mathsf{A}}=-\bm{\mathsf{A}}^\top$, $\bm{\mathsf{C}}=-\bm{\mathsf{C}}^\top$) $\mathsf{N}\times \mathsf{N}$ matrices
while $\bm{\mathsf{B}}$ is an arbitrary real $\mathsf{N}\times \mathsf{N}$ matrix.

Additionally, if we  impose that $\bm{\mathcal{O}}\in \mathsf{U(N)}$, then $\bm{\mathsf{M}}_{\mathsf{SO(2N)}}$ is
also a generator of $\mathsf{U(N)}$: $\bm{\mathsf{M}}_{\mathsf{SO(2N)}}\supset \bm{\mathsf{M}}_{\mathsf{U(N)}}$. Then the
corresponding transformation must also leave the fourth quantity
in Eq.(\ref{u(n)invariant 4 (real)}) invariant: $\vect{\Lambda}'^\top {\bm{\mathsf{J}}}\Sigma'=\vect{\Lambda}^\top
\bm{\mathsf{J}}\vect{\Sigma}~\Rightarrow~\\e^{ia.\bm{\mathsf{M}}_{\mathsf{U(N)}}^\top}\bm{\mathsf{J}}
e^{ia.\bm{\mathsf{M}}_{\mathsf{U(N)}}}= \bm{\mathsf{J}}$, which under infinitesimal transformations takes
the form
\begin{equation}
	\bm{\mathsf{M}}_{\mathsf{U(N)}}^\top \bm{\mathsf{J}}+\bm{\mathsf{J}} \bm{\mathsf{M}}_{\mathsf{U(N)}}=0 \label{u(n)invariant 4 (constraint)}
\end{equation}
Inserting Eq.(\ref{so(2n) matrix in terms of u(n)(initial)}) into Eq.(\ref{u(n)invariant 4 (constraint)}) gives
\begin{eqnarray}
	\bm{\mathsf{M}}_{\mathsf{U(N)}}&=&
	\begin{pmatrix}
		\bm{\mathsf{A}}&\bm{\mathsf{B}}\\
		-\bm{\mathsf{B}}&\bm{\mathsf{A}}
	\end{pmatrix}
\end{eqnarray}
where $\bm{\mathsf{A}}$ is a real $\mathsf{N}\times \mathsf{N}$ antisymmetric matrix and
$\bm{\mathsf{B}}$ is a real $\mathsf{N}\times \mathsf{N}$ symmetric matrix.

The number of independent elements in $\bm{\mathsf{A}}$ and $\bm{\mathsf{B}}$ are
$\frac{1}{2}\mathsf{N}(\mathsf{N}-1)$ and $\frac{1}{2}\mathsf{N}(\mathsf{N}+1)$, respectively, giving
a total of $\mathsf{N}^2$ independent elements in $\bm{\mathsf{M}}_{\mathsf{U(N)}}$. The traceless
matrices $\bm{\mathsf{A}}$ (since $\bm{\mathsf{A}}$ is antisymmetric) and the
traceless part of matrices $i\bm{\mathsf{B}}$:
$i[{\bm{\mathsf{B}}}-\frac{1}{\mathsf{N}}\Tr({\bm{\mathsf{B}}})\bf{1}]$ will form the adjoint
$\mathsf{N}^2-1$ dimensional representation of the $\mathsf{SU(N)}$ group and the
trace of $\bm{\mathsf{B}}$: $\frac{i}{\mathsf{N}}\Tr(\bm{\mathsf{B}})\bf{1}$ will be an
$\mathsf{SU(N)}$ singlet. This term generates the $\mathsf{U(1)}$ group of complex
phase transformations. Thus, we have the decomposition
$\mathsf{SO(2N)}\longrightarrow  \mathsf{U(N)}\longrightarrow \mathsf{SU(N)}\otimes \mathsf{U(1)}$.

However, the adjoint and  singlet representations of $\mathsf{SU(N)}$ is
not the full story. There are other generators of $\mathsf{SO(2N)}$ that
are not in $\bm{\mathsf{M}}_{\mathsf{U(N)}}$. These remaining $\binom{2\mathsf{N}}{2}-\mathsf{N}^2=
\mathsf{N}(\mathsf{N}-1)$ generators of $\mathsf{SO(2N)}$ form two antisymmetric tensor
representations of $\mathsf{U(N)}$: $-{\bm{\mathsf{K}}}\pm i{\bm{\mathsf{L}}}$ each of
dimension $\frac{1}{2}\mathsf{N}(\mathsf{N}-1)$:
\begin{eqnarray}
	\begin{pmatrix}
		\bm{\mathsf{K}}&{\bm{\mathsf{L}}}\\
		\bm{\mathsf{L}}&-\bm{\mathsf{K}}
	\end{pmatrix}
\end{eqnarray}
This can be seen from the following argument. From above we have
learned that $\mathsf{2N}$ dimensional real vector,
$\begin{pmatrix}
	\vect{a}\\ \vect{b}
\end{pmatrix}$, decomposes into $\mathsf{N}\oplus {\overline{
		\mathsf{N}}}$ dimensional vectors of $\mathsf{SU(N)}$ corresponding to $\vect{a}\pm i\vect{b}$, that is, $\left[2\mathsf{N}\right]\longrightarrow \left\{\mathsf{N}\right\}\oplus \left\{{\overline {\mathsf{N}}}\right\}$.
Further, the (antisymmetric) generators of $\mathsf{SO(2N)}$ can be
associated with $2^{nd}$ rank antisymmetric tensors. Thus, under
$\mathsf{SO(2N)}\longrightarrow  \mathsf{SU(N)}\otimes \mathsf{U(1)}$ decomposition: $\left [2\mathsf{N}\otimes 2\mathsf{N}\right]_{\textnormal{antisymmetric}}\longrightarrow \left\{(\mathsf{N}\oplus {\overline{\mathsf{N}}})\otimes (\mathsf{N}\oplus {\overline{\mathsf{N}}})\right\}_{\textnormal{antisymmetric}}\Rightarrow \left[\binom{2\mathsf{N}}{2}\right]\longrightarrow \left \{\mathsf{N}\otimes \mathsf{N}\right\}_{\textnormal{antisymmetric}}\oplus \left \{{\overline {\mathsf{N}}}\otimes {\overline{\mathsf{N}}}\right\}_{\textnormal{antisymmetric}}\oplus \left \{\mathsf{N}\otimes {\overline{ \mathsf{N}}}\right\}$. Thus, $\left[\mathsf{N}(2\mathsf{N}-1)\right]\longrightarrow \left\{\frac{1}{2}\mathsf{N}(\mathsf{N}-1)\right\}\oplus\left\{{\overline{\frac{1}{2}\mathsf{N}(\mathsf{N}-1)}}\right\}\oplus
\left\{\mathsf{N}^2-1\right\}\oplus \left\{1\right\}$.
Altogether,
\begin{eqnarray}\label{so(2n) matrix in terms of u(n)(final)}
	\bm{\mathsf{M}}_{\mathsf{SO(2N)}}&=&
	\begin{pmatrix}
		\bm{\mathsf{A}}+\bm{\mathsf{K}}&\bm{\mathsf{B}}+\bm{\mathsf{L}}\\
		-\bm{\mathsf{B}}+\bm{\mathsf{L}}&\bm{\mathsf{A}}-\bm{\mathsf{K}}
	\end{pmatrix}
\end{eqnarray}

\noindent Summarizing:
\begin{itemize}
	\item$\mathsf{N}^2-1$ dimensional adjoint of $\mathsf{SU(N)}$ is formed from
	$\bm{\mathsf{A}}+i\Big(\bm{\mathsf{B}}-\frac{1}{\mathsf{N}}\Tr({\bm{\mathsf{B}}}){\bf 1}\Big)$
	\item Singlet of $\mathsf{SU(N)}$ is formed from $\frac{1}{\mathsf{N}}\Tr({\bm{\mathsf{B}}}){\bf 1}$
	\item Antisymmetric representations of $\mathsf{SU(N)}$ is formed from
	$-\bm{\mathsf{K}}+ i\bm{\mathsf{L}}$ and $-\bm{\mathsf{K}}- i\bm{\mathsf{L}}$ each of
	dimensionality $\frac{1}{2}\mathsf{N}(\mathsf{N}-1)$
	\item $\bm{\mathsf{A}}$, $\bm{\mathsf{K}}$, $\bm{\mathsf{L}}$ are real $\mathsf{N}\times \mathsf{N}$
	antisymmetric matrices and $\bm{\mathsf{B}}$ is a real $\mathsf{N}\times \mathsf{N}$
	symmetric matrix.
\end{itemize}

\subsection{Generators of $\mathsf{SU(N)}$ in terms of $\mathsf{SO(2N)}$ \cite{Syed:2005gd,GrootNibbelink:2000hu}}
For compactness and clarity we drop the subscript from
$\bm{\mathsf{M}}_{\mathsf{SO(2N)}}$ and write them simply as $\bm{\mathsf{M}}$. Looking at the block
structure of $\bm{\mathsf{M}}$ in Eq.(\ref{so(2n) matrix in terms of u(n)(final)}), we can make the following
assignments ($i,j=1,\dotsc,\mathsf{N}$)
\begin{eqnarray}
	\begin{split}
		\bm{\mathsf{A}}^i_j&=-\frac{1}{2}\Big(\bm{\mathsf{M}}_{ij}+\bm{\mathsf{M}}_{i+\mathsf{N}~j+\mathsf{N}}\Big) ;\\
		\bm{\mathsf{K}}^{ij}&=\frac{1}{2}\Big(\bm{\mathsf{M}}_{ij}-\bm{\mathsf{M}}_{i+\mathsf{N}~j+\mathsf{N}}\Big);
	\end{split}
	\qquad\quad
	\begin{split}
		\bm{\mathsf{B}}^i_j&=\frac{1}{2}\Big(\bm{\mathsf{M}}_{i~j+\mathsf{N}}+\bm{\mathsf{M}}_{j~i+\mathsf{N}}\Big),\\
		\bm{\mathsf{L}}^{ij}&=\frac{1}{2}\Big(\bm{\mathsf{M}}_{i~j+\mathsf{N}}-\bm{\mathsf{M}}_{j~i+\mathsf{N}}\Big).
	\end{split}
\end{eqnarray}

The generators of $\mathsf{U(N)}$ group are $\bm{\mathsf{P}}^i_j$, defined by
\begin{equation}
	\bm{\mathsf{P}}^i_j=\bm{\mathsf{A}}^i_j+i\bm{\mathsf{B}}^i_j,
\end{equation}
and the $\mathsf{SU(N)}$ (traceless) generators, $\bm{\mathsf{Q}}^i_j$, are given by
\begin{equation}
	\bm{\mathsf{Q}}^i_j=\bm{\mathsf{P}}^i_j-\frac{1}{\mathsf{N}}\bm{\mathsf{P}}^k_k\delta^i_j.
\end{equation}
They satisfy $\mathsf{U(N)}$ algebra,
\begin{eqnarray}
	\begin{split}
		\Big[\bm{\mathsf{P}}^i_j,\bm{\mathsf{P}}^k_l\Big] &=\delta^i_l\bm{\mathsf{P}}^k_j-\delta^k_j\bm{\mathsf{P}}^i_l,\\
		\Big[\bm{\mathsf{Q}}^i_j,\bm{\mathsf{Q}}^k_l\Big] &=\delta^i_l\bm{\mathsf{Q}}^k_j-\delta^k_j\bm{\mathsf{Q}}^i_l,\\
	\end{split}
\end{eqnarray}

The  $\mathsf{U(1)}$ generator, $\bm{\mathsf{P}}^k_k$, is given by
\begin{equation}
	\bm{\mathsf{P}}^k_k=i\bm{\mathsf{M}}_{i~i+\mathsf{N}}
\end{equation}

Lastly, the broken generators of $\mathsf{SO(2N)}$ group, $\bm{\mathsf{S}}^{ij}$ and
$\overline {\bm{\mathsf{S}}}_{ij}$ are
\begin{eqnarray}
	\begin{split}
		\bm{\mathsf{S}}^{ij}&=-\bm{\mathsf{K}}^{ij}-i\bm{\mathsf{L}}^{ij},\\
		\overline {\bm{\mathsf{S}}}_{ij}&=-\bm{\mathsf{K}}^{ij}+i\bm{\mathsf{L}}^{ij}.
	\end{split}
\end{eqnarray}

\subsection{Branching rules for $\mathsf{SO(2N)}$ into $\mathsf{SU(N)} \otimes
	\mathsf{U(1)}$ irreducible representations \cite{Syed:2005gd}}
The irreducible tensor representations of $\mathsf{SO(2N)}$ can be
decomposed under $\mathsf{SO(2N)}\longrightarrow  \mathsf{SU(N)}\otimes \mathsf{U(1)}$ by forming
tensor products and using Young tableau.
\begin{itemize}
	\item\textsc{Vector of $\mathsf{SO(2N)}$}\\
	
	This case was already considered in the subsection \ref{embed u(n) into so(2n)}:
	\begin{eqnarray}
		\big[2\mathsf{N}\big]&\longrightarrow & \big\{\mathsf{N}\big\}\oplus \big\{{\overline {\mathsf{N}}}\big\}
	\end{eqnarray}
	\item\textsc{$2^{\textnormal{nd}}-$rank tensors of $\mathsf{SO(2\mathsf{N})}$}\\
	
	 The antisymmetric tensor representation was also considered in the
		subsection \ref{embed u(n) into so(2n)}:
		\begin{eqnarray}
			\big[\mathsf{N}(2\mathsf{N}-1)\big]&\longrightarrow &\left\{\frac{1}{2}\mathsf{N}(\mathsf{N}-1)\right\}\oplus
			\left\{{\overline{\frac{1}{2}\mathsf{N}(\mathsf{N}-1)}}\right\}\oplus \left\{\mathsf{N}^2-1\right\}\nonumber\\
           &&\oplus \left\{1\right\}
		\end{eqnarray}
		In the case of $2^{\textnormal{nd}}-$rank symmetric traceless tensor of
		dimensionality $\binom{2\mathsf{N}+1}{2}-\binom{2\mathsf{N}}{0}$, we get
		$\left[2\mathsf{N}\otimes 2\mathsf{N}\right]_{\textnormal{symmetric}}
		\longrightarrow \left\{(\mathsf{N}\oplus {\overline {\mathsf{N}}})\otimes
		(\mathsf{N}\oplus {\overline{\mathsf{N}}})\right\}_{\textnormal{symmetric}}$, which on simplifying gives
		\begin{eqnarray}
			\big[(\mathsf{N}+1)(2\mathsf{N}-1)\big] &\longrightarrow & \left\{\frac{1}{2}\mathsf{N}(\mathsf{N}+1)\right\}\oplus \left\{ {\overline{\frac{1}{2}\mathsf{N}(\mathsf{N}+1)}}\right\}\oplus \left\{\mathsf{N}^2-1\right\}\nonumber\\
           &&\oplus \left\{1\right\}
		\end{eqnarray}
	
	\item\textsc{$3^{\textnormal{rd}}-$rank tensors of $\mathsf{SO(2\mathsf{N})}$}\\
	
	Here we form an anti-symmetrized and a symmetrized product of
	three vectors and subtract off the trace in the case of a
	symmetric tensor representation.
	 The result for the decomposition of  antisymmetric tensor
		representation with dimensionality $\binom{2\mathsf{N}}{3}$ is given by
		\begin{eqnarray}
			\left[\frac{2}{3}\mathsf{N}(2\mathsf{N}-1)(\mathsf{N}-1)\right]&\longrightarrow & \left\{\frac{1}{6}\mathsf{N}(\mathsf{N}-1)(\mathsf{N}-2)\right\}
			\oplus \left\{{\overline{\frac{1}{6}\mathsf{N}(\mathsf{N}-1)(\mathsf{N}-2)}}\right\}\nonumber\\
           &&\oplus\left\{\frac{1}{2}\mathsf{N}(\mathsf{N}+1)(\mathsf{N}-2)\right\}
           \oplus\left\{{\overline{\frac{1}{2}\mathsf{N}(\mathsf{N}+1)(\mathsf{N}-2)}}\right\}\nonumber\\
			&&\oplus\left\{\mathsf{N}\right\}
			\oplus \left\{{\overline{\mathsf{N}}}\right\}
		\end{eqnarray}
		 The result for the symmetric traceless tensor representation
		of dimensionality $\binom{2\mathsf{N}+2}{3}-\binom{2\mathsf{N}}{1}$ is
		\begin{eqnarray}
			\left[\frac{2}{3}\mathsf{N}(\mathsf{N}+2)(2\mathsf{N}-1)\right]&\longrightarrow &\left\{\frac{1}{5}\mathsf{N}(\mathsf{N}+1)(\mathsf{N}+2)\right\}
			\oplus \left\{{\overline{\frac{1}{6}\mathsf{N}(\mathsf{N}+1)(\mathsf{N}+2)}}\right\}\nonumber\\
			&& \oplus
			\left\{\frac{1}{2}\mathsf{N}(\mathsf{N}-1)(\mathsf{N}+2)\right\}\oplus \left\{{\overline{\frac{1}{2}\mathsf{N}(\mathsf{N}-1)(\mathsf{N}+2)}}\right\}\nonumber\\
		\end{eqnarray}

	\item\textsc{$4^{\textnormal{th}}-$rank tensors of $\mathsf{SO(2\mathsf{N})}$}\\
	
	 Using the technique as before, we have the following
		decomposition of the $\binom{2\mathsf{N}}{4}$ component $4^{\textnormal{th}}-$rank
		antisymmetric tensor representation
		\begin{eqnarray}
			\left[\frac{1}{6}\mathsf{N}(\mathsf{N}-1)(2\mathsf{N}-1)(2\mathsf{N}-3)\right]&\longrightarrow &
			\left\{\frac{1}{24}\mathsf{N}(\mathsf{N}-1)(\mathsf{N}-2)(\mathsf{N}-3)\right\}\nonumber\\
			&&\oplus \left\{{\overline{\frac{1}{24}\mathsf{N}(\mathsf{N}-1)(\mathsf{N}-2)(\mathsf{N}-3)}}\right\}\nonumber\\
			&&\oplus\left\{\frac{1}{6}\mathsf{N}(\mathsf{N}+1)(\mathsf{N}-1)(\mathsf{N}-3)\right\}\nonumber\\
			&&\oplus \left\{{\overline{\frac{1}{6}\mathsf{N}(\mathsf{N}+1)(\mathsf{N}-1)(\mathsf{N}-3)}}\right\} \nonumber\\
			&&\oplus\left\{\frac{1}{2}\mathsf{N}(\mathsf{N}-1)\right\}\oplus\left\{{\overline{\frac{1}{2}\mathsf{N}(\mathsf{N}-1)}}\right\}\nonumber\\
			&&\oplus\left\{\frac{1}{4}\mathsf{N}^2(\mathsf{N}+1)(\mathsf{N}-3)\right\}\oplus \left\{\mathsf{N}^2-1\right\}\nonumber\\
			&&\oplus \left\{1\right\}
		\end{eqnarray}
		 For the case of symmetric traceless tensor representation of
		dimensionality $\binom{2\mathsf{N}+3}{4}-\binom{2\mathsf{N}+1}{2}$, the
		decomposition is
		\begin{eqnarray}
			\left[\frac{1}{6}\mathsf{N}(2\mathsf{N}+1)(\mathsf{N}+3)(2\mathsf{N}-1)\right] &\longrightarrow &
			\left\{\frac{1}{24}\mathsf{N}(\mathsf{N}+1)(\mathsf{N}+2)(\mathsf{N}+3)\right\} \nonumber\\
			&&\oplus\left\{{\overline{\frac{1}{24}\mathsf{N}(\mathsf{N}+1)(\mathsf{N}+2)(\mathsf{N}+3)}}\right\}\nonumber\\
			&&\oplus\left\{\frac{1}{6}\mathsf{N}(\mathsf{N}+1)(\mathsf{N}-1)(\mathsf{N}+3)\right\}\nonumber\\
			&& \oplus\left\{{\overline{\frac{1}{6}\mathsf{N}(\mathsf{N}+1)(\mathsf{N}-1)(\mathsf{N}+3)}}\right\} \nonumber\\
			&& \oplus \left\{\frac{1}{4}\mathsf{N}^2(\mathsf{N}-1)(\mathsf{N}+3)\right\}
		\end{eqnarray}

	\item\textsc{$5^{\textnormal{th}}-$rank tensors of $\mathsf{SO(2\mathsf{N})}$}\\
	
	 The $5^{\textnormal{th}}-$rank
		antisymmetric tensor representation has dimensionality is
		$\binom{2\mathsf{N}}{5}$ and can be decomposed as follows
		\begin{eqnarray}
			\left[\frac{1}{15}\mathsf{N}(2\mathsf{N}-1)(\mathsf{N}-1)(2\mathsf{N}-3)(\mathsf{N}-2)\right] &\longrightarrow &
			\left\{\frac{1}{120}\mathsf{N}(\mathsf{N}-1)(\mathsf{N}-2)(\mathsf{N}-3)(\mathsf{N}-4)\right\}\nonumber\\
			&&
			\oplus\left\{{\overline{\frac{1}{120}\mathsf{N}(\mathsf{N}-1)(\mathsf{N}-2)(\mathsf{N}-3)(\mathsf{N}-4)}}\right\}\nonumber\\
			&&
			\oplus\left\{\frac{1}{24}\mathsf{N}(\mathsf{N}+1)(\mathsf{N}-1)(\mathsf{N}-2)(\mathsf{N}-4)\right\}\nonumber\\
			&&
			\oplus\left\{{\overline{\frac{1}{24}\mathsf{N}(\mathsf{N}+1)(\mathsf{N}-1)(\mathsf{N}-2)(\mathsf{N}-4)}}\right\} \nonumber\\
			&&
			\oplus\left\{\frac{1}{12}\mathsf{N}^2(\mathsf{N}+1)(\mathsf{N}-1)(\mathsf{N}-4)\right\} \nonumber\\
			&&\oplus
			\left\{{\overline{\frac{1}{12}\mathsf{N}^2(\mathsf{N}+1)(\mathsf{N}-1)(\mathsf{N}-4)}}\right\}\nonumber\\
			&&\oplus\left\{\frac{1}{6}\mathsf{N}(\mathsf{N}-1)(\mathsf{N}-2)\right\}\nonumber\\
			&&\oplus\left\{{\overline{\frac{1}{6}\mathsf{N}(\mathsf{N}-1)(\mathsf{N}-2)}}\right\}\nonumber\\
			&&\oplus\left\{\frac{1}{2}\mathsf{N}(\mathsf{N}+1)(\mathsf{N}-2)\right\} \nonumber\\
			&&\oplus \left\{{\overline
				{\frac{1}{2}\mathsf{N}(\mathsf{N}+1)(\mathsf{N}-2)}}\right\}\nonumber\\
			&&\oplus \left\{\mathsf{N}\right\}\oplus \left\{{\overline {\mathsf{N}}}\right\}
		\end{eqnarray}
		 For the case of symmetric traceless tensor representation of
		dimensionality $\binom{2\mathsf{N}+4}{5}-\binom{2\mathsf{N}+2}{3}$, the
		decomposition is
		\begin{eqnarray}
			\left[\frac{1}{15}\mathsf{N}(\mathsf{N}+1)(2\mathsf{N}+1)(\mathsf{N}+4)(2\mathsf{N}-1)\right] &\longrightarrow &
			\left\{\frac{1}{120}\mathsf{N}(\mathsf{N}+1)(\mathsf{N}+2)(\mathsf{N}+3)(\mathsf{N}+4)\right\}\nonumber\\
			&&
			\oplus\left\{{\overline{\frac{1}{120}\mathsf{N}(\mathsf{N}+1)(\mathsf{N}+2)(\mathsf{N}+3)(\mathsf{N}+4)}}\right\}\nonumber\\
			&&
			\oplus\left\{\frac{1}{24}\mathsf{N}(\mathsf{N}+1)(\mathsf{N}-1)(\mathsf{N}+2)(\mathsf{N}+4)\right\}\nonumber\\
			&&
			\oplus\left\{{\overline{\frac{1}{24}\mathsf{N}(\mathsf{N}+1)(\mathsf{N}-1)(\mathsf{N}+2)(\mathsf{N}+4)}}\right\} \nonumber\\
			&&
			\oplus \left\{\frac{1}{4}\mathsf{N}^2(\mathsf{N}-1)(\mathsf{N}+3)\right\} \nonumber\\
			&&\oplus
			\left\{{\overline{\frac{1}{4}\mathsf{N}^2(\mathsf{N}-1)(\mathsf{N}+3)}}\right\}
		\end{eqnarray}

	\item\textsc{Specializing to $\mathsf{SO(10)}$ gauge group}
	
	\begin{align}
		\textnormal{Under}\quad \mathsf{SO(10)}&\longrightarrow  \mathsf{SU(5)} \otimes \mathsf{U(1)}\nonumber\\
		\nonumber \\
		\vect{{\phi}}^{(10)}_{\mu}:\quad \left[\mathsf{10}\right]&\longrightarrow \left\{\mathsf{5}\right\}\oplus \left\{{\overline{\mathsf{5}}}\right\}             \label{eq:1}    \\
		\nonumber \\
		\stags
		\boldsymbol{\varPhi}_{\mu\nu}^{(45)({A})}:\quad \left[\mathsf{45}\right]&\longrightarrow  \left\{\mathsf{24}\right\}\oplus\left\{\mathsf{10}\right\}\oplus \left\{{\overline{\mathsf{10}}}\right\}\oplus\left\{\mathsf{1}\right\} \tag{\stag}      \label{eq:2a}   \\
		\boldsymbol{\varPhi}_{\mu\nu}^{({S})}:\quad \left[\mathsf{54}\right]&\longrightarrow\left\{\mathsf{24}\right\}\oplus\left\{\mathsf{15}\right\}\oplus
		\left\{{\overline{\mathsf{15}}}\right\}    \tag{\stag}      \label{eq:2b}   \\
		\nonumber \\
		\stags
		\boldsymbol{\varPhi}_{\mu\nu\rho}^{{(120)}({A})}:\quad \left[\mathsf{120}\right]&\longrightarrow \left\{\mathsf{5}\right\}\oplus\left\{{\overline{\mathsf{5}}}\right\}\oplus\left\{\mathsf{10}\right\}
		\oplus \left\{{\overline{\mathsf{10}}}\right\}\oplus\left\{\mathsf{45}\right\}\oplus\left\{{\overline{\mathsf{45}}}\right\}     \tag{\stag}      \label{eq:3a}   \\
		\boldsymbol{\varPhi}_{\mu\nu\rho}^{{(210')}({\cal S})}:\quad \left[\mathsf{210'}\right]&\longrightarrow \left\{\mathsf{35}\right\}\oplus\left\{{\overline{\mathsf{35}}}\right\}\oplus\left\{\mathsf{70}\right\}
		\oplus \left\{{\overline{\mathsf{70}}}\right\}   \tag{\stag}      \label{eq:3b}\\
		\nonumber \\
		\stags
		\boldsymbol{\varPhi}_{\mu\nu\rho\sigma}^{{(210)}({A})}:\quad \left[\mathsf{210}\right]&\longrightarrow \left\{\mathsf{1}\right\}\oplus\left\{\mathsf{5}\right\}\oplus\left\{{\overline{\mathsf{5}}}\right\}\oplus\left\{\mathsf{10}\right\} \oplus\left\{{\overline{\mathsf{10}}}\right\}\oplus\left\{\mathsf{24}\right\}\oplus\left\{\mathsf{40}\right\}\nonumber\\ &\qquad\oplus\left\{{\overline{\mathsf{40}}}\right\} \oplus\left\{\mathsf{75}\right\}     \tag{\stag}      \label{eq:4a}   \\
		\boldsymbol{\varPhi}_{\mu\nu\rho\sigma}^{{(660)}({\cal S})}:\quad \left[\mathsf{660}\right]&\longrightarrow \left\{\mathsf{70}\right\}\oplus\left\{{\overline{\mathsf{70}}}\right\}\oplus\left\{\mathsf{160}\right\}\oplus \left\{{\overline{\mathsf{160}}}\right\}\oplus\left\{\mathsf{200}\right\}  \tag{\stag}      \label{eq:4b}\\
		\nonumber \\
		\stags
		\boldsymbol{\varXi}_{\mu\nu\rho\sigma\lambda}^{{(252)}({A})}:\quad \left[\mathsf{252}\right]&\longrightarrow \left\{\mathsf{1}\right\}\oplus\left\{\mathsf{1}\right\}\oplus\left\{\mathsf{5}\right\}\oplus\left\{{\overline{\mathsf{5}}}\right\}\oplus\left\{\mathsf{10}\right\}
		\oplus \left\{{\overline{\mathsf{10}}}\right\}\oplus\left\{\mathsf{15}\right\}\nonumber\\ &\qquad\oplus\left\{{\overline{\mathsf{15}}}\right\}\oplus\left\{\mathsf{45}\right\}
		\oplus\left\{{\overline{\mathsf{45}}}\right\}\oplus\left\{\mathsf{50}\right\}\oplus\left\{{\overline{\mathsf{50}}}\right\}     \tag{\stag}      \label{eq:5a}   \\
		\boldsymbol{\varPhi}_{\mu\nu\rho\sigma\lambda}^{{(1782)}({\cal S})}:\quad \left[\mathsf{1782}\right]&\longrightarrow \left\{\mathsf{126}\right\}\oplus\left\{{\overline{\mathsf{126}}}\right\}\oplus\left\{\mathsf{315}\right\} \oplus \left\{{\overline{\mathsf{315}}}\right\}\oplus\left\{\mathsf{450}\right\} \nonumber\\ &\qquad\oplus \left\{{\overline{\mathsf{450}}}\right\}  \tag{\stag}      \label{eq:5b}
	\end{align}
	Recall from Eqs.(\ref{decompositions})-(\ref{duality conditions}) that the real $\mathsf{252}$-dimensional $5^{\textnormal{th}}-$rank tensor of $\mathsf{SO(10)}$,
	$\boldsymbol{\varXi}_{\mu\nu\rho\sigma\lambda}^{{(252)}({A})}$ decomposes  into two
	complex $5^{\textnormal{th}}-$rank  tensors,
	${\boldsymbol{\varPhi}}_{\mu\nu\lambda\rho\sigma}^{{(126)}({A})}\left(\equiv \boldsymbol{\varOmega}_{\mu\nu\rho\sigma\lambda}^{({-})}\right)$ and
	$\boldsymbol{\varPhi}_{\mu\nu\lambda\rho\sigma}^{{(\overline{126})}({A})}\left(\equiv \boldsymbol{\varOmega}_{\mu\nu\rho\sigma\lambda}^{({+})}\right)$ each
	of dimensionality $\mathsf{126}$: $\boldsymbol{\varXi}^{{(252)}({A})}_{\mu\nu\lambda\rho\sigma}$=$\boldsymbol{\varPhi}_{\mu\nu\lambda\rho\sigma}^{{(\overline{126})}({A})}+\boldsymbol{\varPhi}_{\mu\nu\lambda\rho\sigma}^{{(126)}({A})}$,
	where
	\begin{equation}
		\begin{pmatrix}
			\boldsymbol{\varPhi}_{\mu\nu\lambda\rho\sigma}^{{(\overline{126})}({A})}\\
			\boldsymbol{\varPhi}_{\mu\nu\lambda\rho\sigma}^{(126)({A})}
		\end{pmatrix}
		=
		\frac{1}{2}\left(\delta_{\mu\alpha}
		\delta_{\nu\beta}\delta_{\rho\gamma}\delta_{\lambda\delta}\delta_{\sigma\theta}
		\pm
		\frac{i}{5!}\epsilon_{\mu\nu\rho\lambda\sigma\alpha\beta\gamma\delta\theta}\right)
		\boldsymbol{\varXi}_{\alpha\beta\gamma\delta\theta}^{{(252)}({A})}
	\end{equation}
	
	\begin{subequations}
		\begin{align}
			\boldsymbol{\varPhi}_{\mu\nu\lambda\rho\sigma}^{{(\overline{126})}({A})}:\quad\left[{\overline{\mathsf{126}}}\right]&\longrightarrow \left\{\mathsf{1}\right\}\oplus\left\{{{\mathsf{5}}}\right\}\oplus\left\{{\overline{\mathsf{10}}}\right\}\oplus\left\{{{\mathsf{15}}}\right\}
			\oplus\left\{{\overline{\mathsf{45}}}\right\}
			\oplus\left\{{{\mathsf{50}}}\right\}\\
			{\boldsymbol{\varPhi}}_{\mu\nu\lambda\rho\sigma}^{(126)({A})}:
			\quad \left[\mathsf{126}\right]&\longrightarrow \left\{\mathsf{1}\right\}\oplus\left\{{\overline{\mathsf{5}}}\right\}\oplus\left\{\mathsf{10}\right\}\oplus\left\{{\overline{\mathsf{15}}}\right\}\oplus\left\{\mathsf{
				45}\right\}
			\oplus\left\{{\overline{\mathsf{50}}}\right\}
		\end{align}
	\end{subequations}
\end{itemize}

\subsection{Technique for the evaluation of $\mathsf{SO(2N)}$ invariant tensor couplings. The Basic Theorem \cite{Nath:2001uw,Nath:2001yj,Syed:2004if,Syed:2005gd}}\label{su(n) reducible tensors}
Here we discuss a  technique  for the analysis of $\mathsf{SO(2N)}$ invariant couplings which allows a full exhibition of the $\mathsf{SU(N)}$ invariant content of the tensor representations. The technique utilizes a basis consisting of a specific set of reducible $\mathsf{SU(N)}$ tensors in terms of which the $\mathsf{SO(2N)}$ invariant couplings have a simple expansion.

We begin with the identification that the natural basis \cite{*}
for the expansion of the $\mathsf{SO(2N)}$ invariants is in terms of a specific
set of $\mathsf{SU(N)}$ reducible tensors, $\vect{{\phi}}_{c_k}$ and
$\vect{{\phi}}_{\overline c_k}$ which we define as
\begin{eqnarray}
	\begin{split}
		\vect{{A}}^k&\equiv{\vect{{\phi}}}_{c_k}\equiv{\vect{{\phi}}}_{2k}+i{\vect{{\phi}}}_{2k-1},\\
		\vect{{A}}_k&\equiv{\vect{{\phi}}}_{\overline c_k}\equiv{\vect{{\phi}}}_{2k}-i{\vect{{\phi}}}_{2k-1}
	\end{split}
\end{eqnarray}
Inversely,
\begin{equation}
	{\vect{{\phi}}}_{\mu}=
	\left\{\begin{aligned}
		&{\vect{{\phi}}}_{2k}\quad=\frac{1}{2}\left({\vect{{\phi}}}_{c_k}+{\vect{{\phi}}}_{\overline c_k}\right),   \\
		&{\vect{{\phi}}}_{2k-1}=-\frac{i}{2}\left({\vect{{\phi}}}_{c_k}-{\vect{{\phi}}}_{\overline c_k}\right),
	\end{aligned}\right.
\end{equation}
where $i,j,k,\dotso=1,\dotso,\mathsf{N}$ are $\mathsf{SU(N)}$ indices and $\mu,\nu,\rho,\dotso=1,\dotso,\mathsf{2N}$ are $\mathsf{SO(2N)}$ indices.
This can be extended immediately
to define the quantity $\boldsymbol{\varPhi}_{c_ic_j\bar c_k\dotso}$
with an arbitrary number of unbarred and barred indices where each
$c$ index can be expanded out so that
\begin{eqnarray}
	\vect{{A}}^i\vect{{A}}^j\vect{{A}}_k\dotsm={\boldsymbol{\varPhi}}_{c_ic_j\overline
		c_k\dotso}={\boldsymbol{\varPhi}}_{2ic_j\overline c_k\dotso}+i{\boldsymbol{\varPhi}}_{2i-1c_j\overline
		c_k\dotso}\quad{\textnormal{etc..}}
\end{eqnarray}
Thus, for example, the quantity  $\boldsymbol{\varPhi}_{c_{i_1}c_{i_2}\overline c_{i_3}\dotso c_{i_r}}$ is a sum of
$2^r$ terms gotten by expanding all the c indices.
${\boldsymbol{\varPhi}}_{c_i\overline c_jc_k\dotso\overline c_r}$ is completely antisymmetric in
the interchange of its $c-$indices whether unbarred or barred:
\begin{equation}
	{\boldsymbol{\varPhi}}_{c_i\overline c_jc_k\dotso\overline c_r}=-{\boldsymbol{\varPhi}}_{c_k\overline c_jc_i\dotso\overline c_r}\quad{\textnormal{etc..}}\nonumber\\
	\nonumber
\end{equation}
Further,
\begin{equation}
	{\boldsymbol{\varPhi}}^*_{c_i\overline c_jc_k\dotso\overline c_r}={\boldsymbol{\varPhi}}_{\overline c_ic_j\overline
		c_k\dotso c_r}\quad{\textnormal{etc..}}\nonumber\\
	\nonumber
\end{equation}

Use of quantities like ${\boldsymbol{\varPhi}}^*_{c_i\overline c_jc_k\dotso\overline
	c_r}$ are also useful in evaluating kinetic energy like terms such
as:
$-\frac{1}{2}\partial_{\mathpzc{A}}\vect{{\phi}}_{\mu}\partial^{\mathpzc{A}}\vect{{\phi}}_{\mu}^{\dagger}$,
$\frac{1}{4}\boldsymbol{\varPhi}_{\mu\nu}^{{\mathpzc{AB}}}\boldsymbol{\varPhi}_{\mu\nu{\mathpzc{AB}}}$, etc., where ${\mathpzc{A}}$ and $\mathpzc{B}$ are Dirac indices.

As a first example, consider  the $\mathsf{SO(10)}$ bilinear invariant $\mathsf{10}\cdot\mathsf{10}$: $\vect{{I}}_{\mu}\vect{\mathit{J}}_{\mu}$. We evaluate this invariant in terms of the specific set of $\mathsf{SU(5)}$ reducible tensors. To that end, write  $\vect{\mathit{I}}_{\mu}=\left(\vect{\mathit{I}}_{2i},\vect{\mathit{I}}_{2i-1}\right),~ \vect{\mathit{J}}_{\mu}=\left(\vect{\mathit{J}}_{2i},\vect{\mathit{J}}_{2i-1}\right)$, where $i=1, 2,\dotso, 5$ and $\mu=1, 2,\dotso, 5$. Then $\vect{\mathit{I}}_{\mu}\vect{\mathit{J}}_{\mu}=\vect{\mathit{I}}_{2i}\vect{\mathit{J}}_{2i}+\vect{\mathit{I}}_{2i-1}\vect{\mathit{J}}_{2i-1}$. Writing, $\vect{\mathit{I}}_{c_i}\equiv \vect{\mathit{I}}_{2i}+iX_{2i-1}$ and $\vect{\mathit{J}}_{c_i}\equiv \vect{\mathit{J}}_{2i}+iY_{2i-1}$, we get $\vect{\mathit{I}}_{c_i}\vect{\mathit{J}}_{\overline c_i}
=\vect{\mathit{I}}_{2i}\vect{\mathit{J}}_{2i}+\vect{\mathit{I}}_{2i-1}\vect{\mathit{J}}_{2i-1}+i\left(\vect{\mathit{I}}_{2i-1}\vect{\mathit{J}}_{2i}-\vect{\mathit{I}}_{2i}\vect{\mathit{J}}_{2i-1}\right)$ and $\vect{\mathit{I}}_{\overline c_i}\vect{\mathit{J}}_{c_i}
=\vect{\mathit{I}}_{2i}\vect{\mathit{J}}_{2i}+\vect{\mathit{I}}_{2i-1}\vect{\mathit{J}}_{2i-1}+i\left(\vect{\mathit{I}}_{2i}\vect{\mathit{J}}_{2i-1}-\vect{\mathit{I}}_{2i-1}\vect{\mathit{J}}_{2i}\right)$. Adding the last two equations give $\vect{\mathit{I}}_{c_i}\vect{\mathit{J}}_{\overline c_i}+\vect{\mathit{I}}_{\overline c_i}\vect{\mathit{J}}_{c_i}=
2\left(\vect{\mathit{I}}_{2i}\vect{\mathit{J}}_{2i}+\vect{\mathit{I}}_{2i-1}\vect{\mathit{J}}_{2i-1}\right)$ and hence, $$\mathsf{10}\cdot\mathsf{10}\Big|_{\mathsf{SO(10)}}:\quad\vect{\mathit{I}}_{\mu}\vect{\mathit{J}}_{\mu}
=\displaystyle\frac{1}{2^1}\Big(\vect{\mathit{I}}_{c_i}\vect{\mathit{J}}_{\overline c_i}+\vect{\mathit{I}}_{\overline c_i}\vect{\mathit{J}}_{c_i}\Big).$$

One can now exploit the last result to compute other $\mathsf{SO(10)}$ bilinear invariants such as $\mathsf{120}\cdot\mathsf{120}$: $\vect{\mathit{I}}_{\mu\nu\rho}\vect{\mathit{J}}_{\mu\nu\rho}$. Therefore, $\vect{\mathit{I}}_{\mu\nu\rho}\vect{\mathit{J}}_{\mu\nu\rho}=\frac{1}{2}(\vect{\mathit{I}}_{c_i\nu\rho}\vect{\mathit{J}}_{\overline c_i\nu\rho}
+\vect{\mathit{I}}_{\overline c_i\nu\rho}\vect{\mathit{J}}_{c_i\nu\rho})$. Repeating the process, we get $\vect{\mathit{I}}_{c_i\nu\rho}\vect{\mathit{J}}_{\overline c_i\nu\rho}=\frac{1}{2}(\vect{\mathit{I}}_{c_ic_j\rho}\vect{\mathit{J}}_{\overline c_i\overline c_j\rho}
+\vect{\mathit{I}}_{c_i\overline c_j \rho}\vect{\mathit{J}}_{\overline c_ic_j\rho})=\frac{1}{2}[\frac{1}{2}(\vect{\mathit{I}}_{c_ic_jc_k}\vect{\mathit{J}}_{\overline c_i\overline c_j\overline c_k}
+\vect{\mathit{I}}_{c_ic_j \overline c_k}\vect{\mathit{J}}_{\overline c_i\overline c_jc_k})+\frac{1}{2}(\vect{\mathit{I}}_{c_i\overline c_jc_k}\vect{\mathit{J}}_{\overline c_ic_j\overline c_k}
+\vect{\mathit{I}}_{c_i\overline c_j \overline c_k}\vect{\mathit{J}}_{\overline c_ic_jc_k})]$ and similarly, $\vect{\mathit{I}}_{\overline c_i\nu\rho}\vect{\mathit{J}}_{c_i\nu\rho}=\frac{1}{2}[\frac{1}{2}(\vect{\mathit{I}}_{\overline c_ic_jc_k}\vect{\mathit{J}}_{c_i\overline c_j\overline c_k}
+\vect{\mathit{I}}_{\overline c_ic_j \overline c_k}\vect{\mathit{J}}_{c_i\overline c_jc_k})+\frac{1}{2}(\vect{\mathit{I}}_{\overline c_i\overline c_jc_k}\vect{\mathit{J}}_{c_ic_j\overline c_k}
+\vect{\mathit{I}}_{\overline c_i \overline c_j \overline c_k}\vect{\mathit{J}}_{c_ic_jc_k})]$. Thus, $\vect{\mathit{I}}_{\mu\nu\rho}\vect{\mathit{J}}_{\mu\nu\rho}=
\frac{1}{2}(\frac{1}{4})(\vect{\mathit{I}}_{c_ic_jc_k}\vect{\mathit{J}}_{\overline c_i\overline c_j\overline c_k}
+\vect{\mathit{I}}_{c_ic_j \overline c_k}\vect{\mathit{J}}_{\overline c_i\overline c_jc_k}+\vect{\mathit{I}}_{c_i\overline c_jc_k}\vect{\mathit{J}}_{\overline c_ic_j\overline c_k}
+\vect{\mathit{I}}_{c_i\overline c_j \overline c_k}\vect{\mathit{J}}_{\overline c_ic_jc_k}+\vect{\mathit{I}}_{\overline c_ic_jc_k}\vect{\mathit{J}}_{c_i\overline c_j\overline c_k}
+\vect{\mathit{I}}_{\overline c_ic_j \overline c_k}\vect{\mathit{J}}_{c_i\overline c_jc_k}+ \vect{\mathit{I}}_{\overline c_i\overline c_jc_k}\vect{\mathit{J}}_{c_ic_j\overline c_k}
+\vect{\mathit{I}}_{\overline c_i \overline c_j \overline c_k}\vect{\mathit{J}}_{c_ic_jc_k})$. Finally, using the antisymmetry of the $c-$indices, we get
\begin{eqnarray*}
\mathsf{120}\cdot\mathsf{120}\Big|_{\mathsf{SO(10)}}:\quad\vect{\mathit{I}}_{\mu\nu\rho}\vect{\mathit{J}}_{\mu\nu\rho}&=&\frac{1}{2^3}
\Big(\vect{\mathit{I}}_{c_ic_jc_k}\vect{\mathit{J}}_{\overline c_i\overline c_j\overline c_k}+\vect{\mathit{I}}_{\overline c_i \overline c_j \overline c_k}\vect{\mathit{J}}_{c_ic_jc_k}+3\vect{\mathit{I}}_{c_ic_j \overline c_k}\vect{\mathit{J}}_{\overline c_i\overline c_jc_k}+3\vect{\mathit{I}}_{c_i\overline c_j \overline c_k}\vect{\mathit{J}}_{\overline c_ic_jc_k}\\
	&&\qquad+\vect{\mathit{I}}_{\overline c_ic_jc_k}\vect{\mathit{J}}_{c_i\overline c_j\overline c_k}\Big).
\end{eqnarray*}

Higher order $\mathsf{SO(10)}$ invariants in terms of specific set of $\mathsf{SU(N)}$ reducible tensors can also be easily computed. As a third example, consider the $\mathsf{SO(10)}$ trilinear invariant $\mathsf{10}\cdot\mathsf{10}\cdot\mathsf{45}$: $\vect{\mathit{I}}_{\mu}\vect{\mathit{J}}_{\nu} \vect{\mathit{L}}_{\mu\nu}$. Expanding,
$\vect{\mathit{I}}_{\mu}\vect{\mathit{J}}_{\nu} \vect{\mathit{L}}_{\mu\nu}=\frac{1}{2}(\vect{\mathit{I}}_{c_i}\vect{\mathit{J}}_{\nu} \vect{\mathit{L}}_{\overline c_i\nu}+
\vect{\mathit{I}}_{\overline c_i}\vect{\mathit{J}}_{\nu} \vect{\mathit{L}}_{c_i\nu})=\frac{1}{2}[\frac{1}{2}(\vect{\mathit{I}}_{c_i}\vect{\mathit{J}}_{c_j} \vect{\mathit{L}}_{\overline c_i\overline c_j}+
\vect{\mathit{I}}_{c_i}\vect{\mathit{J}}_{\overline c_j} \vect{\mathit{L}}_{\overline c_ic_j})
+\frac{1}{2}(\vect{\mathit{I}}_{\overline c_i}\vect{\mathit{J}}_{c_j} \vect{\mathit{L}}_{c_i\overline c_j}+
\vect{\mathit{I}}_{\overline c_i}\vect{\mathit{J}}_{\overline c_j} \vect{\mathit{L}}_{c_ic_j})]$. Rearranging,
\begin{eqnarray*}
\mathsf{10}\cdot\mathsf{10}\cdot\mathsf{45}\Big|_{\mathsf{SO(10)}}:\quad\vect{\mathit{I}}_{\mu}\vect{\mathit{J}}_{\nu} \vect{\mathit{L}}_{\mu\nu}&=&\frac{1}{2^2}\Big[\vect{\mathit{I}}_{c_i}\vect{\mathit{J}}_{c_j} \vect{\mathit{L}}_{\overline c_i\overline c_j}+\vect{\mathit{I}}_{\overline c_i}\vect{\mathit{J}}_{\overline c_j} \vect{\mathit{L}}_{c_ic_j}
+\big(\vect{\mathit{I}}_{c_i}\vect{\mathit{J}}_{\overline c_j}- \vect{\mathit{I}}_{\overline c_j}\vect{\mathit{J}}_{c_i}\big)\vect{\mathit{L}}_{\overline c_ic_j}
\Big].
\end{eqnarray*}

For our last example, we compute the quartic $\mathsf{SO(10)}$ invariant, $\mathsf{45}\cdot\mathsf{45}\cdot\mathsf{45}\cdot\mathsf{45}$: $\vect{\mathit{H}}_{\alpha\beta}\vect{\mathit{I}}_{\alpha\gamma} \vect{\mathit{J}}_{\beta\delta}\vect{\mathit{L}}_{\gamma\delta}$. The result is
\begin{eqnarray*}
	\mathsf{45}\cdot\mathsf{45}\cdot\mathsf{45}\cdot\mathsf{45}\Big|_{\mathsf{SO(10)}}:\quad\vect{\mathit{H}}_{\alpha\beta}\vect{\mathit{I}}_{\alpha\gamma} \vect{\mathit{J}}_{\beta\delta}\vect{\mathit{L}}_{\gamma\delta}&=&
	\frac{1}{2^4}\bigg[\Big(\vect{\mathit{H}}_{\bar c_i\bar c_j}\vect{\mathit{I}}_{c_ic_k}+\vect{\mathit{H}}_{c_i\bar c_j}\vect{\mathit{I}}_{\bar c_ic_k}\Big)
	\Big(\vect{\mathit{J}}_{c_j\bar c_l}\vect{\mathit{L}}_{\bar c_k c_l}+\vect{\mathit{J}}_{c_jc_l}\vect{\mathit{L}}_{\bar c_k \bar c_l}\Big)\\
	&&\qquad+\Big(\vect{\mathit{H}}_{\bar c_i\bar c_j}\vect{\mathit{I}}_{c_i\bar c_k}+\vect{\mathit{H}}_{c_i\bar c_j}\vect{\mathit{I}}_{\bar c_i\bar c_k}\Big)
	\Big(\vect{\mathit{J}}_{c_j\bar c_l}\vect{\mathit{L}}_{c_k c_l}+\vect{\mathit{J}}_{c_jc_l}\vect{\mathit{L}}_{c_k \bar c_l}\Big)\\
	&&\qquad+\Big(\vect{\mathit{H}}_{c_ic_j}\vect{\mathit{I}}_{\bar c_i\bar c_k}+\vect{\mathit{H}}_{\bar c_ic_j}\vect{\mathit{I}}_{c_ic_k}\Big)
	\Big(\vect{\mathit{J}}_{\bar c_j\bar c_l}\vect{\mathit{L}}_{c_k c_l}+\vect{\mathit{J}}_{\bar c_jc_l}\vect{\mathit{L}}_{c_k \bar c_l}\Big)\\
	&&\qquad+\Big(\vect{\mathit{H}}_{c_ic_j}\vect{\mathit{I}}_{\bar c_ic_k}+\vect{\mathit{H}}_{\bar c_ic_j}\vect{\mathit{I}}_{c_i\bar c_k}\Big)
	\Big(\vect{\mathit{J}}_{\bar c_j\bar c_l}\vect{\mathit{L}}_{\bar c_k c_l}+\vect{\mathit{J}}_{\bar c_jc_l}\vect{\mathit{L}}_{\bar c_k \bar c_l}\Big)
	\bigg]
\end{eqnarray*}

We now give some general results here,
\begin{subequations}
	\begin{align}
		\boldsymbol{\varPhi}_{\mu_1\mu_2\mu_3\dotso\mu_r}\boldsymbol{\varPhi}_{\mu_1\mu_2\mu_3\dotso\mu_r}
		&=\frac{1}{2^r}\left[\boldsymbol{\varPhi}_{c_{i_1}c_{i_2}c_{i_3}\dotso c_{i_{r-1}}c_{i_r}}
		\boldsymbol{\varPhi}_{\overline c_{i_1}\overline c_{i_2}\overline
			c_{i_3}\dotso\overline c_{i_{r-1}}\overline c_{i_r}}\right.\nonumber\\
		&\left.
		\qquad\quad+ \boldsymbol{\varPhi}_{\overline
			c_{i_1}c_{i_2}c_{i_3}\dotso c_{i_{r-1}}c_{i_r}}\boldsymbol{\varPhi}_{c_{i_1}\overline
			c_{i_2}\overline c_{i_3}\dotso\overline c_{i_{r-1}}\overline c_{i_r}}+\dotsb\right.\nonumber\\
		&\left.\qquad\quad+\boldsymbol{\varPhi}_{ c_{i_1}c_{i_2} c_{i_3}\dotso c_{i_{r-1}}\overline
			c_{i_r}}\boldsymbol{\varPhi}_{\overline c_{i_1}\overline c_{i_2}\overline c_{i_3}
			\dotso\overline c_{i_{r-1}}c_{i_r}}\right.\nonumber\\
		&\left.
		\qquad\quad+\boldsymbol{\varPhi}_{\overline c_{i_1}\overline c_{i_2}
			c_{i_3}\dotso c_{i_{r-1}} c_{i_r}}\boldsymbol{\varPhi}_{c_{i_1} c_{i_2}\overline
			c_{i_3}\dotso\overline c_{i_{r-1}}\overline c_{i_r}}+\dotsb\right.\nonumber\\
		&\left.
		\qquad\quad+\boldsymbol{\varPhi}_{c_{i_1} c_{i_2} c_{i_3}\dotso\overline c_{i_{r-1}}\overline
			c_{i_r}}\boldsymbol{\varPhi}_{\overline c_{i_1} \overline
			c_{i_2} \overline c_{i_3}\dotso c_{i_{r-1}}c_{i_r}}+\dotsb\right.\nonumber\\
		&\left.
		\qquad\quad+\boldsymbol{\varPhi}_{\overline c_{i_1}\overline c_{i_2} \overline
			c_{i_3}\dotso\overline c_{i_{r-1}} \overline c_{i_r}}\boldsymbol{\varPhi}_{c_{i_1}
			c_{i_2} c_{i_3}\dotso
			c_{i_{r-1}}c_{i_r}}\right],\\
\nonumber\\
		\boldsymbol{\varPhi}_{\mu_1\mu_2\mu_3\dotso\mu_r}\boldsymbol{\varPhi}_{\mu_1\mu_2\mu_3\dotso\mu_r}^{\dagger}
		&=\frac{1}{2^r}\left[\boldsymbol{\varPhi}_{c_{i_1}c_{i_2}c_{i_3}...c_{i_{r-1}}c_{i_r}}
		\boldsymbol{\varPhi}_{ c_{i_1} c_{i_2}
			c_{i_3}\dotso c_{i_{r-1}} c_{i_r}}^{\dagger}\right.\nonumber\\
		&\left.
		\qquad\quad+ \boldsymbol{\varPhi}_{\overline
			c_{i_1}c_{i_2}c_{i_3}\dotso c_{i_{r-1}}c_{i_r}}\boldsymbol{\varPhi}_{\overline c_{i_1}
			c_{i_2} c_{i_3}\dotso c_{i_{r-1}} c_{i_r}}^{\dagger}+\dotsb\right.\nonumber\\
		&\left.
		\qquad\quad+\boldsymbol{\varPhi}_{
			c_{i_1}c_{i_2} c_{i_3}\dotso c_{i_{r-1}}\overline c_{i_r}}\boldsymbol{\varPhi}_{
			c_{i_1} c_{i_2}c_{i_3}\dotso c_{i_{r-1}}\overline c_{i_r}}^{\dagger}\right.\nonumber\\
		&\left.
		\qquad\quad+\boldsymbol{\varPhi}_{\overline c_{i_1}\overline c_{i_2}
			c_{i_3}\dotso c_{i_{r-1}} c_{i_r}}\boldsymbol{\varPhi}_{\overline c_{i_1} \overline
			c_{i_2} c_{i_3}\dotso c_{i_{r-1}}c_{i_r}}^{\dagger} +\dotso \right.\nonumber\\
		&\left.
		\qquad\quad+\boldsymbol{\varPhi}_{
			c_{i_1} c_{i_2} c_{i_3}\dotso\overline c_{i_{r-1}}\overline
			c_{i_r}}\boldsymbol{\varPhi}_{c_{i_1}
			c_{i_2} c_{i_3}\dotso\overline c_{i_{r-1}}\overline c_{i_r}}^{\dagger}+\dotsb\right.\nonumber\\
		&\left.
		\qquad\quad+\boldsymbol{\varPhi}_{\overline c_{i_1}\overline c_{i_2} \overline
			c_{i_3}\dotso\overline c_{i_{r-1}} \overline c_{i_r}}\boldsymbol{\varPhi}_{\overline
			c_{i_1} \overline c_{i_2} \overline c_{i_3}\dotso
			\overline c_{i_{r-1}}\overline c_{i_r}}^{\dagger}\right]
	\end{align}
\end{subequations}

Finally, we make the observation that the object $\boldsymbol{\varPhi}_{c_ic_j\overline c_k\dotso c_r}$ transforms like a reducible representation of $\mathsf{SU(N)}$. Thus if we are able to compute the $\mathsf{SO(2N)}$ invariant couplings in terms of these reducible tensors of $\mathsf{SU(N)}$ then
there remains only the further step of decomposing the reducible tensors into their irreducible parts. See subsection \ref{so(10) tensors in terms of su(5) tensors}.

\subsection{Explicit decomposition of irreducible $\mathsf{SO(10)}$ tensors in terms of $\mathsf{SU(5)}$ irreducible tensors with canonically normalized kinetic energy terms \cite{Nath:2001uw,Nath:2001yj}}\label{so(10) tensors in terms of su(5) tensors}
\begin{itemize}
	\item \textsc{$\mathsf{SU(5)}$ tensors in the $\mathsf{10}-$plet of
		$\mathsf{SO(10)}$}\\
	
	The $\mathsf{10}-$plet of $\mathsf{SO(10)}$, $\vect{{\phi}}_{\mu}^{(10)}$, can  be decomposed
	in $\mathsf{SU(5)}$ components as follows
	\begin{eqnarray}
		\vect{{\phi}}^{(10)}_{\bar c_i}=h^{(10)}_{i};\qquad\vect{{\phi}}^{(10)}_{c_i}={h}^{(10)i}.
	\end{eqnarray}
	The tensors are already in their irreducible form and one can
	identify $\vect{{\phi}}_{c_i}^{(10)}$ with the $\mathsf{5}-$plet of Higgs and $\vect{{\phi}}_{\bar
		c_i}^{(10)}$ with the $\mathsf{\overline{5}}-$plet of Higgs. Now the kinetic energy for the $\mathsf{10}-$dimensional Higgs
	field is
	\begin{eqnarray}
		\begin{split}
			{\mathcal L}_{\textnormal{KE}}^{(10)}&= -\partial_{\mathpzc{A}}\vect{{\phi}}_{\mu}^{(10)}\partial^{\mathpzc{A}} \vect{{\phi}}_{\mu}^{(10)\dagger}\nonumber\\
			&=-\frac{1}{2}\left(\partial_{\mathpzc{A}}\vect{{\phi}}_{c_i}^{(10)}\partial^{\mathpzc{A}}\vect{{\phi}}_{c_i}^{(10)\dagger}+
			\partial_{\mathpzc{A}}\vect{{\phi}}_{\overline{c}_i}^{(10)}\partial^{\mathpzc{A}}\vect{{\phi}}_{\overline{c}_i}^{(10)\dagger}\right)\nonumber\\
			&=-\frac{1}{2}\left(\partial_{{\mathpzc{A}}}h^{(10)}_{i}\partial^{{\mathpzc{A}}}h_{i}^{(10)\dagger}
			+\partial_{{\mathpzc{A}}}h^{(10)i}\partial^{{\mathpzc{A}}}{h}^{(10)i\dagger}\right)\nonumber\\
			&\equiv-\partial_{{\mathpzc{A}}}H^{(10)}_{i}\partial^{{\mathpzc{A}}}H_{i}^{(10)\dagger}-\partial_{{\mathpzc{A}}}H^{(10)i}\partial^{{\mathpzc{A}}}{H}^{(10)i\dagger}.
		\end{split}
	\end{eqnarray}
	Therefore, we normalize the fields according to
	\begin{eqnarray}
		h^{(10)}_{i}={\sqrt{2}}H^{(10)}_{i};\qquad{
			h}^{(10)i}={\sqrt{2}}{H}^{(10)i}.
	\end{eqnarray}

	\item \textsc{$\mathsf{SU(5)}$ tensors in the $\mathsf{45}-$plet of
		$\mathsf{SO(10)}$}\\
	
	The $\mathsf{45-}$plet of $\mathsf{SO(10)}$ Higgs, $\boldsymbol{\varPhi}_{\mu\nu}^{(45)}$, can be
	decomposed in $\mathsf{SU(5)}$ multiplets as follows
	\begin{equation}
		\begin{split}
			\boldsymbol{\varPhi}_{\overline c_i\overline c_j}^{(45)}&={\mathsf h}_{ij}^{(45)};\\
			\boldsymbol{\varPhi}_{c_n\overline c_n}^{(45)}&=h^{(45)};
		\end{split}
		\qquad
		\begin{split}
			\boldsymbol{\varPhi}_{c_ic_j}^{(45)}&=h^{(45)ij};\\
			\boldsymbol{\varPhi}_{c_i\overline c_j}^{(45)}&=h_{j}^{(45)i}+ \frac{1}{5}\delta_j^ih^{(45)},
		\end{split}
	\end{equation}
	where $h^{(45)}$, $h^{(45)ij}$, $h_{ij}^{(45)}$ and $h_{j}^{(45)i}$ are the $\mathsf{1-}$plet, $\mathsf{10-}$plet, $\overline{\mathsf{10}}-$plet
	and $\mathsf{24-}$plet representations of $\mathsf{SU(5)}$, respectively. To normalize these $\mathsf{SU(5)}$ Higgs fields, we carry out a field redefinition:
	\begin{equation}
		\begin{split}
			h^{(45)}&=\sqrt {10} H^{(45)};\\
			h_{j}^{(45)i}&=\sqrt{2}H_{j}^{(45)i};
		\end{split}
		\qquad
		\begin{split}
			h_{ij}^{(45)}&=\sqrt 2 H_{ij}^{(45)};\\
			h^{(45)ij}&=\sqrt 2H^{(45)ij}.
		\end{split}
	\end{equation}
	In terms of the normalized fields, the kinetic energy of the
	$\mathsf{45-}$plet of Higgs,
	\begin{eqnarray}
		\begin{split}
			{\mathcal L}_{\textnormal{KE}}^{(45)}&=-\partial_{\mathpzc{A}}\boldsymbol{\varPhi}_{\mu\nu}^{(45)}\partial^{\mathpzc{A}}\boldsymbol{\varPhi}_{\mu\nu}^{(45)\dagger}\nonumber\\
			&=-\frac{1}{4}\left(\partial_{\mathpzc{A}}\boldsymbol{\varPhi}_{c_ic_j}^{(45)}\partial^{\mathpzc{A}}\boldsymbol{\varPhi}_{c_ic_j}^{(45)\dagger}+
			\partial_{\mathpzc{A}}\boldsymbol{\varPhi}_{\overline{c}_i\bar c_j}^{(45)}\partial^{\mathpzc{A}}\boldsymbol{\varPhi}_{\overline{c}_i\bar c_j}^{(45)\dagger}
			+2 \partial_{\mathpzc{A}}\boldsymbol{\varPhi}_{{c}_i\bar c_j}^{(45)}\partial^{\mathpzc{A}}\boldsymbol{\varPhi}_{{c}_i\bar c_j}^{(45)\dagger}\right)\nonumber\\
			&=-\frac{1}{4}\Bigg[\partial_{{\mathpzc{A}}}h^{(45)ij}\partial^{{\mathpzc{A}}}h^{(45)ij\dagger}+\partial_{{\mathpzc{A}}}h_{ij}^{(45)}\partial^{{\mathpzc{A}}}
			{h}_{ij}^{(45)\dagger}
			+2\bigg(\partial_{{\mathpzc{A}}}h_{j}^{(45)i}\partial^{\mathpzc{A}}
			h_{j}^{(45)i\dagger}\nonumber\\
			&~~~~\quad\quad+\frac{1}{5}\partial_{{\mathpzc{A}}}h^{(45)}\partial^{\mathpzc{A}}h^{(45)\dagger}\bigg)\Bigg]\nonumber\\
			&\equiv-\partial^{\mathpzc{A}} H^{(45)}\partial_{\mathpzc{A}} H^{(45)\dagger}-\frac{1}{2!}\partial^{\mathpzc{A}} H^{(45)}_{ij}\partial_{\mathpzc{A}}H^{(45)\dagger}_{ij}-\frac{1}{2!}
			\partial^{\mathpzc{A}} H^{(45)ij}\partial_{\mathpzc{A}} H^{(45)ij\dagger}\nonumber\\
			&~~~~\quad-\partial_{\mathpzc{A}} H^{(45)i}_j\partial^{\mathpzc{A}} H^{(45)i\dagger}_j.
		\end{split}
	\end{eqnarray}

	\item \textsc{$\mathsf{SU(5)}$ tensors in the $\mathsf{120}-$plet of
		$\mathsf{SO(10)}$}\\
	
	The $\mathsf{120-}$plet of $\mathsf{SO(10)}$, $\boldsymbol{\varPhi}_{\mu\nu\rho}^{(120)}$, can be decomposed in $\mathsf{SU(5)}$ components
	as follows
	\begin{equation}
		\begin{split}
			\boldsymbol{\varPhi}_{c_ic_j\bar c_k}^{(120)}&=h^{(120)ij}_k+\frac{1}{4}\left[\delta^i_k h^{(120)j}- \delta^j_k h^{(120)i}\right];\\
			\boldsymbol{\varPhi}_{c_i\bar c_j\bar c_k}^{(120)}&=h^{(120)i}_{jk}+\frac{1}{4}\left[\delta^i_j h^{(120)}_k-\delta^i_k h^{(120)}_j\right];\\
\boldsymbol{\varPhi}_{\bar c_nc_n c_i}^{(120)}&=h^{(120)i};\\
		\end{split}
		\qquad
		\begin{split}
			\boldsymbol{\varPhi}_{c_ic_jc_k}^{(120)}&=\epsilon^{ijklm}h^{(120)}_{lm},\\
			\boldsymbol{\varPhi}_{\bar c_i\bar c_j\bar c_k}^{(120)}&=\epsilon_{ijklm}h^{(120)lm},\\
		\boldsymbol{\varPhi}_{\bar c_nc_n \bar c_i}^{(120)}&=h^{(120)}_i,
		\end{split}
	\end{equation}
	where $h^{(120)}_{i}$, $h^{(120)i}$, $h^{(120)}_{ij}$, $h^{(120)ij}$, $h^{(120)ij}_k$, $h^{(120)i}_{jk}$
	are the $\mathsf{\overline{5}}-$plet , $\mathsf{5}-$plet , $\mathsf{\overline{10}}-$plet, $\mathsf{10}-$plet, $\mathsf{45}-$plet  and $\mathsf{\overline{45}}-$plet representations of $\mathsf{SU(5)}$.
	To normalize them we make the following redefinition of fields
	\begin{equation}
		\begin{split}
			h^{(120)i}&=\frac{4}{\sqrt 3} H^{(120)i};\\
			h^{(120)}_{i}&=\frac{4}{\sqrt 3} H^{(120)}_{i};
		\end{split}
		\qquad
		\begin{split}
			h^{(120)ij}&=\frac{1}{\sqrt 3} H^{(120)ij};\\
			h^{(120)ij}_k&=\frac{2}{\sqrt 3} H^{(120)ij}_k;
		\end{split}
		\qquad
		\begin{split}
			h^{(120)}_{ij}&=\frac{1}{\sqrt 3} H^{(120)}_{ij},\\
			h^{(120)i}_{jk}&=\frac{2}{\sqrt 3} H^{(120)i}_{jk}.
		\end{split}
	\end{equation}
	In terms of the redefined fields the kinetic energy term for the
	$\mathsf{120}$ multiplet takes on the form
	\begin{eqnarray}
		\begin{split}
			{\mathcal L}_{\textnormal{KE}}^{(120)}&=-\partial_{{\mathpzc{A}}}\boldsymbol{\varPhi}_{\mu\nu\lambda}^{(120)}\partial^{{\mathpzc{A}}}\boldsymbol{\varPhi}_{\mu\nu\lambda}^{(120)\dagger}\nonumber\\
			&=-\frac{1}{8}\left(\partial_{\mathpzc{A}}\boldsymbol{\varPhi}^{(120)}_{c_ic_jc_k}\partial^{\mathpzc{A}}\boldsymbol{\varPhi}_{c_ic_jc_k}^{(120)\dagger}
+\partial_{\mathpzc{A}}\boldsymbol{\varPhi}^{(120)}
			_{\overline{c}_i\overline{c}_j\overline{c}_k}\partial^{\mathpzc{A}}\boldsymbol{\varPhi}_{\overline{c}_i\overline{c}_j\overline{c}_k}^{(120)\dagger}
			+3\partial_{\mathpzc{A}}\boldsymbol{\varPhi}^{(120)}_{c_ic_j\overline{c}_k}\partial^{\mathpzc{A}}\boldsymbol{\varPhi}_{c_ic_j\overline{c}_k}^{(120)\dagger}\right.\\
			&\left.\qquad\quad+3\partial_{\mathpzc{A}}\boldsymbol{\varPhi}^{(120)}_{c_i\overline{c}_j\overline{c}_k}\partial^{\mathpzc{A}}
\boldsymbol{\varPhi}_{c_i\overline{c}_j\overline{c}_k}^{(120)\dagger}\right)\nonumber\\
			&=-\frac{1}{8}\left[12\partial_{\mathpzc{A}}h_{ij}\partial^{\mathpzc{A}}h_{ij}^{\dagger}+12\partial_{\mathpzc{A}}h^{ij}\partial^{\mathpzc{A}}h^{ij\dagger}
			+3\left(\partial_{\mathpzc{A}}h^{(120)ij}_k\partial^{\mathpzc{A}}h^{(120)ij\dagger}_k\right.\right.\nonumber\\
			&\left.\left.\qquad\quad+\frac{1}{2}\partial_{\mathpzc{A}}h^{(120)i} \partial^{\mathpzc{A}}h^{(120)i\dagger}\right)+3\left(\partial_{\mathpzc{A}}h_{ij}^{(120)k}\partial^{\mathpzc{A}}h_{ij}^{(120)k\dagger}+\frac{1}{2}\partial_{\mathpzc{A}}h_{i}^{(120)}
			\partial^{\mathpzc{A}}h_{i}^{(120)\dagger}\right)\right]\nonumber\\
			&\equiv-\frac{1}{2!}\partial_{{\mathpzc{A}}}H^{(120)ij}\partial^{{\mathpzc{A}}} H^{(120)ij\dagger}
			-\frac{1}{2!}\partial_{{\mathpzc{A}}}  H^{(120)}_{ij}\partial^{{\mathpzc{A}}}  H^{(120)\dagger}_{ij}
			-\frac{1}{2!}\partial_{{\mathpzc{A}}} H^{(120)ij}_k \partial^{{\mathpzc{A}}}H^{(120)ij\dagger}_k\\
&\quad-\frac{1}{2!}\partial_{{\mathpzc{A}}} H^{(120)i}_{jk}\partial^{{\mathpzc{A}}}H^{(120)i\dagger}_{jk}
			-\partial_{{\mathpzc{A}}} H^{(120)i}\partial^{{\mathpzc{A}}} H^{(120)i\dagger}
			-\partial_{{\mathpzc{A}}} H^{(120)}_{i}\partial^{{\mathpzc{A}}} H^{(120)\dagger}_{i}.
		\end{split}
	\end{eqnarray}

	\item \textsc{$\mathsf{SU(5)}$ tensors in the $\mathsf{210}-$plet of
		$\mathsf{SO(10)}$}\\
	
	The $\mathsf{210}-$plet of $\mathsf{SO(10)}$, $\boldsymbol{\varPhi}_{\mu\nu\rho\sigma}^{(210)}$, has
	the following decomposition in $\mathsf{SU(5)}$ multiplets
\begin{eqnarray}
\begin{split}
\boldsymbol{\varPhi}_{c_ic_j\overline c_k\overline c_l}^{(210)}&=h^{(210)ij}_{kl}+\frac{1}{3} \left[\delta_l^ih^{(210)j}_{k}-\delta_k^ih^{(210)j}_{l}+ \delta_k^jh^{(210)i}_{l}-\delta_l^jh^{(210)i}_{k}\right]\\
&\quad+\frac{1}{20}\left(\delta_l^i \delta_k^j-\delta_k^i\delta_l^j\right)h^{(210)},\\
\boldsymbol{\varPhi}_{c_ic_jc_k\overline c_l}^{(210)}&=h^{(210)ijk}_{l}+\frac{1}{3}\left(\delta_l^kh^{(210)ij}-\delta_l^jh^{(210)ik}+\delta_l^ih^{(210)jk}\right),\\
\boldsymbol{\varPhi}_{\overline c_i\overline c_j\overline c_kc_l}^{(210)}&=h^{(210)l}_{ijk} +\frac{1}{3}\left(\delta_k^lh^{(210)}_{ij}-\delta_j^lh^{(210)}_{ik}+ \delta_i^lh^{(210)}_{jk}\right),\\
\boldsymbol{\varPhi}_{c_i\overline c_jc_m\overline c_m}^{(210)}&=h^{(210)i}_{j}+\frac{1}{5}\delta_j^ih^{(210)},\\
\boldsymbol{\varPhi}_{\overline c_i\overline c_j \overline c_k\overline c_l}^{(210)}&=\frac{1}{24}\epsilon_{ijklm}h^{(210)m},\\
\boldsymbol{\varPhi}_{c_ic_jc_kc_l}^{(210)}&=\frac{1}{24}\epsilon^{ijklm}h^{(210)}_{m},\\
\boldsymbol{\varPhi}_{c_ic_jc_m\overline c_m}^{(210)}&=h^{(210){ij}},\\
\boldsymbol{\varPhi}_{\overline c_i\overline c_j \overline c_mc_m}^{(210)}&=h^{(210)}_{ij},\\
\boldsymbol{\varPhi}_{c_m\overline c_mc_n\overline c_n}^{(210)}&=h^{(210)},
\end{split}
\end{eqnarray}
	where $h^{(210)}$, $h^{(210)i}$, $h^{(210)}_{i}$, $h^{(210)ij}$, $h^{(210)}_{ij}$, $h_{j}^{(210)i}$, $h_{l}^{(210)ijk}$; $h_{jkl}^{(210)i}$ and $h_{kl}^{(210)ij}$
	are the $\mathsf{1}-$plet, $\mathsf{5}-$plet, $\overline {\mathsf{5}}-$plet, $\mathsf{10}-$plet, $\overline{\mathsf{10}}-$plet, $\mathsf{24}-$plet, $\mathsf{40}-$plet, $\overline {\mathsf{40}}-$plet and $\mathsf{75}-$plet
	representations of $\mathsf{SU(5)}$, respectively. To normalize these fields, we carry out a field redefinition
	\begin{equation}
		\begin{split}
			h^{(210)}&=4\sqrt{\frac{5}{3}}H^{(210)};\\
			h^{(210)i}&=8\sqrt{6}H^{(210)i};\\
			h_{i}^{(210)}&=8\sqrt{6}H_{i}^{(210)};
		\end{split}
		\quad
		\begin{split}
			h^{(210)ij}&=\sqrt{2}H^{(210)ij};\\
			h_{ij}^{(210)}&=\sqrt{2}H_{ij}^{(210)};\\
			h_{j}^{(210)i}&=\sqrt{2}H_{j}^{(210)i};
		\end{split}
		\quad
		\begin{split}
			h_{l}^{(210)ijk}&=\sqrt{\frac{2}{3}}H_{l}^{(210)ijk},\\
			h_{jkl}^{(210)i} &=\sqrt{\frac{2}{3}}H_{jkl}^{(210)i},\\
			h_{kl}^{(210)ij} &=\sqrt{\frac{2}{3}}H_{kl}^{(210)ij}.
		\end{split}
	\end{equation}
	Now the kinetic energy for the $\mathsf{210}-$dimensional Higgs field  in terms of the
	redefined fields takes the form
	\begin{eqnarray}
		\begin{split}
			{\mathcal L}_{\textnormal{KE}}^{(210)}&= -\partial_{\mathpzc{A}}\boldsymbol{\varPhi}_{\mu\nu\rho\lambda}^{(210)}\partial^{\mathpzc{A}} \boldsymbol{\varPhi}_{\mu\nu\rho\lambda}^{(210)\dagger}\nonumber\\
			&=-\frac{1}{16}\left(\partial_{\mathpzc{A}}\boldsymbol{\varPhi}_{c_ic_jc_kc_l}^{(210)}\partial^{\mathpzc{A}}\boldsymbol{\varPhi}_{c_ic_jc_kc_l}^{(210)\dagger}+
			\partial_{\mathpzc{A}}\boldsymbol{\varPhi}_{\overline{c}_i\overline{c}_j\overline{c}_k\overline{c}_l}^{(210)}\partial^{\mathpzc{A}}\boldsymbol{\varPhi}_{\overline{c}_i\overline{c}_j\overline{c}_k\overline{c}_l}^{(210)\dagger}
			+4\partial_{\mathpzc{A}}\boldsymbol{\varPhi}_{c_ic_jc_k\overline{c}_l}^{(210)}\partial^{\mathpzc{A}}\boldsymbol{\varPhi}_{c_ic_jc_k\overline{c}_l}^{(210)\dagger}\right.\nonumber\\
			&\left.\quad\qquad\quad+4\partial_{\mathpzc{A}}\boldsymbol{\varPhi}_{c_i\overline{c}_j\overline{c}_k\overline{c}_l}^{(210)}\partial^{\mathpzc{A}}
\boldsymbol{\varPhi}_{c_i\overline{c}_j\overline{c}_k\overline{c}_l}^{(210)\dagger}
			+6\partial_{\mathpzc{A}}\boldsymbol{\varPhi}_{c_i{c}_j\overline{c}_k\overline{c}_l}^{(210)}\partial^{\mathpzc{A}}\boldsymbol{\varPhi}
			_{c_i{c}_j\overline{c}_k\overline{c}_l}^{(210)\dagger}\right)\nonumber\\
			&\equiv-\partial_{\mathpzc{A}} H^{(210)}\partial^{\mathpzc{A}}
			H^{(210)\dagger} -\partial_{\mathpzc{A}}H^{(210)i}\partial^{\mathpzc{A}} H^{(210)i\dagger}-\partial_{\mathpzc{A}} H_i^{(210)}\partial^{\mathpzc{A}} H_{i}^{(210)\dagger}\nonumber\\
			&\quad-\frac{1}{2!}\partial_{\mathpzc{A}} H_{ij}^{(210)}\partial^{\mathpzc{A}}H_{ij}^{(210)\dagger}-\frac{1}{2!}\partial_{\mathpzc{A}}H^{(210)ij}\partial^{\mathpzc{A}}H^{(210)ij\dagger}
-\partial_{\mathpzc{A}}H_j^{(210)i}\partial^{\mathpzc{A}}H_j^{(210)i\dagger} \nonumber\\
			&
			\quad-\frac{1}{3!}\partial_{\mathpzc{A}}H^{(210)l}_{ijk}\partial^{\mathpzc{A}}H^{(210)l\dagger}_{ijk}-\frac{1}{3!}\partial_{\mathpzc{A}} H_l^{(210)ijk} \partial^{\mathpzc{A}}H_l^{(210)ijk\dagger}-\frac{1}{2!}\frac{1}{2!}\partial_{\mathpzc{A}}H_{kl}^{(210)ij}\partial^{\mathpzc{A}}H_{kl}^{(210)ij\dagger}.
		\end{split}
	\end{eqnarray}

	\item \textsc{$\mathsf{SU(5)}$ tensors in the $\mathsf{\overline{126}+126}-$plet of
		$\mathsf{SO(10)}$}\\
	
	The $\mathsf{252}-$dimensional tensor of $\mathsf{SO(10)}$, $\boldsymbol{\varXi}_{\mu\nu\lambda\rho\sigma}^{(252)}$, has
	the following decomposition in $\mathsf{SU(5)}$ multiplets
	\begin{eqnarray}
		\begin{split}
			\boldsymbol{\varXi}_{c_ic_jc_k\bar c_l\bar c_m}^{(252)}&=h^{{(\overline{126})} ijk}_{lm}+\frac{1}{2}\bigg(\delta^i_lh^{{(126)}jk}_m -\delta^j_lh^{{(126)}ik}_m+\delta^k_lh^{{(126)}ij}_m-\delta^i_mh^{{(126)}jk}_l+\delta^j_mh^{{(126)}ik}_l \\
&\quad-\delta^k_mh^{{(126)}ij}_l\bigg)+\frac{1}{12}\bigg(\delta^i_l\delta^j_mh^{{(\overline{126})}k}
-\delta^j_l\delta^i_mh^{{(\overline{126})}k}-\delta^i_l\delta^k_mh^{{(\overline{126})}j}
+\delta^k_l\delta^i_mh^{{(\overline{126})}j}\\
&\quad+\delta^j_l\delta^k_mh^{{(\overline{126})}i}-\delta^k_l\delta^j_mh^{{(\overline{126})}i}\bigg);\\
			\boldsymbol{\varXi}_{c_ic_j\bar c_k\bar c_l\bar c_m}^{(252)}&=h^{{(126)}ij}_{klm}+\frac{1}{2} \bigg(\delta^i_kh^{{(\overline{126})}j}_{lm}-\delta^i_lh^{{(\overline{126})}j}_{km}  +\delta^i_mh^{{(\overline{126})}j}_{kl}-\delta^j_kh^{{(\overline{126})}i}_{lm}+\delta^j_lh^{{(\overline{126})}i}_{km} \\
&\quad -\delta^j_mh^{{(\overline{126})}i}_{kl}\bigg)+\frac{1}{12}\bigg(\delta^i_k\delta^j_lh_m^{(126)}-\delta^i_k\delta^j_mh_l^{(126)}
-\delta^i_l\delta^j_kh_m^{(126)}+\delta^i_l\delta^j_mh_l^{(126)}\\
&\quad+\delta^i_m\delta^j_kh_l^{(126)}-\delta^i_m\delta^j_lh_k^{(126)}\bigg).\\
	\boldsymbol{\varXi}_{c_i\bar c_j\bar c_k\bar c_l\bar c_m}^{(252)}&=\epsilon_{jklmn}h^{{(\overline{126})}ni}_{(S)}+\frac{1}{2} \bigg(\delta^i_j\epsilon_{klmpq}-\delta^i_k\epsilon_{jlmpq}+\delta^i_l\epsilon_{jkmpq}
			-\delta^i_m\epsilon_{jklpq} \bigg)h^{{(126)}pq},\\
\boldsymbol{\varXi}_{c_ic_jc_kc_l\bar c_m}^{(252)}&=\epsilon^{ijkln} h_{(S)nm}^{{(126)} }+\frac{1}{2}\bigg(\delta^i_m\epsilon^{jklpq}-\delta^j_m\epsilon^{iklpq}+\delta^k_m\epsilon^{ijlpq}-\delta^l_m\epsilon^{ijkpq} \bigg)h^{(\overline{126})}_{pq};\\
\boldsymbol{\varXi}_{c_i\bar c_j\bar c_k\bar c_nc_n}^{(252)}&=h^{{(\overline{126})}i}_{jk}+\frac{1}{4}\left(\delta^i_kh_j^{(126)} -\delta^i_jh_k^{(126)}\right);\\
\boldsymbol{\varXi}_{\bar c_i c_j c_k c_n\bar c_n}^{(252)}&=h_{i}^{{(126)}jk}+\frac{1}{4}\left(\delta_i^kh^{{(\overline{126})}j}-\delta_i^jh^{{(\overline{126})}k}\right),\\
	\boldsymbol{\varXi}_{c_ic_jc_kc_lc_m}^{(252)}&=\epsilon^{ijklm} h^{(\overline{126})};\\		
\boldsymbol{\varXi}_{\bar c_i\bar c_j\bar c_k\bar c_l\bar c_m}^{(252)}&= \epsilon_{ijklm}h^{(126)} ,\\
\boldsymbol{\varXi}_{c_ic_jc_k\bar c_nc_n}^{(252)}&=\epsilon^{ijklm}h^{(\overline{126})}_{lm};\\
\boldsymbol{\varXi}_{\bar c_i\bar c_j\bar c_kc_n\bar c_n}^{(252)}&=\epsilon_{ijklm}h^{{(126)}lm},\\
			\boldsymbol{\varXi}_{c_i\bar c_nc_n\bar c_pc_p}^{(252)}&=h^{{(\overline{126})}i};\\
			\boldsymbol{\varXi}_{\bar c_i\bar c_nc_n\bar c_pc_p}^{(252)}&=h_{i}^{(126)}.
		\end{split}
	\end{eqnarray}
	The fields that appear above are not yet properly normalized. To normalize the fields we carry out a field redefinition, so that,
	\begin{eqnarray}
		\begin{split}
			h^{(\overline{126})} &=\frac{2}{\sqrt {15}}H^{(\overline{126})} ;\\
			h^{{(\overline{126})}i}&=4\sqrt{\frac{ 2}{5}}H^{{(\overline{126})}i};\\
			h^{(\overline{126})}_{lm}&=\sqrt{\frac{ 2}{15}}H^{(\overline{126})}_{lm};\\
			h^{{(\overline{126})}ni}_{(S)}&=\sqrt{\frac{2}{15}}H^{{(\overline{126})}ni}_{(S)};\\
			h^{{(\overline{126})}i}_{jk}&=2\sqrt{\frac{2}{15}} H^{{(\overline{126})}i}_{jk};\\
			h^{{(\overline{126})} ijk}_{lm}&=\frac{2}{\sqrt {15}} H^{{(\overline{126})} ijk}_{lm};
		\end{split}
		\qquad\quad
		\begin{split}
			h^{(126)} &=\frac{2}{\sqrt {15}}H^{(126)} ,\\
			h_i^{(126)}&=4\sqrt{\frac{ 2}{5}}H_i^{(126)},\\
			h^{{(126)}lm}&=\sqrt{\frac{ 2}{15}}H^{{(126)}lm},\\
			h_{(S)ni}^{{(126)}}&= \sqrt{\frac{2}{15}}H_{(S)ni}^{{(126)} },\\
			h_i^{{(126)}jk}&=2\sqrt{\frac{2}{15}} H_i^{{(126)}jk},\\
			h_{ijk}^{{(126)}lm}&=\frac{2}{\sqrt {15}} H_{ijk}^{{(126)}lm}.
		\end{split}
	\end{eqnarray}
	The  kinetic energy for the $\mathsf{252}-$plet field ${\mathcal L}_{\textnormal{KE}}^{(252)}=-\partial_{{\mathpzc{A}}}\boldsymbol{\varXi}_{\mu\nu\lambda\rho\sigma}^{(252)}
	\partial^{{\mathpzc{A}}}\boldsymbol{\varXi}_{\mu\nu\lambda\rho\sigma}^{(252)\dagger}$ in terms of the normalized fields is then given by
	\begin{eqnarray}
		\begin{split}
			{\mathcal L}_{\textnormal{KE}}^{(252)}&=-\partial_{{\mathpzc{A}}}\boldsymbol{\varXi}_{\mu\nu\lambda\rho\sigma}^{(252)}
			\partial^{{\mathpzc{A}}}\boldsymbol{\varXi}_{\mu\nu\lambda\rho\sigma}^{(252)\dagger}\nonumber\\
			&=-\frac{1}{32}\bigg(\partial_{\mathpzc{A}}\boldsymbol{\varXi}_{c_ic_jc_kc_lc_m}^{(252)}\partial^{\mathpzc{A}}\boldsymbol{\varXi}_{c_ic_jc_kc_lc_m}^{(252)\dagger}
			+\partial_{\mathpzc{A}}\boldsymbol{\varXi}_{\overline{c}_i\overline{c}_j\overline{c}_k\overline{c}_l\overline{c}_m}^{(252)}\partial^{\mathpzc{A}}
\boldsymbol{\varXi}_{\overline{c}_i\overline{c}_j\overline{c}_k\overline{c}_l\overline{c}_m}^{(252)\dagger}\nonumber\\
			&\quad\qquad\quad+5\partial_{\mathpzc{A}}\boldsymbol{\varXi}_{c_ic_jc_kc_l\overline{c}_m}^{(252)}\partial^{\mathpzc{A}}
\boldsymbol{\varXi}_{c_ic_jc_kc_l\overline{c}_m}^{(252)\dagger}
			+5\partial_{\mathpzc{A}}\boldsymbol{\varXi}_{{c}_i\overline{c}_j\overline{c}_k\overline{c}_l\overline{c}_m}^{(252)}\partial^{\mathpzc{A}}\boldsymbol{\varXi}
			_{{c}_i\overline{c}_j\overline{c}_k\overline{c}_l\overline{c}_m}^{(252)\dagger}\nonumber\\
			&\quad\qquad\quad +10\partial_{\mathpzc{A}}\boldsymbol{\varXi}_{c_ic_jc_k\overline{c}_l\overline{c}_m}^{(252)}\partial^{\mathpzc{A}}
\boldsymbol{\varXi}_{c_ic_jc_k\overline{c}_l\overline{c}_m}^{(252)\dagger}
			+10\partial_{\mathpzc{A}}\boldsymbol{\varXi}_{{c}_i{c}_j\overline{c}_k\overline{c}_l\overline{c}_m}^{(252)}\partial^{\mathpzc{A}}\boldsymbol{\varXi}
			_{{c}_i{c}_j\overline{c}_k\overline{c}_l\overline{c}_m}^{(252)\dagger}\bigg)\nonumber\\
			&\equiv-\partial_{{\mathpzc{A}}} H^{(126)}\partial^{{\mathpzc{A}}} H^{{(126)}\dagger}-\partial_{{\mathpzc{A}}}H^{(\overline{126})}\partial^{{\mathpzc{A}}} H^{(\overline{126})\dagger}
			- \partial_{{\mathpzc{A}}} H_{i}^{(126)}\partial^{{\mathpzc{A}}} H^{{(126)}\dagger}_i\nonumber\\
			&~\quad- \partial_{{\mathpzc{A}}} H^{{(\overline{126})} i}\partial^{{\mathpzc{A}}} H^{{(\overline{126})} i\dagger}- \frac{1}{2!}\partial_{{\mathpzc{A}}} H^{{(126)} ij}\partial^{{\mathpzc{A}}} H^{{(126)} ij\dagger}
			- \frac{1}{2!}\partial_{{\mathpzc{A}}} H_{ij}^{(\overline{126})} \partial^{{\mathpzc{A}}} H_{ij}^{{(\overline{126})}\dagger}\nonumber\\
			&~\quad- \frac{1}{2!}\partial_{{\mathpzc{A}}} H_{(S)ij}^{{(126)}}
			\partial^{{\mathpzc{A}}} H^{({(126)}\dagger}_{(S)ij}- \frac{1}{2!}\partial_{{\mathpzc{A}}} H^{{(\overline{126})} ij}_{(S)}\partial^{{\mathpzc{A}}} H^{{(\overline{126})}ij\dagger}_{(S)}-\frac{1}{2!}\partial_{{\mathpzc{A}}} H^{{(126)}jk}_i\partial^{{\mathpzc{A}}} H_{i}^{{(126)}jk\dagger}\nonumber\\
			&~\quad
			-\frac{1}{2!}\partial_{{\mathpzc{A}}} H_{jk}^{(\overline{126})i}\partial^{{\mathpzc{A}}} H_{jk}^{(\overline{126})i\dagger}-\frac{1}{3!2!}\partial_{{\mathpzc{A}}} H_{ijk}^{{(\overline{126})}lm}
			\partial^{{\mathpzc{A}}} H^{{(\overline{126})} lm\dagger}_{ijk}\nonumber\\
			&~\quad-\frac{1}{3!2!}\partial_{{\mathpzc{A}}} H^{(\overline{126})ijk}_{lm}
			\partial^{{\mathpzc{A}}} H^{(\overline{126})ijk\dagger}_{lm},
		\end{split}
	\end{eqnarray}
	where $H^{(126)}(H^{(\overline{126})})$, $H_{i}^{(126)}$,  $H^{{(\overline{126})}i}$, $H^{{(126)}ij}$, $H_{ij}^{(\overline{126})}$, $H_{(S)ij}^{{(126)}}$, $H^{{(\overline{126})}ij}_{(S)}$, $H^{{(126)}jk}_i$, $H_{jk}^{{(\overline{126})}i}$, $H_{ijk}^{{(126)}lm}$, $H^{{(\overline{126})}ijk}_{lm}$
	are the $\mathsf{1},~ \mathsf{\overline{5}},~ \mathsf{5},~ \mathsf{10}, ~\mathsf{\overline{10}},~
	\mathsf{\overline{15}}, ~\mathsf{15}, ~\mathsf{45}, ~\mathsf{\overline{45}}, ~\mathsf{\overline{50}}, ~\mathsf{50}$
	representations of $\mathsf{SU(5)}$, respectively.
\end{itemize}

\subsection{Extraction and normalization of $\mathsf{SU(3)_C}$ triplets, $\mathsf{SU(2)_L}$ doublets and $\mathsf{SU(3)_C}\times \mathsf{SU(2)_L}\times\mathsf{U(1)_Y}$ singlets in $\mathsf{SU(5)}$ fields \cite{Nath:2015kaa}}\label{singlets,doublets and triplets of SM}
 SM doublets, triplets and singlets contained in the $\mathsf{SU(5)}$ fields are needed in the spontaneous breaking of $\mathsf{SO(10)}-$GUT  and electroweak symmetry. Let's first identify them:
\[
\mathsf{SU{(2)}}~ {\textnormal{weak doublets}}:
 \left\{
\begin{matrix}
{}^{({5}_{{10}})}\!{\mathsf D}^{a},~ {}^{(\overline{5}_{{10}})}\!{\mathsf D}_{a}\\
{}^{({5}_{{120}})}\!{\mathsf D}^{a},~ {}^{(\overline{5}_{{120}})}\!{\mathsf D}_{a},~{}^{({45}_{{120}})}\!{\mathsf D}^{a},~{}^{(\overline{45}_{{120}})}\!{\mathsf D}_{a}\\
         {}^{(\overline{5}_{{126}})}\!{\mathsf D}_{a},~ {}^{({45}_{{126}})}\!{\mathsf D}^{a}\\
            {}^{({5}_{\overline{126}})}\!{\mathsf D}^{a},~ {}^{(\overline{45}_{\overline{126}})}\!{\mathsf D}_{a}\\
            {}^{({5}_{{210}})}\!{\mathsf D}^{a},~ {}^{(\overline{5}_{{210}})}\!{\mathsf D}_{a},
\end{matrix}
\right.
\]
\[
\mathsf{SU{(3)}}~ {\textnormal{color triplets}}:
 \left\{
\begin{matrix}
{}^{({5}_{{10}})}\!{\mathsf T}^{\alpha},~ {}^{(\overline{5}_{{10}})}\!{\mathsf T}_{\alpha}\\
{}^{({5}_{{120}})}\!{\mathsf T}^{\alpha},~ {}^{(\overline{5}_{{120}})}\!{\mathsf T}_{\alpha},~{}^{({45}_{{120}})}\!{\mathsf T}^{\alpha},~{}^{(\overline{45}_{{120}})}\!{\mathsf T}_{\alpha}\\
         {}^{(\overline{5}_{{126}})}\!{\mathsf T}_{\alpha},~ {}^{({45}_{{126}})}\!{\mathsf T}^{\alpha},~{}^{(\overline{50}_{{126}})}\!{\mathsf T}_{\alpha}\\
            {}^{({5}_{\overline{126}})}\!{\mathsf T}^{\alpha},~ {}^{(\overline{45}_{\overline{126}})}\!{\mathsf T}_{\alpha},~{}^{({50}_{\overline{126}})}\!{\mathsf T}_{\alpha}\\
            {}^{({5}_{{210}})}\!{\mathsf T}^{\alpha},~ {}^{(\overline{5}_{{210}})}\!{\mathsf T}_{\alpha}
\end{matrix}
\right.
\]
and
\[
\mathsf{SU(3)_C}\times \mathsf{SU(2)_L}\times\mathsf{U(1)_Y}~ {\textnormal{singlets}}:
 \left\{
\begin{matrix}
\mathsf S_{1_{_{{45}}}} ,~ \mathsf S_{24_{_{{45}}}}\\
\mathsf S_{1_{_{{126}}}}\\
\mathsf S_{1_{_{\overline{126}}}}\\
\mathsf S_{1_{_{{210}}}} ,~ \mathsf S_{24_{_{{210}}}},~ \mathsf S_{75_{_{{210}}}},
\end{matrix}
\right.
\]
where for example, the notation ${}^{(\overline{50}_{{126}})}\!{\mathsf T}_{\alpha}$ means the triplet field with canonically normalized kinetic energy term and residing in the $\mathsf{SU(5)}$'s $\overline{50}-$plet of $\mathsf{SO(10)}$'s $\mathsf{126}-$plet. Here $\alpha,~ \beta, ~\gamma=1,~2,~3$ are $\mathsf{SU(3)}$ color indices, while $a,~b=4,~5$ are $\mathsf{SU(2)}$ weak indices.\\

\begin{itemize}
\item \textsc{SM doublets and triplets in $\mathsf{5}-$plet and $\overline{\mathsf{5}}-$plet of $\mathsf{SU(5)}$}\\

$\mathsf{5}$ and $\overline{\mathsf{5}}$ fields of $\mathsf{SU(5)}$  have the following  $\mathsf{SU(3)_C\times SU(2)_L \times U(1)_Y}$ decomposition \cite{Slansky:1981yr}
     \begin{eqnarray*}
{H}^{(\#)i}~(\mathsf{5})&=&(\mathsf{1,2},3)~ {}^{({5}_{{\#}})}\!{\mathcal D}^{a}+(\mathsf{3,1},-2)~{}^{({5}_{{\#}})}\!{\mathcal T}^{\alpha},\\
{H}^{(\#)}_i~(\overline{\mathsf{5}})&=&(\mathsf{1,2},-3) ~{}^{(\overline{5}_{{\#}})}\!{\mathcal D}_{a}+(\mathsf{\overline{3},1},2)~{}^{(\overline{5}_{{\#}})}\!{\mathcal T}_{\alpha},
\end{eqnarray*}
where $\#$ appearing in the subscript refers to $\mathsf{10}$, $\mathsf{120}$, $\mathsf{126}$, $\overline{\mathsf{126}}$, $\mathsf{210}$ fields of $\mathsf{SO(10)}$ and we have defined
\begin{equation}
		\begin{split}
			{H}^{(\#)a}&\equiv{}^{({5}_{{\#}})}\!{\mathcal D}^{a};\\
			{H}^{(\#)\alpha}&\equiv{}^{({5}_{{\#}})}\!{\mathcal T}^{\alpha};\\
		\end{split}
		\qquad
		\begin{split}
			{H}^{(\#)}_a&\equiv{}^{(\overline{5}_{{\#}})}\!{\mathcal D}_{a};\\
			{H}^{(\#)}_{\alpha}&\equiv{}^{(\overline{5}_{{\#}})}\!{\mathcal T}_{\alpha};\\
		\end{split}
	\end{equation}
Here, ${\mathcal D}$'s and ${\mathcal T}$'s represent unnormalized SM doublet and triplet fields. The kinetic energy of the  $\mathsf{5}-$ and $\mathsf{5}-$ plets are given by
\begin{eqnarray*}
-\partial_A{H}^{(\#)i}~\partial^A{H}^{(\#)i\dagger}&=&
-\left[\partial_A{}^{({5}_{{\#}})}\!{\mathsf D}^{a}~\partial^A{}^{({5}_{{\#}})}\!{\mathsf D}^{a\dagger}+\partial_A{}^{({5}_{{\#}})}\!{\mathsf T}^{\alpha}~\partial^A{}^{({5}_{{\#}})}\!{\mathsf T}^{\alpha\dagger}\right],\\
-\partial_A{H}^{(\#)}_i~\partial^A{H}^{(\#)\dagger}_i&=&
-\left[\partial_A{}^{(\overline{5}_{{\#}})}\!{\mathsf D}_{a}~\partial^A{}^{(\overline{5}_{{\#}})}\!{\mathsf D}^{\dagger}_a+\partial_A{}^{({5}_{{\#}})}\!{\mathsf T}_{\alpha}~\partial^A{}^{({5}_{{\#}})}\!{\mathsf T}^{\dagger}_{\alpha}\right],
\end{eqnarray*}
so that the SM fields are normalized according to
\begin{equation}
		\begin{split}
			{}^{({5}_{{\#}})}\!{\mathcal D}^{a}&={}^{({5}_{{\#}})}\!{\mathsf D}^{a};\\
			{}^{({5}_{{\#}})}\!{\mathcal T}^{\alpha}&={}^{({5}_{{\#}})}\!{\mathsf T}^{\alpha};\\
		\end{split}
		\qquad
		\begin{split}
			{}^{(\overline{5}_{{\#}})}\!{\mathcal D}_{a}&={}^{(\overline{5}_{{\#}})}\!{\mathsf D}_{a};\\
			{}^{(\overline{5}_{{\#}})}\!{\mathcal T}_{\alpha}&={}^{(\overline{5}_{{\#}})}\!{\mathsf T}_{\alpha};\\
		\end{split}
	\end{equation}\\

\item \textsc{SM doublets and triplets in $\mathsf{45}-$plet and $\overline{\mathsf{45}}-$plet of $\mathsf{SU(5)}$}\\

$\mathsf{45}-$dimensional field of $\mathsf{SU(5)}$  have the following  $\mathsf{SU(3)_C\times SU(2)_L \times U(1)_Y}$ decomposition \cite{Slansky:1981yr}
     \begin{eqnarray*}
 {{H}^{(\#)ij}_{k}}(\mathsf{45})&=&(\mathsf{1,2},3)~{{}^{({45}_{\#})}\!{\mathcal D}^{a}}+(\mathsf{3,1},-2)~{{}^{({45}_{\#})}\!{\mathcal T}^{\alpha}}+(\mathsf{3,3},-2){\mathcal W}_{b}^{a\alpha}+(\mathsf{\overline{3},1},8){\mathcal W}_{\alpha}\\
 &&~+(\mathsf{\overline{3},2},-7){\mathcal W}_{a\alpha}
+(\mathsf{\overline{6},1},-2){\mathcal W}^{\alpha\beta}_{\gamma}+(\mathsf{{8},2},3){\mathcal W}^{\alpha a}_{\beta},
\end{eqnarray*}
where $\#$ refers to $\mathsf{120}$ and $\mathsf{126}$ fields of $\mathsf{SO(10)}$. The traceless condition on the $\mathsf{SU(5)}$ tensor $\mathsf{H}^{(\#)ij}_{k}$
leads to the definitions
\begin{equation}
{{H}^{(\#)\beta a}_{\beta}}=-{{H}^{(\#)b a}_{b}}
\equiv {{}^{({45}_{\#})}\!{\mathcal D}^{a}};\qquad{{H}^{(\#)\beta \alpha}_{\beta}}=-{{H}^{(\#)b \alpha}_{b}}
\equiv {{}^{({45}_{\#})}\!{\mathcal T}^{\alpha}}.
\end{equation}
We now express all the reducible tensors of the $\mathsf{45}$-plet in terms of the irreducible ones as follows:
\begin{eqnarray}\label{su(5)'s 45 in irred SM states}
\begin{split}
{{H}^{(\#)a\alpha}_{b}}={\mathcal W}_{b}^{a\alpha}-\frac{1}{2}\delta^a_b~{{}^{({45}_{\#})}\!{\mathcal T}^{\alpha}};&\quad{{H}^{(\#)\alpha a}_{\beta}}=
{\mathcal W}_{\beta}^{\alpha a}+\frac{1}{3}\delta^{\alpha}_{\beta}~{{}^{({45}_{\#})}\!{\mathcal D}^{a}};\\
{{H}^{(\#)ab}_{\alpha}}=\epsilon^{ab}{\mathcal W}_{\alpha};&\quad{\mathsf{H}^{(\#)\alpha\beta}_{a}}=
\epsilon^{\alpha\beta\gamma}{\mathcal W}_{a\gamma};\\
{{H}^{(\#)ab}_{c}}=\delta^{b}_{c}~{{}^{({45}_{\#})}\!{\mathcal D}^{a}}-\delta^{a}_{c}~{{}^{({45}_{\#})}\!{\mathcal D}^{b}};&\quad
{{H}^{(\#)\alpha\beta}_{\gamma}}={\mathcal W}_{\gamma}^{\alpha\beta}+\frac{1}{2}\left[\delta^{\alpha}_{\gamma}~{{}^{({45}_{\#})}\!{\mathcal T}^{\beta}}-    \delta^{\beta}_{\gamma}{{}^{({45}_{\#})}\!{\mathcal T}^{\alpha}}\right].
\end{split}
\end{eqnarray}

The kinetic energy of the $\mathsf{45}$-plet is given by
\begin{eqnarray*}
-\partial_A{{H}^{(\#)ij}_{k}}\partial^A{{H}^{(\#)ij\dagger}_{k}}&=&
-\left[\partial_A{{}^{({45}_{\#})}\!{\mathsf D}^{a}}\partial^A{{}^{({45}_{\#})}\!{\mathsf D}^{a\dagger}}+\partial_A{{}^{({45}_{\#})}\!{\mathsf T}^{\alpha}}\partial^A{{}^{({45}_{\#})}\!{\mathsf T}^{\alpha\dagger}}\right.\\
&&\left.~~~~+\partial_A{\mathsf W}_{b}^{a\alpha}\partial^A{\mathsf W}_{b}^{a\alpha\dagger}+\partial_A{\mathsf W}_{\alpha}\partial^A{\mathsf W}_{\alpha}^{\dagger} +\partial_A{\mathsf W}_{a\alpha}\partial^A{\mathsf W}_{a\alpha}^{\dagger}\right.\\
&&\left.~~~~+\partial_A{\mathsf W}^{\alpha\beta}_{\gamma}\partial^A{\mathsf W}^{\alpha\beta\dagger}_{\gamma}+\partial_A{\mathsf W}^{\alpha a}_{\beta}\partial^A{\mathsf W}^{\alpha a\dagger}_{\beta}\right],
\end{eqnarray*}
 so that the $\mathsf{SU(3)_C\times SU(2)_L \times U(1)_Y}$ fields are normalized according to
\begin{eqnarray}\label{normailzed irred SM states in su(5)'s 45}
\begin{split}
 {{}^{({45}_{\#})}\!{\mathcal D}^{a}}=\frac{1}{2}\sqrt{\frac{3}{2}}~{{}^{({45}_{\#})}\!{\mathsf D}^{a}};&\qquad{{}^{({45}_{\#})}\!{\mathcal T}^{\alpha}}=\frac{1}{\sqrt{2}}~{{}^{({45}_{\#})}\!{\mathsf T}^{\alpha}};\\
 {\mathcal W}_{\alpha}=\frac{1}{\sqrt{2}} {\mathsf W}_{\alpha};&\qquad{\mathcal W}_{a\alpha}=\frac{1}{\sqrt{6}}{\mathsf W}_{a\alpha};\\
 {\mathcal W}_{\beta}^{\alpha a}=\frac{1}{\sqrt{2}}{\mathsf W}_{\beta}^{\alpha a};&\qquad
 {\mathcal W}_{b}^{a\alpha}=\frac{1}{\sqrt{2}}{\mathsf W}_{b}^{a\alpha};\\
 {\mathcal W}_{\gamma}^{\alpha\beta}&=\frac{1}{\sqrt{2}}{\mathsf W}_{\gamma}^{\alpha\beta}.
 \end{split}
 \end{eqnarray}
 One can now easily extend Eqs.(\ref{su(5)'s 45 in irred SM states}) and (\ref{normailzed irred SM states in su(5)'s 45}) to $\mathsf{SU(5)}$'s $\mathsf{\overline{45}}$. field contained in $\mathsf{120}-$ and $\mathsf{\overline{126}}-$plets:
\begin{eqnarray}\label{normailzed irred SM states in su(5)'s overline{45}}
\begin{split}
{{H}^{(\#)b}_{a\alpha}}={\mathcal W}_{a\alpha}^{b}-\frac{1}{2}\delta^b_b~{{}^{({45}_{\#})}\!{\mathcal T}_{\alpha}};&\quad{{H}^{(\#)\beta}_{\alpha a}}=
{\mathcal W}_{\alpha a}^{\beta}+\frac{1}{3}\delta^{\beta}_{\alpha}~{{}^{(\overline{45}_{\#})}\!{\mathcal D}_{a}};\\
{{H}^{(\#)\alpha}_{ab}}=\epsilon_{ab}{\mathcal W}^{\alpha};&\quad{\mathsf{H}^{(\#)a}_{\alpha\beta}}=
\epsilon_{\alpha\beta\gamma}{\mathcal W}^{a\gamma};\\
{{H}^{(\#)c}_{ab}}=\delta^{c}_{b}~{{}^{(\overline{45}_{\#})}\!{\mathcal D}_{a}}-\delta^{c}_{a}~{{}^{(\overline{45}_{\#})}\!{\mathcal D}_{b}};&\quad
{{H}^{(\#)\gamma}_{\alpha\beta}}={\mathcal W}^{\gamma}_{\alpha\beta}+\frac{1}{2}\left[\delta_{\alpha}^{\gamma}~{{}^{(\overline{45}_{\#})}\!{\mathcal T}_{\beta}}-    \delta_{\beta}^{\gamma}{{}^{(\overline{45}_{\#})}\!{\mathcal T}_{\alpha}}\right].\\
\\
 {{}^{(\overline{45}_{\#})}\!{\mathcal D}_{a}}=\frac{1}{2}\sqrt{\frac{3}{2}}~{{}^{(\overline{45}_{\#})}\!{\mathsf D}_{a}};&\qquad{{}^{(\overline{45}_{\#})}\!{\mathcal T}_{\alpha}}=\frac{1}{\sqrt{2}}~{{}^{(\overline{45}_{\#})}\!{\mathsf T}_{\alpha}};\\
 {\mathcal W}^{\alpha}=\frac{1}{\sqrt{2}} {\mathsf W}^{\alpha};&\qquad{\mathcal W}^{a\alpha}=\frac{1}{\sqrt{6}}{\mathsf W}^{a\alpha};\\
 {\mathcal W}^{\beta}_{\alpha a}=\frac{1}{\sqrt{2}}{\mathsf W}^{\beta}_{\alpha a};&\qquad
 {\mathcal W}^{b}_{a\alpha}=\frac{1}{\sqrt{2}}{\mathsf W}^{b}_{a\alpha};\\
 {\mathcal W}^{\gamma}_{\alpha\beta}&=\frac{1}{\sqrt{2}}{\mathsf W}^{\gamma}_{\alpha\beta}.
 \end{split}
\end{eqnarray}
 In Eq.(\ref{normailzed irred SM states in su(5)'s overline{45}}), $\#$ refers to $\mathsf{120}-$ and $\mathsf{\overline{126}}-$plets.\\

\item \textsc{SM triplets in $\mathsf{50}-$plet and $\overline{\mathsf{50}}-$plet of $\mathsf{SU(5)}$}\\

$\mathsf{50}-$dimensional  field of $\mathsf{SU(5)}$  have the following  $\mathsf{SU(3)_C\times SU(2)_L \times U(1)_Y}$ decomposition \cite{Slansky:1981yr}
 \begin{eqnarray*}
{H}^{(\overline{126})ijk}_{lm}(\mathsf{50})&=&(\mathsf{1,1},-12){\mathcal X}+(\mathsf{3,1},-2)~{}^{({50}_{\overline{126}})}\!{\mathcal T}^{\alpha}+(\mathsf{\bar{3},2},-7){\mathcal X}^{\alpha\beta}_a
+(\mathsf{\bar{6},3},-2){\mathcal X}^{\alpha\beta a}_{\gamma b}\\
&&~+
(\mathsf{6,1},8){\mathcal X}^{\alpha}_{\beta\gamma}+(\mathsf{8,2},3){\mathcal  X}^{\alpha a}_{\beta},
 \end{eqnarray*}
where we have defined
\begin{equation}
		\begin{split}
			{H}^{(\overline{126})ab \gamma}_{ab}={H}^{(\overline{126})\alpha\beta \gamma}_{\alpha\beta}
&=-{H}^{(\overline{126})a\alpha \gamma}_{a\alpha}\equiv{}^{({50}_{\overline{126}})}\!{\mathcal T}^{\alpha};\\
			{H}^{(\overline{126})\gamma \alpha\beta}_{\gamma a}&\equiv{\mathcal X}^{\alpha\beta}_{a};\\
\overline{\Delta}^{\gamma\alpha a}_{\gamma \beta}&\equiv{\mathcal  X}^{\alpha a}_{\beta}.
		\end{split}
	\end{equation}
The first relationship above  follows from the traceless condition on the $\mathsf{SU(5)}$ irreducible tensor ${H}^{(\overline{126})ijk}_{lm}$:
\[
{H}^{(\overline{126})a\alpha i}_{a\alpha}=-\frac{1}{2}\left[{H}^{(\overline{126})\alpha\beta i}_{\alpha\beta}+{H}^{(\overline{126})abi}_{ab}\right].
\]
We now express the reducible tensors of the $\mathsf{50}-$plet in terms of the SM irreducible ones as follows:
\begin{eqnarray}
\begin{split}
{H}^{(\overline{126})\alpha\beta a}_{\gamma\sigma}=\delta^{\alpha}_{\gamma}{\mathcal X}^{\beta a}_{\sigma}-
\delta^{\alpha}_{\sigma}{\mathcal X}^{\beta a}_{\gamma}+\delta^{\beta}_{\sigma}{\mathcal X}^{\alpha a}_{\gamma}
-\delta^{\beta}_{\gamma}{\mathcal X}^{\alpha a}_{\sigma};&\quad{H}^{(\overline{126})\alpha\beta\gamma}_{ab}=\epsilon^{\alpha\beta\gamma}\epsilon_{ab}{\mathcal X}\\
{H}^{(\overline{126})\alpha\beta\gamma}_{\sigma a}=\delta^{\gamma}_{\sigma}{\mathcal X}^{\alpha\beta}_{a}-
\delta^{\beta}_{\sigma}{\mathcal X}^{\alpha\gamma}_{a}+\delta^{\alpha}_{\sigma}{\mathcal X}^{\beta \gamma}_{a}
;&\quad{H}^{(\overline{126})\alpha ab}_{\beta\gamma}=\epsilon^{ab}{\mathcal X}^{\alpha}_{\beta\gamma};\\
{H}^{(\overline{126})\alpha\beta a}_{\gamma b}={\mathcal X}^{\alpha\beta a}_{\gamma b}+\frac{1}{4}\delta_b^a\left[
\delta^{\alpha}_{\gamma}{}^{({50}_{\overline{126}})}\!{\mathcal T}^{\beta}-\delta^{\beta}_{\gamma}{}^{({50}_{\overline{126}})}\!{\mathcal T}^{\alpha}\right];&\quad
{H}^{(\overline{126})ab \alpha}_{c\beta}=\delta^{a}_{c}{\mathcal X}^{\alpha b}_{\beta}-\delta^{b}_{c}{\mathcal X}^{\alpha a}_{\beta}\\
{H}^{(\overline{126})ab \alpha}_{cd}=\frac{1}{2}\left[\delta_c^a\delta_d^b-\delta_d^a\delta_c^b\right]{}^{({50}_{\overline{126}})}\!{\mathcal T}^{\alpha}
;&\quad\mathsf{H}^{(\overline{126})\alpha\beta a}_{bc}=\delta^{a}_{c}{\mathcal X}^{\alpha \beta}_b-\delta^{a}_{b}{\mathcal X}^{\alpha\beta}_{c}.
\end{split}
\end{eqnarray}
\begin{eqnarray}
{H}^{(\overline{126})\alpha\beta\gamma}_{\rho\sigma}&=&\frac{1}{2}\left[\delta^{\alpha}_{\rho}\delta^{\beta}_{\sigma}~{}^{({50}_{\overline{126}})}\!{\mathcal T}^{\gamma}-
\delta^{\beta}_{\rho}\delta^{\alpha}_{\sigma}~{}^{({50}_{\overline{126}})}\!{\mathcal T}^{\gamma}-
\delta^{\alpha}_{\rho}\delta^{\gamma}_{\sigma}~{}^{({50}_{\overline{126}})}\!{\mathcal T}^{\beta}+
\delta^{\gamma}_{\rho}\delta^{\alpha}_{\sigma}~{}^{({50}_{\overline{126}})}\!{\mathcal T}^{\beta}\right.\nonumber\\
&&\left.+\delta^{\beta}_{\rho}\delta^{\gamma}_{\sigma}~{}^{({50}_{\overline{126}})}\!{\mathcal T}^{\alpha}-
\delta^{\gamma}_{\rho}\delta^{\beta}_{\sigma}~{}^{({50}_{\overline{126}})}\!{\mathcal T}^{\alpha}
\right].
\end{eqnarray}

The kinetic energy of the $\mathsf{50}$-plet is given by
\begin{eqnarray*}
-\partial_A{H}^{(\overline{126})ijk}_{lm}\partial^A{H}^{(\overline{126})ijk\dagger}_{lm}&=&
-\left[\partial_A{\mathsf X}\partial^A{\mathsf X}^{\dagger}+\partial_A~{}^{({50}_{\overline{126}})}\!{\mathsf T}^{\alpha}
\partial^A~{}^{({50}_{\overline{126}})}\!{\mathsf T}^{\alpha\dagger}+\frac{1}{2!}\partial_A{\mathsf X}^{\alpha\beta}_{a}
\partial^A{\mathsf X}^{\alpha\beta\dagger}_{a}\right.\nonumber\\
&&\left.~~~~+\frac{1}{2!}\partial_A{\mathsf X}^{\alpha a}_{\beta}
\partial^A{\mathsf X}^{\alpha a\dagger}_{\beta}+\frac{1}{3!}\frac{1}{2!}\partial_A{\mathsf X}^{\alpha\beta a}_{\gamma b}
\partial^A{\mathsf X}^{\alpha\beta a\dagger}_{\gamma b}\right.\nonumber\\
&&\left.~~~~+\frac{1}{2!}\partial_A{\mathsf X}^{\alpha}_{\beta\gamma}
\partial^A{\mathsf X}^{\alpha\dagger}_{\beta\gamma}\right],
\end{eqnarray*}
so that the SM fields are normalized according to
\begin{eqnarray}
\begin{split}
 {\mathcal X}=\frac{1}{2\sqrt{3}}{\mathsf X};&\qquad{}^{({50}_{\overline{126}})}\!{\mathcal T}^{\alpha}
=\frac{1}{{3}}~{}^{({50}_{\overline{126}})}\!{\mathsf T}^{\alpha}\\
{\mathcal X}^{\alpha\beta}_{a}=\frac{1}{2\sqrt{6}}{\mathsf X}^{\alpha\beta}_{a};&\qquad
 {\mathcal X}^{\alpha a}_{\beta}=\frac{1}{4\sqrt{3}}{\mathsf X}^{\alpha a}_{\beta};\\
  {\mathcal X}^{\alpha\beta a}_{\gamma b}=\frac{1}{6\sqrt{2}}{\mathsf X}^{\alpha\beta a}_{\gamma b};&\qquad{\mathcal X}^{\alpha}_{\beta\gamma}=\frac{1}{2
 \sqrt{3}}{\mathsf X}^{\alpha}_{\beta\gamma}.
\end{split}
 \end{eqnarray}
One can now extend the above results to $\mathsf{\overline{50}}$ of $\mathsf{SU(5)}$ contained in ${\mathsf{126}}$ plet.\\

\item \textsc{SM singlet in $\mathsf{24}-$plet of $\mathsf{SU(5)}$}\\

$\mathsf{24}-$dimensional field of $\mathsf{SU(5)}$  have the following  $\mathsf{SU(3)_C\times SU(2)_L \times U(1)_Y}$ decomposition \cite{Slansky:1981yr}
\[
{H}^{(\#)i}_j(\mathsf{24})=(\mathsf{1,1},0)\mathcal S_{24_{_{\#}}}+(\mathsf{1,3},0){\mathcal Y}_{b}^{a}+(\mathsf{8,1},0){\mathcal Y}_{\beta}^{\alpha}+[(\mathsf{3,2},-5){\mathcal Y}^{\alpha}_a + c.c.],
\]
where $\#$ refers to $\mathsf{45}$ and $\mathsf{210}$ fields of $\mathsf{SO(10)}$. The tracelessness condition on the tensor ${H}^{(\#)i}_{j}$ gives the following definition
\begin{equation}
{H}^{(\#)\alpha}_{\alpha}=-{H}^{(\#)a}_{a}
\equiv \mathcal S_{24_{_{210}}}.
\end{equation}
The reducible tensors of the  $\mathsf{24}$-plet can be expressed in terms of the irreducible ones as follows:
\begin{eqnarray}
{H}^{(\#)a}_{b}={\mathcal Y}_{b}^{a}-\frac{1}{2}\delta^a_b\mathcal S_{24_{_{\#}}};~~~&&~~~{H}^{(\#)\alpha}_{\beta}=
{\mathcal Y}_{\beta}^{\alpha}+\frac{1}{3}\delta^{\alpha}_{\beta}\mathcal S_{24_{_{\#}}}.
\end{eqnarray}

The kinetic energy of the  $\mathsf{24}$-plet is given by
\begin{eqnarray*}
-\partial_A{H}^{(\#)i}_{j}\partial^A{H}^{(\#)i\dagger}_{j}&=&-\left[\partial_A\mathsf S_{24_{_{\#}}}\partial^A\mathsf S_{24_{_{\#}}}^{\dagger}+\partial_A{\mathsf Y}_{\beta}^{\alpha}
\partial^A{\mathsf Y}_{\beta}^{\alpha\dagger}+\partial_A{\mathsf Y}^{a}_{b}
\partial^A{\mathsf Y}^{a\dagger}_{b}\right.\nonumber\\
&&\left.~~~~+\partial_A{\mathsf Y}^{\alpha}_{a}
\partial^A{\mathsf Y}^{\alpha\dagger}_{a}+\partial_A{\mathsf Y}^{a}_{\alpha}\partial^A{\mathsf Y}_{\alpha}^{a\dagger}\right],
\end{eqnarray*}
 so that the SM fields are normalized according to
 \begin{eqnarray}
 \begin{split}
 \mathcal S_{24_{_{\#}}}=\sqrt{\frac{6}{5}}\mathsf S_{24_{_{\#}}};&\quad{\mathcal Y}_{\beta}^{\alpha}={\mathsf  Y}_{\beta}^{\alpha};\quad
{\mathcal Y}_{b}^{a}={\mathsf Y }^{a}_{b};\\
 {\mathcal Y}^{\alpha}_{a}= {\mathsf  Y}^{\alpha}_{a};&\quad{\mathcal Y}^{a}_{\alpha}={\mathsf  Y}^{a}_{\alpha}.
 \end{split}
 \end{eqnarray}\\

 \item \textsc{SM singlet in $\mathsf{75}-$plet of $\mathsf{SU(5)}$}\\

$\mathsf{75}-$dimensional  field of $\mathsf{SU(5)}$  have the following  $\mathsf{SU(3)_C\times SU(2)_L \times U(1)_Y}$ decomposition \cite{Slansky:1981yr}
\begin{eqnarray}
{H}^{(210)ij}_{kl}(\mathsf{75})&=&(\mathsf{1,1},0){\mathcal S_{75_{_{210}}}}+(\mathsf{8,1},0){\mathcal Z}_{\beta}^{\alpha}+(\mathsf{8,3},0){\mathcal Z}_{\beta b}^{\alpha a}\nonumber\\
&&~+[(\mathsf{3,2},-5){\mathcal Z}^{\alpha}_a+(\mathsf{\bar{6},2},-5){\mathcal Z}_{\gamma a}^{\alpha\beta}+(\mathsf{\bar{3},1},-10){\mathcal Z}_{\alpha}\nonumber\\
&&~+c.c.],
  \end{eqnarray}
where we have defined
\begin{equation}
		\begin{split}
{H}^{(210)ab}_{ab}={H}^{(210)\alpha\beta}_{\alpha\beta}
&=-{H}^{(210)\alpha a}_{\alpha a}\equiv \mathcal S_{75_{_{210}}};\\
{H}^{(210)\alpha b}_{a b}&\equiv {\mathbf X}_{a}^{\alpha};\\
{\mathcal Z}_{\beta}^{\alpha}\equiv {H}^{(210)\alpha a}_{\beta a}&+\frac{1}{3}\delta^{\alpha}_{\beta}\mathcal S_{75_{_{210}}}.
\end{split}
	\end{equation}
Again the first relationship above follows from the double tracelessness condition on the tensor ${H}^{(210)ij}_{kl}$:
\begin{equation*}
{H}^{(210)\alpha a}_{\alpha a}=-\frac{1}{2}\left({H}^{(210)\alpha\beta}_{\alpha\beta}+{H}^{(210)ab}_{ab}\right).
\end{equation*}
The reducible tensors of the $\mathsf{75}-$plet can be expressed in terms of the irreducible SM states as follows:
 \begin{eqnarray}
 \begin{split}
{H}^{(210)ab}_{cd}&=\frac{1}{2}\left(\delta^a_c\delta^b_d-\delta^a_d\delta^b_c\right)\mathcal S_{75_{_{210}}};\\
{H}^{(210)\alpha \beta}_{\gamma \sigma}&=
\frac{1}{2}\left(\delta^{\alpha}_{\sigma}{\mathcal Z}_{\gamma}^{\beta}-\delta^{\alpha}_{\gamma}{\mathcal Z}_{\sigma}^{\beta}\right)
+\frac{1}{6}\left(\delta^{\alpha}_{\gamma}\delta^{\beta}_{\sigma}-\delta^{\alpha}_{\sigma}\delta^{\beta}_{\gamma}\right)\mathcal S_{75_{_{210}}}\\
{H}^{(210)\alpha a}_{\beta b}&={\mathcal Z}_{\beta b}^{\alpha a}+\frac{1}{2}\delta^a_b{\mathcal Z}_{\beta }^{\alpha}-\frac{1}{6}\delta^a_b\delta^{\alpha}_{\beta}\mathcal S_{75_{_{210}}};\\
{H}^{(210)\alpha\beta}_{\gamma a}&={\mathcal Z}_{\gamma a}^{\alpha \beta}-\frac{1}{2}\left(\delta^{\alpha}_{\gamma}{\mathcal Z}^{\beta}_a-\delta^{\beta}_{\gamma}{\mathcal Z}^{\alpha}_a\right);\\
{H}^{(210)\alpha \beta}_{ab}&=\epsilon_{ab}\epsilon^{\alpha\beta\gamma}{\mathcal Z}_{\gamma};\\
{H}^{(210)a\alpha}_{bc}&=\delta^a_b{\mathcal Z}^{\alpha}_c-
\delta^a_c{\mathcal Z}^{\alpha}_b.
 \end{split}
 \end{eqnarray}
  The kinetic energy of the $\mathsf{75}-$plet is given by
\begin{eqnarray*}
-\partial_A{H}^{(210)ij}_{kl}\partial^A{H}^{(210)ij\dagger}_{kl}&=&-\left[\partial_A\mathsf S_{75_{_{210}}}\partial^A\mathsf S_{75_{_{210}}}^{\dagger}+\partial_A{\mathsf Z}_{\alpha}
\partial^A{\mathsf Z}_{\alpha}^{\dagger}+\partial_A{\mathsf Z}^{\alpha}
\partial^A{\mathsf Z}^{\alpha\dagger}\right.\\
&&\left.~~~~+\partial_A{\mathsf Z}^{\alpha}_{\beta}
\partial^A{\mathsf Z}^{\alpha\dagger}_{\beta}+\partial_A{\mathsf Z}^{\alpha}_{a}
\partial^A{\mathsf Z}^{\alpha\dagger}_{a}+\partial_A{\mathsf Z}_{\alpha}^{a}
\partial^A{\mathsf Z}^{a\dagger}_{\alpha}+
\right.\\
&&\left.~~~~+\frac{1}{2!}\frac{1}{2!}\partial_A{\mathsf Z}^{\alpha\beta}_{\gamma a}
\partial^A{\mathsf Z}^{\alpha\beta\dagger}_{\gamma a}+\frac{1}{2!}\frac{1}{2!}\partial_A{\mathsf Z}_{\alpha\beta}^{\gamma a}
\partial^A{\mathsf Z}_{\alpha\beta}^{\gamma a\dagger}\right.\\
&&\left.~~~~+\frac{1}{2!}\frac{1}{2!}\partial_A{\mathsf Z}^{\alpha a}_{\beta b}
\partial^A{\mathsf Z}^{\alpha a\dagger}_{\beta b}\right],
\end{eqnarray*}
 so that the SM fields are normalized according to
  \begin{eqnarray}
 \begin{split}
 \mathcal S_{75_{_{210}}}&=\frac{1}{\sqrt{2}}\mathsf S_{75_{_{210}}};\\
{\mathcal Z}_{\alpha}&=\frac{1}{{2}}{\mathsf Z}_{\alpha};\\
{\mathcal Z}^{\alpha}&=\frac{1}{{2}}{\mathsf Z}^{\alpha};
 \end{split}
 \quad
  \begin{split}
  {\mathcal Z}^{\alpha}_{\beta}&=\frac{1}{\sqrt{3}}{\mathsf Z}^{\alpha}_{\beta};\\
 {\mathcal Z}^{\alpha}_{a}&=\frac{1}{\sqrt{6}}{\mathsf Z}^{\alpha}_{a};\\
 {\mathcal Z}_{\alpha}^{a}&=\frac{1}{\sqrt{6}}{\mathsf Z}_{\alpha}^{a};
  \end{split}
   \quad
    \begin{split}
 {\mathcal Z}^{\alpha\beta}_{\gamma a}&=\frac{1}{2\sqrt{2}}{\mathsf Z}^{\alpha\beta}_{\gamma a},\\
 {\mathcal Z}_{\alpha\beta}^{\gamma a}&=\frac{1}{2\sqrt{2}}{\mathsf Z}_{\alpha\beta}^{\gamma a},\\
{\mathcal Z}^{\alpha a}_{\beta b}&=\frac{1}{{4}}{\mathsf Z}^{\alpha a}_{\beta b}.
  \end{split}
  \end{eqnarray}

\end{itemize}

\end{appendices}

    \end{document}